\documentclass[prd,aps,10pt,nofootinbib,twocolumn,superscriptaddress,preprintnumbers,balancelastpage,longbibliography,floatfix]{revtex4-2}

\usepackage{amsmath,amssymb}	
\usepackage{mathtools}
\usepackage{fontawesome}
\usepackage[dvipsnames]{xcolor}
\usepackage{hyperref}
\usepackage{xspace}
\usepackage{fancyhdr}
\usepackage{braket}
\usepackage{graphicx}
\usepackage{blindtext}
\usepackage{nicefrac}
\usepackage{lipsum}
\usepackage{bbold}
\usepackage{tabularx}
\usepackage{multirow}
\usepackage{dcolumn}
\usepackage{makecell}
\usepackage{floatrow}
\usepackage{afterpage}
\usepackage{longtable}
\newcommand{\Fermi}{\emph{Fermi}\xspace}
\newcommand{\HEALPix}{\texttt{HEALPix}\xspace}
\newcommand{\dd}{\mathrm{d}}

\definecolor{deepgreen}{rgb}{0.2,0.8,0.2}

\colorlet{linkcolor}{BrickRed}

\hypersetup{colorlinks=true,
linkcolor=linkcolor,
citecolor=linkcolor,
urlcolor=linkcolor,
,linktocpage=true
,pdfproducer=medialab}

\newcommand\Tstrut{\rule{0pt}{2.6ex}}         
\newcommand\Bstrut{\rule[-1.6ex]{0pt}{0pt}}   

\newcolumntype{C}[1]{>{\centering\let\newline\\\arraybackslash\hspace{0pt}}m{#1}}

\begin{document}

\preprint{\hfill MIT-CTP/5337}

\title{A neural simulation-based inference approach for characterizing \\ the Galactic Center $\gamma$-ray excess}

\author{Siddharth Mishra-Sharma}
\email{sm8383@nyu.edu}
\thanks{ORCID: \href{https://orcid.org/0000-0001-9088-7845}{0000-0001-9088-7845}}
\affiliation{Center for Theoretical Physics, Massachusetts Institute of Technology, Cambridge, MA 02139, USA}
\affiliation{The NSF AI Institute for Artificial Intelligence and Fundamental Interactions}
\affiliation{Department of Physics, Massachusetts Institute of Technology, Cambridge, MA 02139, USA}
\affiliation{Department of Physics, Harvard University, Cambridge, MA 02138, USA}
\affiliation{Center for Cosmology and Particle Physics, Department of Physics, New York University, New York, NY 10003, USA}

\author{Kyle Cranmer}
\email{kyle.cranmer@nyu.edu}
\thanks{ORCID: \href{https://orcid.org/0000-0002-5769-7094}{0000-0002-5769-7094}}
\affiliation{Center for Cosmology and Particle Physics, Department of Physics, New York University, New York, NY 10003, USA}
\affiliation{Center for Data Science, New York University, 60 Fifth Ave, New York, NY 10011, USA}

\date{\today}

\begin{abstract}
The nature of the \Fermi $\gamma$-ray Galactic Center Excess (GCE) has remained a persistent mystery for over a decade. Although the excess is broadly compatible with emission expected due to dark matter annihilation, an explanation in terms of a population of unresolved astrophysical point sources \emph{e.g.}, millisecond pulsars, remains viable. The effort to uncover the origin of the GCE is hampered in particular by an incomplete understanding of diffuse emission of Galactic origin. This can lead to spurious features that make it difficult to robustly differentiate smooth emission, as expected for a dark matter origin, from more ``clumpy'' emission expected from a population of relatively bright, unresolved point sources. We use recent advancements in the field of simulation-based inference, in particular density estimation techniques using normalizing flows, in order to characterize the contribution of modeled components, including unresolved point source populations, to the GCE. Compared to traditional techniques based on the statistical distribution of photon counts, our machine learning-based method is able to utilize more of the information contained in a given model of the Galactic Center emission, and in particular can perform posterior parameter estimation while accounting for pixel-to-pixel spatial correlations in the $\gamma$-ray map. This makes the method demonstrably more resilient to certain forms of model misspecification. On application to \Fermi data, the method generically attributes a smaller fraction of the GCE flux to unresolved point sources when compared to traditional approaches. We nevertheless infer such a contribution to make up a non-negligible fraction of the GCE across all analysis variations considered, with at least $38^{+9}_{-19}\%$ of the excess attributed to unresolved point sources in our baseline analysis.
\end{abstract}

\maketitle

\tableofcontents

\section{Introduction}
\label{sec:intro}

Dark matter (DM) represents one of the major unsolved problems in particle physics and cosmology today. The traditional Weakly-Interacting Massive Particle (WIMP) paradigm envisions production of dark matter in the early Universe through freeze-out of dark sector particles weakly coupled to the Standard Model (SM) sector. In this scenario, one of the most promising avenues of detecting a dark matter signal is through an observation of excess $\gamma$-ray photons at $\sim\mathrm{GeV}$ energies from DM-rich regions of the sky produced through the cascade of SM particles resulting from DM self-annihilation. 

The \Fermi $\gamma$-ray Galactic Center Excess (GCE), first identified over a decade ago using data from the \Fermi Large Area Telescope (LAT)~\cite{Atwood:2009ez}, is an excess of photons in the Galactic Center with properties---such as energy spectrum and spatial morphology---broadly compatible with the expectation due to annihilating DM~\cite{Goodenough:2009gk,Hooper:2010mq,Boyarsky:2010dr,Hooper:2011ti,Abazajian:2012pn,Hooper:2013rwa,Gordon:2013vta,Abazajian:2014fta,Daylan:2014rsa,Calore:2014xka,Abazajian:2014hsa,Fermi-LAT:2015sau,Linden:2016rcf,Macias:2016nev,Clark:2016mbb}. The nature of the GCE remains contentious however, as competing explanations in terms of a population of unresolved astrophysical point sources (PSs), in particular millisecond pulsars (MSPs), remaining viable~\cite{Abazajian:2014fta,Abazajian:2010zy,Hooper:2013nhl,Calore:2014oga,Cholis:2014lta,Petrovic:2014xra,Yuan:2014yda,Brandt:2015ula,Gautam:2021wqn,Ploeg:2020jeh}. Analyses of the morphology of the excess have shown it to prefer a spatial distribution correlated with baryonic structures in the Galactic Center region rather than a distribution expected due to DM annihilation~\cite{Macias:2016nev,Macias:2019omb,Bartels:2017vsx}, although these conclusions can depend on details of the modeling~\cite{DiMauro:2020rcr,DiMauro:2021raz}. Studies leveraging the statistical distribution of photon counts in the Galactic Center have shown the $\gamma$-ray data to prefer a point source origin of the excess~\cite{Lee:2015fea,Bartels:2015aea,Buschmann:2020adf,Chang:2019ars}, a conclusion corroborated using wavelet-based techniques~\cite{Bartels:2015aea}. Recent studies have, however, pointed out the potential of unknown systematics, such as the poorly understood morphology of the diffuse foreground emission and the existence of unmodeled point source populations, to affect the conclusions of these analyses~\cite{Leane:2019xiy}. Ref.~\cite{Buschmann:2020adf} showed that many of these issues can be ameliorated through the use of better diffuse foreground models, as well as by augmenting existing models with additional degrees of freedom.

The complexity associated with analyzing high-dimensional $\gamma$-ray maps---typically binned spatially using a pixelization scheme---has motivated the use of approximate likelihoods based on \emph{e.g.}, the statistics of photon counts in individual pixels~\cite{Malyshev:2011zi,Lee:2014mza,Lee:2015fea} or scale decomposition of the photon map using wavelet techniques~\cite{Bartels:2015aea,Balaji:2018rwz,McDermott:2015ydv,Zhong:2019ycb}, in order to enable computationally tractable analyses. Under certain assumptions, using such approximations can capture all of the information contained in a given spatial model of the $\gamma$-ray data. This is the case, \emph{e.g.}, for a likelihood based on the expected probability distribution of photon counts factorized across pixels when pixel-to-pixel correlations can be assumed to be negligible. When such correlations are present, however, the use of such approximations necessarily involves loss of information compared to that contained in the original $\gamma$-ray map.

Recent developments in machine learning have enabled analysis techniques that can extract more information from high-dimensional datasets, and can therefore be used to leverage more of the information contained in models of $\gamma$-ray emission. Machine learning methods have recently shown promise for analyzing $\gamma$-ray data~\cite{Caron:2021map} and specifically for understanding the nature of the \Fermi GCE~\cite{List:2020mzd,List:2021aer,Caron:2017udl}. In particular, Ref.~\cite{List:2020mzd} used a method based on Bayesian neural networks in order to infer the flux fractions associated with various modeled components in the Galactic Center region, finding the GCE to be predominantly smooth in contrast to prior analyses depending the statistics of photon counts. Ref.~\cite{List:2021aer} extended this framework, using a novel non-parametric approach~\cite{List2021} to extract the characteristics of the PS population associated with the GCE, finding a non-negligible portion of the emission to be attributable to a dim PS population. We will show the results of our analysis on \Fermi data to be qualitatively consistent with those obtained in that work. 

\begin{figure*}
\centering
\includegraphics[width=0.99\textwidth]{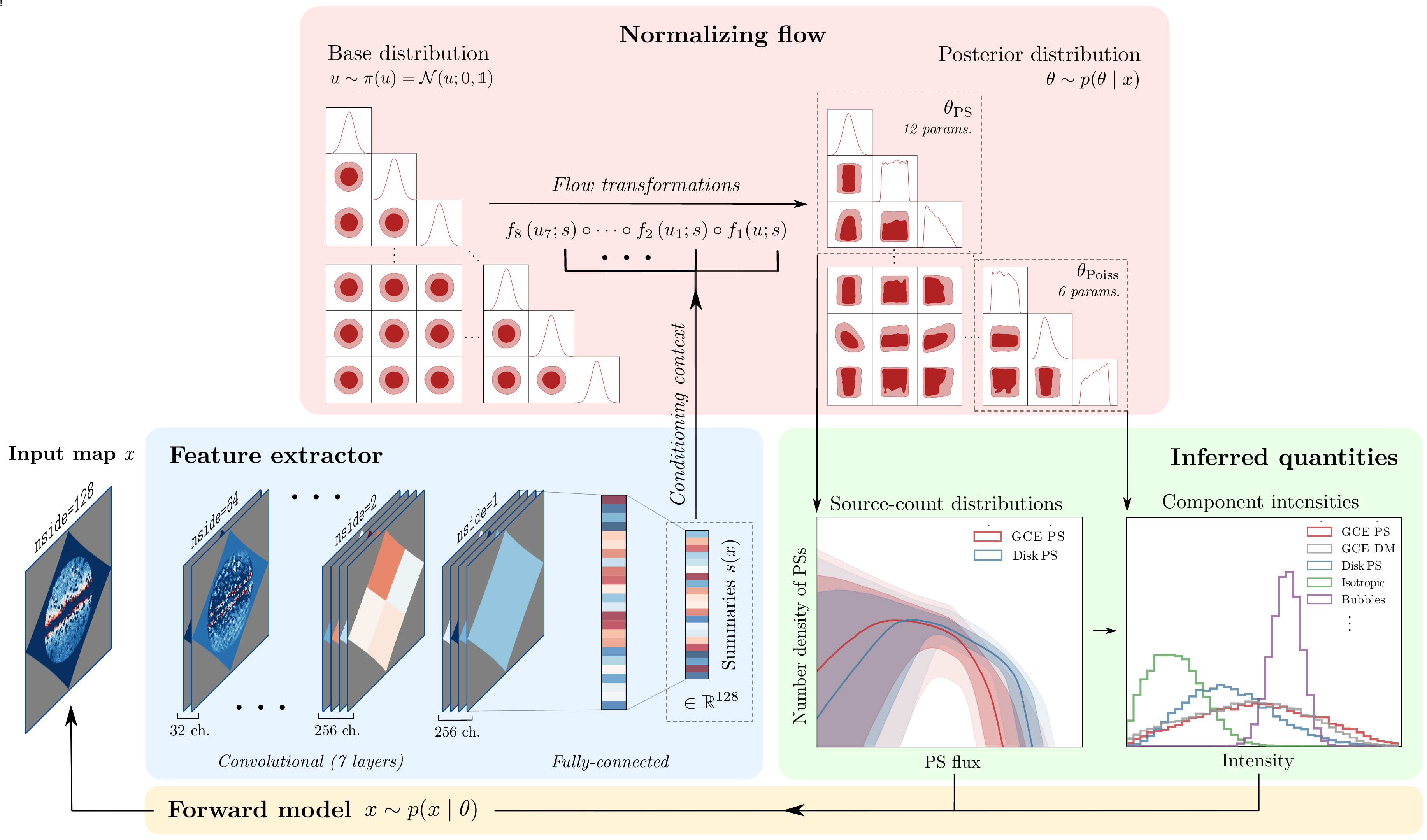}
\caption{A schematic overview of the inference framework used in this work. A normalizing flow is used to model posterior distribution of the parameters of interest characterizing the contribution of point source populations as well as diffuse (``smooth'') components to the $\gamma$-ray data. The flow transformation from the base distribution to the posterior is conditioned on learned summaries of the $\gamma$-ray map extracted using a convolutional neural network. The normalizing flow and feature-extractor neural networks are trained simultaneously using maps simulated from the forward model. Once trained, samples from the flow can be generated conditioned on a new dataset of interest in order to obtain an estimate of the corresponding parameter posteriors, which can be used to infer physical quantities of interest such as source-count distributions of modeled PS populations as well as fluxes associated with the diffuse components. See Sec.~\ref{sec:analysis} for a detailed description of the analysis pipeline.}
\label{fig:figure}
\end{figure*}

In this paper, we present a complementary approach that leverages recent developments in the field of simulation-based inference (SBI, also referred to as likelihood-free inference; see, \emph{e.g.}, Ref.~\cite{Cranmer:2019eaq} for a recent review) in order to weigh in on the nature of the GCE. In particular, we use conditional density estimation techniques based on normalizing flows~\cite{papamakarios2019normalizing,DBLP:conf/icml/RezendeM15} to characterize the contributions of various modeled components, including ``clumpy'' PS-like and ``smooth'' DM-like emission spatially tracing the GCE, to the $\gamma$-ray photon sky at $\sim\mathrm{GeV}$ energies in the Galactic Center region. Rather than using hand-crafted summary statistics, we employ a graph-based spherical convolutional neural network architecture (previously utilized in Refs.~\cite{List:2020mzd,List:2021aer}) in order to extract summaries from $\gamma$-ray maps optimized for the downstream task of estimating the distribution of parameters characterizing the contribution of modeled components to the GCE. Unlike traditional approaches based on the statistics of photon counts, this approach allows us to capture more of the information contained in a model of the Galactic Center emission, and in particular implicitly uses the distribution of pixel-to-pixel correlations as an additional discriminating handle. As we will show, this makes our method more resilient to certain systematic uncertainties compared to these approaches. A schematic illustration of our method is presented in Fig.~\ref{fig:figure}.

This paper is organized as follows. In Sec.~\ref{sec:analysis} we describe our forward model and analysis framework based on neural simulation-based inference. In Sec.~\ref{sec:simulations} we validate our pipeline on mock observations of the \Fermi GCE. Section~\ref{sec:data} presents an application of the method to \Fermi $\gamma$-ray data, including systematic variations on the analysis. In Sec.~\ref{sec:mismodeling} we study the susceptibility of the analysis to known mismodeling of the signal and background templates. We conclude in Sec.~\ref{sec:conclusion}.

\section{Methodology}
\label{sec:analysis}

We begin by describing the various ingredients of our forward model and datasets used. After a brief summary of established methods based on explicit likelihoods, we detail our analysis methodology going over, in turn, the general principles behind simulation-based inference, posterior estimation using normalizing flows, and learning representative summary statistics from high-dimensional $\gamma$-ray maps with neural networks.

\subsection{Datasets and the forward model}
\label{sec:datasets}

\noindent
\textbf{Datasets and region of interest:} We use the datasets and spatial templates from Refs.~\cite{rodd_nicholas_safdi_siddharth_2016,Mishra-Sharma:2016gis} to create simulated maps of \Fermi-LAT data in the Galactic Center region. The templates and data used correspond to 413 weeks of \Fermi-LAT Pass 8 data taken between August 4, 2008 and July 7, 2016. The top quartile of photons as graded by quality of PSF reconstruction in the energy range 2--20~GeV and event class \texttt{ULTRACLEANVETO} are used. The conventional quality cuts are applied: zenith angle less than 90$^\circ$, \texttt{LAT\_CONFIG}==1, and \texttt{DATA\_QUAL}==1.\footnote{\url{https://fermi.gsfc.nasa.gov/ssc/data/analysis/documentation/Cicerone/Cicerone_Data_Exploration/Data_preparation.html}} The maps are binned spatially using the \texttt{HEALPix}~\cite{Gorski:2004by} pixelization scheme with resolution parameter \texttt{nside}=128, roughly corresponding to pixel area $\sim 0.5\,\mathrm{deg}^2$. This dataset has been previously used in the literature for analyses based on explicit likelihoods~\cite{Buschmann:2020adf,Chang:2019ars,Leane:2019xiy} as well as machine learning-based analyses~\cite{List:2020mzd} for characterizing the GCE. All templates are normalized, per-pixel, within a region defined by $r < 30^\circ$.

The inner region of the Galactic plane, where the observed emission is especially difficult to model, is masked at $|b| < 2^\circ$, and a radial cut $r < 25^\circ$ defines the region of interest (ROI) for our analysis. Even though the GCE is spatially confined to the inner $10\mbox{--}15^\circ$ of the Galactic Center~\cite{Daylan:2014rsa,Calore:2014xka}, using a larger ROI improves the ability to constrain other spatially extended templates and helps mitigate spatial degeneracies that would otherwise crop up in a smaller ROI. On the other hand, using a ROI that is too large can exacerbate the effects of misspecified spatial templates~\cite{Chang:2018bpt}. We mask resolved PSs from the 3FGL catalog~\cite{Fermi-LAT:2015bhf} at a radius of $0.8^\circ$, approximately corresponding to 99\% PSF containment for photons in the data type employed~\cite{Fermi-LAT:2015bhf}. \\

\noindent
\textbf{Diffuse emission forward model:} The simulated data maps are a combination of diffuse (alternatively referred to as smooth or Poissonian) and PS contributions. The smooth contributions include \emph{(i)}~the Galactic diffuse foreground emission, \emph{(ii)}~spatially isotropic emission accounting for, \emph{e.g.}, uniform emission from unresolved sources of extragalactic origin, \emph{(iii)}~emission from resolved PSs included in the \Fermi 3FGL catalog~\cite{Fermi-LAT:2015bhf}, and \emph{(iv)}~lobe-like emission associated with the \Fermi bubbles~\cite{Su:2010qj}. Finally, \emph{(v)}~Poissonian DM-like emission is modeled using a line-of-sight integral of the (squared) generalized Navarro-Frenk-White (NFW)~\cite{Navarro:1995iw,Navarro:1996gj} profile,
\begin{equation}
\label{eq:nfw}
\rho_\mathrm{gNFW}(r) \propto \frac{1}{\left(r / r_{\mathrm s}\right)^{\gamma}\left(1+r / r_{\mathrm s}\right)^{3-\gamma}}
\end{equation}
with inner slope $\gamma=1.2$ motivated by previous GCE analyses~\cite{Gordon:2013vta,Daylan:2014rsa,Zhou:2014lva}. Here, $r$ is the radial distance from the Galactic Center, $r_{\mathrm s}=20\,\mathrm{kpc}$ is the Milky Way scale radius, and we take $R_\odot = 8.2\,\mathrm{kpc}$ as the distance to the Galactic Center~\cite{2020arXiv201202169B,2019A&A...625L..10G}. Templates for components \emph{(ii)--(iv)} are obtained from Ref.~\cite{Mishra-Sharma:2016gis}. 

The Galactic foreground component accounts for $\gamma$-rays produced due to cosmic rays interacting with interstellar gas and radiation, which makes up the majority of the observed emission in the Galactic Center region. In particular, bremsstrahlung emission from cosmic-ray electrons scattering off of gas as well as photons produced as a result of the decay of pions produced through cosmic ray protons scattering elastically with the gas both trace the Galactic gas distribution, modulated by the incoming cosmic ray density. These components exhibit structure on smaller angular scales. Additionally, inverse Compton (up-)scattering (ICS) of the interstellar radiation field by cosmic ray electrons produces an important component of the $\gamma$-ray Galactic diffuse emission which spatially traces the Galactic charge carrier density and does not show modulation on small scales. Normalizations of the gas-tracing components, subscripted `brem/$\pi^0$', and the ICS-tracing component, subscripted `ICS', are included separately in our forward model. Templates for these two components are described in our baseline configuration by {Model~O}, introduced in Ref.~\cite{Buschmann:2020adf}. There, it was found to be better fit, as quantified by the likelihood of describing the data up to Poisson noise, to the counts map in the Galactic Center region compared to diffuse foreground templates previously employed in GCE analyses. We explore the effect of variations on the assumed Galactic diffuse model in Sec.~\ref{sec:systematics}. 

The diffuse emission templates have been pre-smoothed with the \Fermi point-spread function (PSF) at 2 GeV for the dataset employed, modeled as a pair of King functions.\footnote{\url{https://fermi.gsfc.nasa.gov/ssc/data/analysis/documentation/Cicerone/Cicerone_LAT_IRFs/IRF_PSF.html}} The total diffuse emission in a given pixel $p$, $x^p$, is modeled as a Poisson realization of a linear combination of the diffuse templates $T^p_i$, where $i$ indexes the individual templates, with their corresponding normalizations $A_i$ regarded as parameters of the forward model; $x^p \sim \mathrm{Pois}\left(x^p\mid\sum_i A_i T^p_i\right)$. \\ 

\noindent
\textbf{PS emission forward model:} Assuming the locations of individual PSs are not known \emph{a-priori}, the statistics of multiple PS populations can be completely specified through \emph{(i)}~their spatial distribution, described by templates $T^p$ discretized over pixels $p$, \emph{(ii)}~the distribution of expected photon counts $S$ contributed by each PS, $p(S)$, and \emph{(iii)}~the distribution of the number of PSs for each population.  Additionally, the modeled instrumental point-spread function quantifies the spatial distribution of photon counts sourced by an individual PS around its location due to the finite angular resolution of the LAT instrument.

Here, we parameterize the distributions of photon counts $S$ contributed by each PS through a doubly-broken power law,
\begin{equation}
\label{eq:scd_bpl}
p(S\mid\theta_\mathrm{PS})\propto \left\{\begin{array}{lc}
\left(\frac{S}{S_{\mathrm b, 1}}\right)^{-n_{1}}, & S \geq S_{\mathrm b, 1} \\
\left(\frac{S}{S_{\mathrm b, 1}}\right)^{-n_{2}}, & S_{\mathrm b, 1}>S \geq S_{\mathrm b, 2} \\
\left(\frac{S_{\mathrm b, 2}}{S_{\mathrm b, 1}}\right)^{-n_{2}}\left(\frac{S}{S_{\mathrm b, 2}}\right)^{-n_{3}}, & S_{\mathrm b, 2}>S
\end{array}\right.
\end{equation}
specified by the break locations $\{S_{\mathrm b, 1}, S_{\mathrm b, 2}\}$, spectral indices (slopes) $\{n_1, n_2, n_3\}$, and appropriately normalized to unity. Higher subscript indices correspond to dimmer parts of the source-count distribution. Together, we denote these parameters by $\theta_\mathrm{PS}$.

The PS component of the simulated \Fermi map is created as follows, practically implemented using the code package \texttt{NPTFit-Sim}~\cite{NPTFit-Sim}. The total number of PSs to be simulated is drawn as $n \sim \mathrm{Pois}(n\mid n_\mathrm{pix}\lambda)$, where $n_\mathrm{pix}$ is the number of pixels in the ROI and $\lambda$ is the mean number of PSs per pixel. The sample of PS angular positions $\{r_n\}$ is drawn from a PDF constructed by linearly interpolating the relevant pixel-wise spatial template $T^p$; $\{r_n\} \sim p(r) \propto T(r)$. The expected number of photons emitted by each PS, indexed by $i$, is drawn by first sampling from the mean PDF of expected photon counts in Eq.~\eqref{eq:scd_bpl}, $S \sim p\left({S}\mid\theta_\mathrm{PS}\right)$, and scaling this as ${S_i} = S \epsilon(r_i) / \langle \epsilon \rangle$ to account for variations in the \Fermi exposure at the sampled PS positions, $\epsilon(r_i)$, over the mean exposure $\langle \epsilon \rangle$ in the ROI. The actual sample of photon counts emitted by the simulated PSs, $\{x_n\}$, is taken to be a Poisson realization of this expectation; ${x_i} \sim \mathrm{Pois}\left({x_i}\mid {S_i}\right)$. Given the angular positions of and photon counts emitted by PSs $\{r_n, x_n\}$, the radial coordinates of photons relative to the positions of PSs are drawn following the modeled \Fermi PSF, with the azimuthal coordinates sampled uniformly assuming a spherically-symmetric PSF. This procedure is repeated for each PS population, and the final simulated PS map is constructed by binning the sampled photon positions within the ROI according to the pixelization scheme used. In practice, in order to avoid computational costs associated with simulating a large number of low-flux PSs, the dim component of the PS population below a specified threshold is partially accounted for in the DM-like component, as described in detail towards the end of this subsection.

In the NPTF literature, modeled PS populations are often compactly described through the so-called source-count distribution (SCD) $\dd^2 N /\dd S \dd\Omega$, which quantifies the differential number density of sources per unit angular area emitting $S$ photons in expectation. The source-count distribution jointly describes the distribution of photon counts from individual PSs $p(S\mid\theta_\mathrm{PS})$ and their mean per-pixel abundance $\lambda$, and is related to these as
\begin{equation}
\label{eq:scd_ps}
\frac{\dd^2 N}{\dd S\dd\Omega}=\lambda \, p(S\mid\theta_\mathrm{PS}) / \Omega_\mathrm{pix}
\end{equation}
where the the pixel area $\Omega_\mathrm{pix}$ is used to convert the per-pixel source count to per-area, rendering it agnostic to pixel size. We will present our results in terms of the source fluxes ($\dd^2 N /\dd F\dd\Omega$) rather than expected counts ($\dd^2 N /\dd S\dd\Omega$), with the conversion $S = \langle \epsilon \rangle F$ where $\langle \epsilon \rangle$ is the mean exposure in the region considered. In the analysis ROI used here, the mean exposure is $\langle \epsilon \rangle \simeq 7\times 10^{10}\,\mathrm{cm}^{2}\,\mathrm{s}$. For brevity, we will denote the distribution as $\dd N /\dd F$, leaving the per-area normalization implicit. 

In this paper, we consider two independent PS populations: \emph{(i)}~those spatially correlated with the GCE, modeled the same as the Poissonian counterpart using a line of sight integral of the (squared) generalized NFW profile in Eq.~\eqref{eq:nfw} with $\gamma = 1.2$, and \emph{(ii)}~those spatially correlated with the Galactic disk, modeled by a doubly-exponential profile motivated by studies of the spatial distribution of Galactic millisecond pulsar populations~\cite{Lorimer:2006qs, Bartels:2018xom},
\begin{equation}
\label{eq:disk_spatial}
\rho_\mathrm{Disk}(R, z) \propto \exp \left(-\frac{R}{R_\mathrm{d}}\right) \, \exp\left(-\frac{|z|}{z_\mathrm{s}}\right)
\end{equation}
where $R$ and $z$ are the radial and vertical Galactic cylindrical coordinates, and the disk scale height and radius are set to $z_\mathrm{s} = 0.3\,\mathrm{kpc}$ and $R_\mathrm{d} = 5\,\mathrm{kpc}$ respectively in the baseline scenario. The final maps are obtained by combining the diffuse and PS emission components of the forward model. \\ 

\noindent
\textbf{Prior specification:} We use uniform priors for the normalization factors of the Poissonian templates. For the PS components, we use uniform priors on the parameters that characterize the broken power-law  distribution of photon counts within the intervals defined below. The break associated with the brighter end of the SCD, $S_{\mathrm{b}, 1}  \in  [5, 40]$\,photons, reflects a `turn-on' associated with the source luminosity function, above which sources are either individually resolved or not inferred to exist. This turn-on is further enforced by specifying a highest slope $n_1 \in [10, 20]$ that is steeply rising with decreasing $S$. The middle slope, $n_2 \in [1.1, 1.99]$, is associated with the physical luminosity function of the source population, typically expected to be in this specified range for a Galactic pulsar population~\cite{Ploeg:2020jeh}. 

Emission from a PS population is nearly degenerate but still statistically distinguishable from that following a Poisson distribution when associated with sources emitting $\sim 0.1$--$1$ counts in expectation~\cite{List:2021aer}; in practice, however, residual effects of model misspecification and degeneracies between multiple PS populations can make characterizing the source-count distribution in this low-photon regime challenging~\cite{Chang:2019ars}. The dimmer break, $S_{\mathrm{b}, 2} \in  [0.1, 4.99]$\,photons, therefore specifies a regime where we do not attempt to explicitly characterize the PS population. This is enforced by allowing for a lowest slope $n_3 \in [-10, 1.99]$ that is steeply falling with decreasing $S$, encouraging the SCD to turn off in this regime. This gives preference to the smooth component in absorbing flux close to and below the single photon regime, and our analysis therefore conservatively aims to estimate a \emph{lower bound} on the contribution of PS emission to the GCE by primarily considering the relatively bright regime of the source-count distribution. In order to quantify the effect of the prior in the low-photon regime, we also explore an alternative specification where the lower range of the upper break prior is brought down to a single photon, $S_{\mathrm{b}, 1}  \in  [1, 30]$\,photons, giving the PS component more overlap closer to the degeneracy regime and thus allowing it to account for more of the dim emission. In App.~\ref{app:priors}, we show how the prior choices map onto the source-count distribution for the baseline and alternative configurations.

The overall abundance of PSs associated with a modeled population is specified as follows. Rather than sampling the expected number of PSs per pixel $\lambda$ with a uniform prior, we instead uniformly sample a related parameter $\langle S^{\rm PS} \rangle = \int\,\dd S\,S\,\lambda \, p(S\mid\theta_\mathrm{PS})$, the expected number of photon counts contributed by the PS population per pixel. Similarly, for the Poissonian GCE component, the template normalization $A_\mathrm{GCE}$ is reparameterized through a constant multiplicative factor into the mean per-pixel expected counts $\langle S^{\rm Poiss}_\mathrm{GCE} \rangle$. This is done in order to place the flux distribution of the PS-like component $\langle S^{\gtrsim 1\,\mathrm{ph}} \rangle$ on the same `footing' as that associate with smooth emission $\langle S^{\lesssim 1\,\mathrm{ph}} \rangle$. Since a uniform prior on $\lambda$ would not correspond to a uniform prior on $\langle S^{\rm PS} \rangle$, these reparameterizations \emph{a-priori} distribute photons approximately uniformly among the regimes $\langle S^{\lesssim 1\,\mathrm{ph}} \rangle$ and $\langle S^{\gtrsim 1\,\mathrm{ph}} \rangle$. We note here the possibility of using other prior prescription proposed in the literature, \emph{e.g.} in Ref.~\cite{Collin:2021ufc} where, in addition to enforcing an equivalence between dim PSs and smooth emission (rather than enforcing a distinction between relatively-bright PSs and smooth emission as done here), the SCD slopes are specified in terms of the angles between adjacent parts of the broken power law and the break positions are specified as a fraction relative to the brightest break.

The forward model is thus specified by a total of 18 parameters---6 for the overall normalizations of the Poissonian templates $\{\langle S_{\rm GCE}^{\rm Poiss} \rangle$, $A_{{\rm brem}/\pi^0}$, $A_{\rm ICS}$, $A_\text{iso}$, $A_\text{bub}$, $A_\text{3FGL}\}$, and $6\times2$ parameters modeling the source-count distributions associated with GCE-correlated and disk-correlated PS populations $\{\langle S^{\rm PS} \rangle, n_1, n_2, n_3, S_{\mathrm b, 1}, S_{\mathrm b, 2}\}$. The priors used in the forward model are summarized in Tab.~\ref{tab:priors}. In order to improve sample efficiency, the priors are motivated by posteriors obtained from a Poissonian template fit to the real \Fermi data. 

\begin{table}[tb]
\small
\begin{center}
\begin{tabular}{C{2cm}C{2cm} | C{2cm}C{2cm}}
\toprule
\multicolumn{2}{c}{\textbf{Poissonian}} & \multicolumn{2}{c}{\textbf{PS-like (GCE and disk)}}\Tstrut\Bstrut	\\   
\Xhline{1\arrayrulewidth}
\textbf{Parameter}	 & \textbf{Prior range}  & \textbf{Parameter}	&  \textbf{Prior range}\Tstrut\Bstrut	\\   
\Xhline{1\arrayrulewidth}
$\langle S_{\rm GCE}^{\rm Poiss} \rangle$ & [0, 2.5]\,ph  & $\langle S^{\rm PS} \rangle$ & [0, 2.5]\,ph\Tstrut\Bstrut \\
$A_{{\rm brem}/\pi^0}$ & [6, 12]  &  $n_1$ & [10, 20]\Tstrut\Bstrut  \\ 
$A_{\rm ICS}$  & [1, 6]  & $n_2$ & [1.1, 1.99]\Tstrut\Bstrut  \\ 
$A_\text{iso}$ & [0, 1.5] &  $n_3$ & [-10, 1.99]\Tstrut\Bstrut \\
$A_\text{bub}$ & [0, 1.5] &  $S_{\mathrm b,1}$ & [5, 40]\,ph\Tstrut\Bstrut \\
$A_\text{3FGL}$ & [0, 1.5] & $S_{\mathrm b,2}$  & [0.1, 4.99]\,ph\Tstrut\Bstrut \\
\botrule
\end{tabular}
\end{center}
\caption{Parameter priors used for the components of the forward model described in Sec.~\ref{sec:datasets}. All priors are uniform within the ranges specified. Priors on the Poissonian components, corresponding to overall normalization, are shown in the left table column, while those of the GCE- and disk-correlated PS components, parameterized according to Eq.~\eqref{eq:scd_bpl}, are shown in the right table column. The overall normalizations of the Poissonian GCE and PS-like components are parameterized through the mean number of photon counts contributed by the respective components in the ROI.}
\label{tab:priors}
\end{table}  

\subsection{Inference with likelihoods based on simplified data representations}
\label{sec:likelihood-methods}

Before discussing the methodology used in this paper in detail, we will provide a brief overview of an established class of techniques---Non-Poissonian template fitting---that have been successfully deployed in order to characterize the contribution of PSs to the GCE. 
We will focus on a schematic description of the method without delving into details of the implementation, aiming to highlight the elements that introduce approximations and where our ML-based approach differs.

A central object in statistical inference is the likelihood $p(x\mid \theta)$, which quantifies the probability of an observation $x$ given parameters of interest $\theta$. In the simplest incarnation of astrophysical template-fitting methods dealing with counts data, the likelihood of the map $x$ in the region of interest is computed as a pixel-wise product of Poisson likelihoods with mean given by a linear combination of spatial templates $T_i^p$, $p(x\mid \theta) = \prod_p \mathrm{Pois}\left(x^p\mid\sum_i A_i T_i^p\right)$, where normalizations $A_i$ of the respective spatial templates are the parameters of interest. This captures the diffuse part of the forward model described in Sec.~\ref{sec:datasets}, and inference here can easily be performed within a frequentist or Bayesian framework. 

In practice, unobserved latent variables $z$ are often involved in the data-generation process, and computing the likelihood involves marginalizing over the latent space, $p(x\mid\theta) = \int \dd z\,p(x\mid\theta, z)$. In typical problems of interest, the high dimensionality of the latent space often means that this integral is intractable, necessitating simplifications in statistical treatment as well as theoretical modeling. 
For the forward model in Sec.~\ref{sec:datasets}, the presence of PS populations 
introduces a large number of latent variables, specifically the position of and counts emitted by each PS. Ignoring the contribution from diffuse components for the moment and considering only a single isotropically-distributed PS population, the likelihood for the map $x$ in the region of interest is given by
\begin{equation}
\label{eq:data_likelihood}
p(x\mid\lambda, \theta_\mathrm{PS}) = \sum_{n = 0}^{\infty} \int \dd^{n} z \, p\left(n\mid\lambda\right)\,p(z\mid\theta_\mathrm{PS})\,p(x|z),
\end{equation}
where $\theta_\mathrm{PS}$ parameterize the distribution of photon counts from individual PSs. $n$ is the total number of PSs in the ROI, with the sum running over all possible number of PSs. This high-dimensional integral is, for all practical purposes, computationally intractable. The presence of a finite instrumental PSF introduces additional latent processes, decoupling the positions of the photons and PSs. Given these difficulties, a simplification of the problem setting is typically required to make further progress.

The 1-point PDF (probability distribution function) framework, first introduced in the context of $\gamma$-ray analyses in Ref.~\cite{Malyshev:2011zi} and extended to allow for non-trivial spatial PS distributions in Refs.~\cite{Lee:2014mza,Lee:2015fea} under the name of non-Poissonian template fitting (NPTF), considers a simplification of the problem by computing the pixel-wise likelihood assuming each pixel to be statistically independent (\emph{1-point} then referring to values over individual, independent spatial positions in the sky). This significantly reduces the latent space dimensionality by eliminating the positions of individual PSs as latent variables. Since non-Poissonian template fitting has been widely used in analyses of the GCE, we briefly outline the basic philosophy behind this method, pointing the interested reader to a more detailed discussion as well as numerical implementations in Refs.~\cite{Lee:2015fea,Mishra-Sharma:2016gis}.

Since emission from each PS can be regarded as independent conditioned on $\theta_\mathrm{PS}$, the probability of a given PS, indexed $i$, emitting $x^p_i$ photons in a pixel $p$ is given by
\begin{equation}
\label{eq:pixel-wise_likelihood}
p(x^p_i\mid\theta_\mathrm{PS}) = \int \dd S_i \,p(S_i\mid\theta_\mathrm{PS})\,p(x^p_i|S_i),
\end{equation}
where $S_i$ are the expected photon counts from the PS following some probability distribution parameterized by $\theta_\mathrm{PS}$, in this case following a doubly-broken power law with parameters $\theta_\mathrm{PS} = \{n_1, n_2, n_3, S_\mathrm{b,1}, S_\mathrm{b,2}\}$, and $p(x^p_i|S_i)$ is the distribution of actual counts given latent $S_i$, assumed to follow a Poisson distribution on $S_i$. The probability of having a total of $x_p$ counts in a pixel from multiple PSs is then described by a multinomial distribution, subject to the constraint that the total number of counts be equal to the observed counts:
\small
\begin{align}
\label{eq:pixel-wise_likelihood_multinomial}
\begin{split}
p(x^p\mid&\lambda,\theta_\mathrm{PS}) =  \sum_{n = 0}^{\infty}  p\left(n \mid \lambda\right) \sum_{n_{j}} \delta\left(\sum_j n_{j}j - x^p\right) \\ 
&\times \delta\left(\sum_j n_{j} - n\right) \frac{n!}{\prod_j n_{j} }\prod_{j=1}^{n} p(x^p_i = j\mid\theta_\mathrm{PS})  ^ {n_{j}},
\end{split}
\end{align}
\normalsize
where $n_j$ is the number of PSs contributing $j$ counts. The distribution of the number of PSs in a pixel is usually assumed to follow a Poisson distribution on the mean expected number of PSs $\lambda$ \emph{i.e.}, $p(n\mid\lambda) = \mathrm{Pois}(n\mid\lambda)$. In this case, the sum over $n$ can be eliminated and the distribution of observed counts is given by
\begin{align}
\label{eq:pixel-wise_likelihood_poisson}
\begin{split}
p(x^p\mid&\lambda, \theta_\mathrm{PS}) = \sum_{n_j = 0}^{\infty} \delta\left(\sum_j n_{j}j - x^p\right) \\ & \times \prod_j \mathrm{Pois}\left(n_{j}\mid\lambda \, p(x^p_i = j\mid\theta_\mathrm{PS})\right).
\end{split}
\end{align}
where $p(x^p_i = j\mid\theta_\mathrm{PS})$ is given by Eq.~\eqref{eq:pixel-wise_likelihood}. 
While not immediately obvious from this expression, eliminating the positions of individual PSs as latent parameters as well as the sum over the possible number of PSs $n$ renders the per-pixel likelihood tractable, and the total data likelihood can then be computed as a product over pixels, $p(x\mid\lambda,\theta_\mathrm{PS}) = \prod_{p} p(x^p\mid\lambda,\theta_\mathrm{PS})$.

We emphasize that we have only provided a brief overview of the NPTF method here, with further analytic simplifications, extensions to approximately incorporate the effect of non-trivial instrumental point-spread function and exposure, as well as a numerical recipe for evaluating the likelihood described in detail in Ref.~\cite{Mishra-Sharma:2016gis}. We note that including the effect of a finite point-spread function in the NPTF framework renders the per-pixel likelihood only approximately correct, since this introduces correlations across pixels over the scale of the PSF size. Previous studies have shown this approximation to be accurate enough for the present problem when using a pixel size of the order of the PSF size itself~\cite{Chang:2019ars}. Further generalizations of the method that can account for more extreme variations in the instrumental point-spread function and exposure without resorting to an approximate treatment---necessary for application to \emph{e.g.} X-ray data---were introduced and studied in Ref.~\cite{Collin:2021ufc}. 

Probabilistic cataloging~\cite{2013AJ....146....7B,2021arXiv210202409L} is another method that has been proposed for characterizing the sub-threshold contribution of PS populations in counts data and found application in $\gamma$-ray analyses~\cite{Daylan:2016tia}. This technique keeps the latent variables in Eq.~\eqref{eq:data_likelihood} \emph{i.e.}, the positions and expected fluxes of individual PSs, as parameters of interest, and uses trans-dimensional sampling techniques to obtain the distribution over possible catalogs of unresolved PS populations. For computational reasons, probabilistic cataloging techniques generally require a strong assumption on the nature of the putative PS population and can thus produce highly prior-dependent results.

In this paper, we show results on \Fermi data using the NPTF algorithm in order to establish a comparison point to previous GCE studies employing the method. We perform these analyses within a Bayesian framework, obtaining an approximation to the posterior distribution $p(\theta\mid x) = p(\theta)\, p(x\mid\theta) / \mathcal Z$, where $\mathcal Z \equiv p(x)$ the Bayesian evidence. 
We use the NPTF likelihood implemented in \texttt{NPTFit}~\cite{Mishra-Sharma:2016gis} and obtain representative posterior samples over the parameters of interest described in Sec.~\ref{sec:datasets} using nested sampling~\cite{Feroz:2013hea,skilling2006} implemented in \texttt{dynesty}~\cite{Speagle_2020}. The static variant of the nested sampling algorithm is run in its default configuration with 1000 live points, stopping when the estimated contribution of the remaining posterior volume to the log-evidence falls below $\Delta \log \mathcal Z < 0.1$. Although it's possible to correct for non-uniform exposure within the NPTF framework by considering independent sub-regions with different exposure values, given the fairly uniform \Fermi exposure in the Galactic Center region we use the mean exposure in our NPTF benchmarks for simplicity.

1-point PDF-based techniques, and in particular NPTF, have been widely applied for characterizing $\gamma$-ray PS populations below the \Fermi detection threshold, both in relation to the GCE~\cite{Lee:2015fea,Leane:2020pfc,Leane:2020nmi,Buschmann:2020adf,Calore:2021bty} and more generally \emph{e.g.}, for characterizing the contribution of extragalactic PSs at high latitudes~\cite{Lisanti:2016jub,Zechlin:2016pme,Zechlin:2015wdz} and for searching for a DM annihilation signal from Galactic subhalos~\cite{Somalwar:2020awt}. It has recently been pointed out, however, that signal and foreground mismodeling associated in particular with the emission in the Galactic Center region can hamper the ability to accurately characterize the contribution of PSs to the GCE~\cite{Leane:2019xiy,Leane:2020pfc}. In particular, Refs.~\cite{Lee:2015fea,Leane:2019xiy,Chang:2019ars} pointed out that spurious residuals associated with foreground mismodeling can lead to the mischaracterization of a purely DM signal as a population of PSs. Ref.~\cite{Buschmann:2020adf} recently showed that many of the issues associated with the expression of such effects in \Fermi data could be mitigated through the use of better Galactic foreground models along with affording them more degrees of freedom on large angular scales.
Refs.~\cite{Leane:2020pfc,Leane:2020nmi} further showed and described analytically how mismodeling, in particular an unmodeled asymmetry in a DM signal, could lead to the spurious inference of PSs in NPTF analyses of the GCE.

The fact that NPTF analyses rely on a simplified per-pixel likelihood can make them especially susceptible to the effects of model misspecification (alternatively referred to as mismodeling)---systematic departures of the forward model from the true data-generating process. This can be intuited from the fact that, assuming a corresponding permutation of template pixel labels, the NPTF likelihood is invariant to a permutation of pixels within the analysis ROI. This means that residuals associated with a misspecified background model can mimic the effect of PSs through the distribution of their photon counts, disregarding the specific spatial structure associated with a PS population. The full likelihood sketched out in Eq.~\eqref{eq:data_likelihood} and implicitly defined by the forward model described in Sec.~\ref{sec:datasets} contains significantly more spatial structure than is encoded in the distribution of photon counts, and in particular accounts for the distribution of pixel-to-pixel correlations in the $\gamma$-ray map; see also Ref.~\cite{List:2021aer} for an extended discussion on this point.
In the rest of this section, we will describe the building blocks of our machine learning-based method that, in contrast to NPTF, aims to estimate the likelihood implicitly associated with the $\gamma$-ray forward model, leveraging pixel-to-pixel spatial correlations with the overall aim of more robustly characterizing the PS contribution to the GCE.

\subsection{Simulation-based inference}

Simulation-based inference (SBI) refers to a class of methods for performing inference when the data-generating process does not have a tractable likelihood. This is the case for the model described in Sec.~\ref{sec:datasets}, where the likelihood in Eq.~\eqref{eq:data_likelihood} cannot be used explicitly for practical purposes without further simplifications. The model is then defined through a simulator as a probabilistic program, often knows as a forward model. Samples ${x}$ from the simulator then \emph{implicitly} define a likelihood, $\{x\}\sim p(x\mid\theta)$. In the simplest existing realizations of SBI, simulated samples $\{x\}$ can be compared to a given dataset of interest $x'$, with the approximate posterior defined by parameter values whose corresponding samples most closely resemble $x'$ according to some distance metric. Such methods---usually grouped under the umbrella of Approximate Bayesian Computation (ABC)~\cite{10.1214/aos/1176346785}---are not uncommon in astrophysics and cosmology. Nevertheless, they suffer from several downsides. The curse of dimensionality usually necessitates reduction of data to representative, hand-crafted, lower-dimensional summary statistics $s(x)$, resulting in loss of information. A notion of distance in the lower-dimensional summaries domain as well as a tolerance threshold, $||s(x) - s(x')|| < \epsilon$, is necessary to trade off between precision and sample efficiency, leading to inexact inference. Additionally, the ABC analysis must be performed anew for each new target dataset.

Recent advances in machine learning, particularly the proliferation of neural network architectures suited to a variety of data structures and the development of algorithms that can efficiently approximate functions and distributions in high dimensions, have galvanized the field of simulation-based inference, substantially increasing its domain of applicability; see Ref.~\cite{Cranmer:2019eaq} for a review of recent developments. In the following subsections, we will describe the specific SBI methods employed in this work for parameter estimation on the forward model described in Sec.~\ref{sec:datasets}.

\subsection{Conditional density estimation with normalizing flows}

We approximate the joint posterior $p(\theta\mid x)$ over the parameters of interest $\theta$ through a distribution $\hat p_\phi(\theta\mid s)$ conditioned on summaries $s=s(x)$ from simulated samples $\{x\}$, parameterized by $\phi$ and modeled by a neural network. This class of simulation-based inference techniques, known as conditional density estimation~\cite{10.5555/3157096.3157212,cranmer_kyle_2016_198541}, directly models the posterior distribution given a set of samples $\{x\}\sim p(x\mid\theta)$ produced from the forward model, where parameters $\theta$ are sampled according to some prior proposal distribution $\{\theta\}\sim p(\theta)$. We note that, given the absence of explicit labels associated with the sampled parameters of interest, estimating the probability density is an example of an unsupervised learning problem. \\

\noindent
\textbf{Normalizing flows:}
In this paper we employ normalizing flows~\cite{papamakarios2019normalizing,DBLP:conf/icml/RezendeM15}, a class of models that provide an efficient way of constructing flexible and expressive high-dimensional probability distributions. Normalizing flows model the (conditional) distribution over the parameters of interest $\hat p_\phi(\theta\mid s)$ as a series of transformations, denoted by $f$ such that $\theta = f(u)$, from a simple base distribution $\pi({u})$ to the target distribution. Suppressing the conditional dependence on $s$ for the moment for simplicity, we have
\begin{equation}
\label{eq:flow_transformation}
\hat{p}({\theta})=\pi(u)\left|\operatorname{det}\left(\frac{\partial u}{\partial {\theta}}\right)\right|=\pi(f^{-1}({\theta}))\left|\operatorname{det}J_{f^{-1}}(\theta)\right|
\end{equation}
where $\operatorname{det}J_{f^{-1}}$ is the Jacobian of the inverse transformation $f^{-1}$.

The defining characteristic of transformations in flow-based models is that they be diffeomorphic \emph{i.e.}, $f$ be differentiable and invertible with a differentiable inverse. This renders the Jacobian and inverse in Eq.~\eqref{eq:flow_transformation} computable, allowing for the evaluation of the probability density of the target distribution $\hat{p}({\theta})$ at a given parameter point $\theta$ once the transformation is defined. In practice, the transformation $f$ (or $f^{-1}$) is chosen such that $\operatorname{det}J$ can be efficiently computed and is usually defined by a neural network, and the base distribution $\pi(u)$ is chosen to be a standard Gaussian $u\sim \mathcal N(u; 0, \mathbb{1})$, which we follow here. 

A crucial property of diffeomorphic transformation such as those that define normalizing flows is that multiple transformations can be chained together through composition. Given two transformations $f_1$ and $f_2$, their composition will also be differentiable and invertible: $\operatorname{det}J_{f_{1}\circ f_2}(\theta) = \operatorname{det}J_{f_{2}}\left(f_1(\theta)\right)\operatorname{det}J_{f_{1}}(\theta)$ and $(f_2 \circ f_1)^{-1} = f_1^{-1} \circ f_2^{-1}$. This can be used to define more expressive probability distributions by chaining together several flow transformation. `Flow' thus refers to the trajectory through which parameters in the simple base distribution are transformed into the target parameter space, and `normalizing' refers to the inverse transformation into the base distribution. Flow-based models are \emph{generative}---given a new dataset $x'$, it is easy to sample from the base distribution and then run the forward transformation conditioned on $x'$, obtaining a set of parameter samples representative of the posterior distribution, $\{\theta\}\sim\hat p(\theta|x')$.

A number of methods have been proposed for defining the flow transformation \emph{e.g.}, based on affine transformations~\cite{10.5555/3294771.3294994,10.5555/3157382.3157627,DBLP:conf/iclr/DinhSB17,DBLP:journals/corr/DinhKB14}, spline-based transformations~\cite{DBLP:conf/nips/DurkanB0P19,durkan2019cubic}, and continuous-time transformations~\cite{DBLP:conf/iclr/GrathwohlCBSD19}. We refer to Ref.~\cite{papamakarios2019normalizing} for a recent review of normalizing flows, including details of practical implementations as well as an overview of proposed methods. \\

\noindent
\textbf{Masked autoregressive flows for (conditional) density estimation:}
In this paper we use Masked Autoregressive Flows (MAFs)~\cite{10.5555/3294771.3294994} to define the flow transformation. Autoregressive models can be used to learn a complex joint probability density $p(\theta)$ as a product of one-dimensional conditional densities where each $\theta_i$ depends only on the previous $\theta_{1:i-1}$ in the parameter sequence: $p(\theta) = \prod_i p(\theta_i\mid \theta_{1:i-1})$. The MAF is built using blocks of affine transformations subject to the autoregressive constraint; for a single block, the affine transformation from $u$ to $\theta$ is expressed as
\begin{equation}
\label{eq:maf_z}
\theta_{i}=u_{i}\cdot \exp \alpha_{i}+\mu_{i} 
\end{equation}
where $\mu_{i}=g_{\mu_{i}}\left({\theta}_{1: i-1} ; {s}\right)$ and $\alpha_i = g_{\alpha_{i}}\left({\theta}_{1: i-1} ; {s}\right)$ are scaling and shift factors modeled by neural networks and additionally parameterized by summaries $s$ from the forward model. The autoregressive property is enforced by masking out connections between network layers using the recipe introduced in Ref.~\cite{DBLP:conf/icml/GermainGML15}. The inverse transformation is easily identified from Eq.~\eqref{eq:maf_z}. This allows for an analytically tractable Jacobian determinant, for an $N$-dimensional distribution given by
\begin{equation}
\label{eq:det}
\left|\operatorname{det}J_{f^{-1}}(\theta)\right|=\exp \left(-\sum_{i=1}^N \alpha_{i}\right)
\end{equation}
and a forward pass through the flow according to Eq.~\eqref{eq:maf_z}.
Multiple transformations $f_j$ can be composed together in order to model more expressive posteriors,
\begin{equation}
\hat{p}({\theta}\mid s)=\pi\left(f^{-1}({\theta})\right) \prod_{j=1}^{K}\left|\operatorname{det}J_{f^{-1}_j}(u_{j-1})\right|
\end{equation}
where we have reinstated the conditional dependence on data summaries $s$, keeping it implicit in the transformations on the right hand side. The log-probability of the posterior can then be computed using Eq.~\eqref{eq:det}:
\begin{equation}
\label{eq:objective}
\log \hat{p}({\theta}\mid s) = \log \left[\pi\left(f^{-1}({\theta})\right)\right]-\sum_{j=1}^{K} \sum_{i=1}^{N} \alpha_{i}^{j},
\end{equation}
which acts as the optimization objective during training. Here, we use 8 MAF transformations, each made up of a 2-layer masked neural network with 128 hidden units and $\tanh$ activations. The ordering of parameters in the autoregressive sequence is randomly permuted between successive transformations in order to reduce dependence on the specific ordering of input variables. Each transformation is conditioned on summaries $s(x)$ extracted from the $\gamma$-ray maps $x$ (described in the next section below) by including these as additional inputs into the transformation block \emph{i.e.}, the scaling and shift factors in Eq.~\eqref{eq:maf_z} can be expressed as $\mu_{i}=g_{\mu_{i}}\left({\theta}_{1: i-1} ; {s(x)}\right)$ and $\alpha_i = g_{\alpha_{i}}\left({\theta}_{1: i-1} ; {s(x)}\right)$.

\subsection{Learning summary statistics with neural networks}

The curse of dimensionality makes it computationally inefficient to condition the density estimation task on the raw dataset $x$ \emph{i.e.}, the $\gamma$-ray pixel counts map in the region of interest (ROI). Representative summaries $s = s_\varphi(x)$ of the data can therefore be used in order to enable a tractable analysis, where $\varphi$ parameterizes the data-to-summary transformation. Although many choices for data summaries are possible---\emph{e.g.}, a Principal Component Analysis (PCA) or angular power spectrum decomposition of the photon counts map, or simply a histogram of the photon counts---in this paper, we use a neural network to automatically learn low-dimensional summaries that are optimized for the specific downstream task at hand of estimating the posterior distributions of the parameters associated with the forward model. \\

\noindent
\textbf{Graph construction and network architecture:}
The \texttt{DeepSphere} architecture~\cite{DBLP:conf/iclr/DefferrardMGP20,Perraudin:2018rbt,deepsphere_rlgm}, with a configuration similar to and inspired by that employed in Ref.~\cite{List:2020mzd}, is used to extract representative summaries from $\gamma$-ray maps and is briefly outlined here. \texttt{DeepSphere} is a graph-based spherical convolutional neural network (CNN) architecture tailored to data sampled on a sphere, and in particular is able to leverage the hierarchical structure of data in the \texttt{HEALPix} representation. This makes it well-suited for our purposes.

The \texttt{HEALPix} sphere can be represented in terms of a weighted undirected graph $\mathcal G = (\mathcal V, \mathcal E, A)$ where $\mathcal V$ is the set of $N_\mathrm{pix} = |\mathcal V|$ vertices, $\mathcal E$ is the set of edges connecting pixels, and $A$ is the weighted adjacency matrix. Each pixel $i$ is represented by a vertex $v_i \in \mathcal V$ and is connected to the 8 (or 7, depending on the pixel)
vertices $v_j$ which represent the neighboring pixels $j$ of pixel $i$, forming edges $(v_i
, v_j) \in \mathcal E$. The weights of the adjacency matrix over neighboring pixels $(i, j)$ are given by $A_{ij} = \exp \left(-{\left\|r_{i}-r_{j}\right\|_{2}^{2}}/{\rho^{2}}\right)$ where $r_i$ specifies the 3-dimensional coordinates of pixel $i$. The kernel widths $\rho$ at a given \HEALPix resolution are obtained from Ref.~\cite{DBLP:conf/iclr/DefferrardMGP20}, which used empirical measures of rotational equivariance in order to optimize for this hyperparameter.

We use the combinatorial graph Laplacian, defined as $ L = D - A$, where $ D$ is the diagonal degree matrix, and which can be used to define a Fourier basis on a graph. By construction being symmetric and positive semi-definite, the graph Laplacian can be decomposed as~\cite{DBLP:conf/nips/DefferrardBV16} $L =  U  \Lambda  U^T$, where $ U$ is an orthonormal eigenvector matrix and $ \Lambda$ is a diagonal eigenvalue matrix. The Laplacian eigenvectors then define the graph Fourier basis, with the Fourier transform $\tilde{x}$ of a signal $ x$ on a graph being its projection $\tilde{x} = U^T x$.
Given a convolutional kernel $h$, graph convolutions can be efficiently performed in the Fourier basis as $h({L}) {x}={U} h({\Lambda}) {U}^{T} {x}$~\cite{DBLP:conf/nips/DefferrardBV16}.

The isotropic \texttt{DeepSphere} convolutional kernel $h$ is defined as a linear combination of Chebychev polynomials, $h({{L}}) = \sum_{k=0}^{K} c_{k} T_{k}({{L}})$ where $T_k$ are the order-$k$ Chebyshev polynomials and $c_k$ are the $K + 1$ filter coefficients which are the trainable parameters to be learned during model optimization. The graph filtering operation can then be expressed as
\begin{equation}
h({L}) {x}={U}\left(\sum_{k=0}^{K} c_{k} T_k({\Lambda})\right) {U}^{T} {x}=\sum_{k=0}^{K} c_{k} T_k({L}) {x}.
\end{equation}
We set $K=5$, having checked that larger values do not quantitatively affect the results of the analysis. $T_k({\Lambda})$ acts on the diagonal eigenvalue matrix, $T_k({\Lambda_{ii}}) = T_k({\Lambda})_{ii}$. 

Following Refs.~\cite{Perraudin:2018rbt,List:2020mzd}, the feature extraction architecture is built out of graph convolutional layers which involve progressively coarsening the pixel representation of the $\gamma$-ray maps while increasing the number of filter channels at each step. The input map corresponds to the 16,384 pixels at \HEALPix resolution \texttt{nside}=128 in the nested pixel ordering within the single pixel corresponding to \texttt{nside}=1 covering the Galactic Center region, with the masked pixels set to zero. Each graph convolution operation is followed by a batch normalization, a ReLU nonlinearity, and a max pooling operation which downsamples the representation by a factor of 4 into the next coarser \HEALPix resolution, starting with input maps at \texttt{nside}=128 until a single pixel channel at \texttt{nside}=1 remains after the final convolutional layer. All together, 7 layers of this kind are employed. The number of filter channels is doubled at each convolutional layer until a maximum of 256. 

The output of the final convolutional layer is augmented with 2 additional auxiliary variables---the log-mean and log-standard deviation of the $\gamma$-ray map within the region of interest---and passed, via a ReLU nonlinearity, through a fully-connected layer with 1024 hidden units outputting a desired number of summary features, which we take as 128 in our baseline configuration. Pixels outside of the ROI as well as masked PSs are set to zero in the input maps. All input maps are standardized to zero mean and unit variance across the training sample.

Using a convolutional neural network-based feature extractor, we implicitly use an approximation to the full data likelihood in Eq.~\eqref{eq:data_likelihood} associated with our forward model of emission in the Galactic Center region. The method is thus able to capture pixel-to-pixel correlations in the $\gamma$-ray map, mitigating some of the limitations of approximate likelihood-based methods described in Sec.~\ref{sec:likelihood-methods}. \\

\noindent
\textbf{Optimization, training, and evaluation:} The optimization objective in Eq.~\eqref{eq:objective}, $\log \hat{p}_\phi(\theta\mid s_\varphi(x))$, is used to train the graph convolutional and normalizing flow neural networks simultaneously, optimizing their respective parameters $\{\varphi, \phi\}$. $10^{6}$ samples are generated using the prior proposal distribution of parameters given in Tab.~\ref{tab:priors}, and models are optimized with batch size 256 using the \texttt{AdamW}~\cite{DBLP:journals/corr/KingmaB14,DBLP:conf/iclr/LoshchilovH19} optimizer with initial learning rate $10^{-3}$ and weight decay $10^{-5}$, using cosine annealing to decay the learning rate across epochs. Training proceeds for up to 30 epochs with early stopping if the validation loss, evaluated on 15\% of samples held out, has not improved after 8 epochs. 

After training, given a new dataset of either real or simulated \Fermi data in our ROI, the posterior is obtained by drawing samples from the flow within the prior distribution using rejection sampling, conditioning each flow transformation on summaries extracted by the convolutional neural network with the new dataset as input. The model is \emph{amortized}, which means that after the upfront cost of training the neural network, the required number of posterior samples corresponding to a new dataset can be obtained on a few-second timescale. This makes it efficient to validate the performance of a trained model using mock data, which we do in the following section before applying the method to \Fermi data.

\begin{figure*}[!htbp]
\centering
\includegraphics[width=0.95\textwidth]{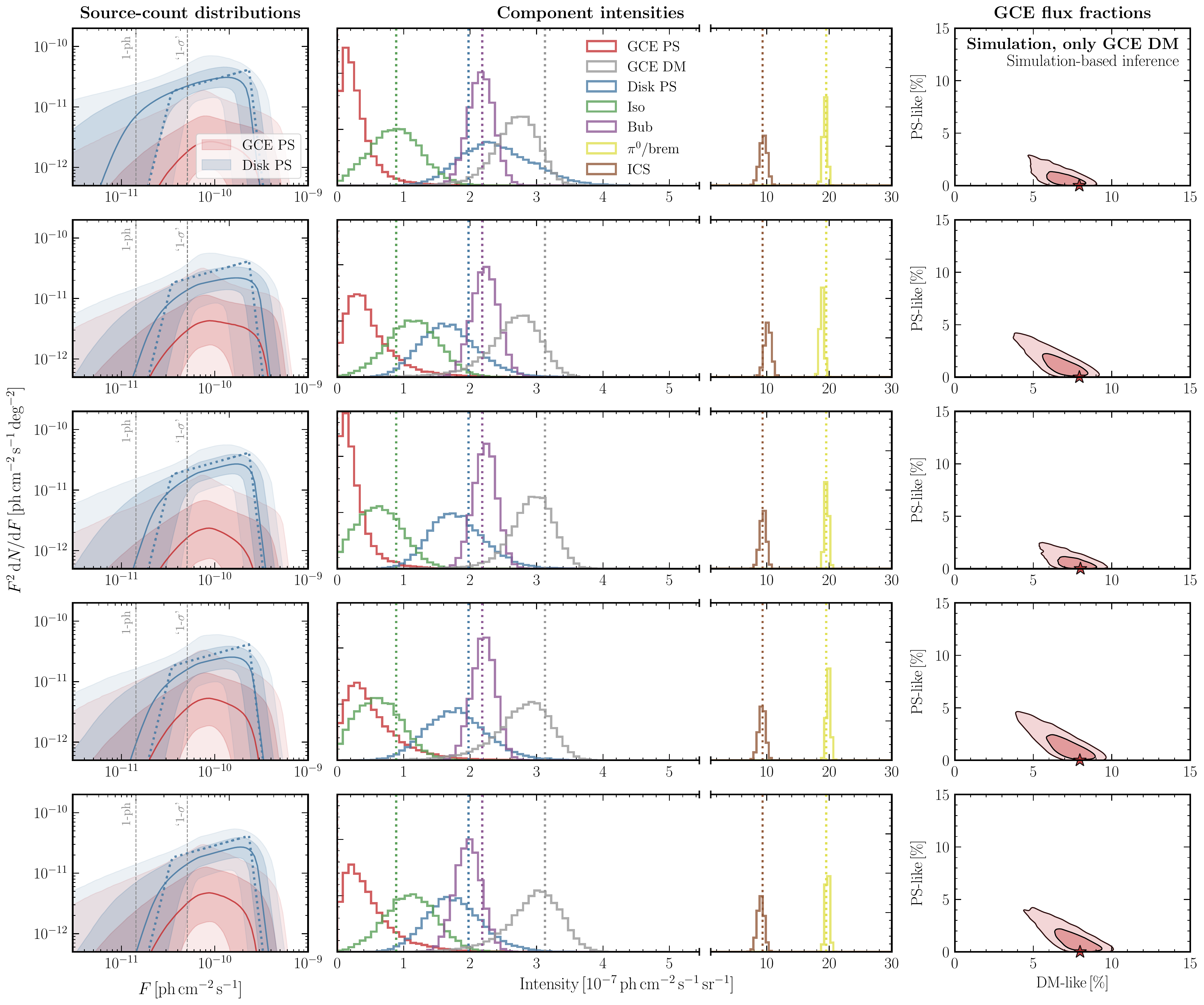}
\caption{Results of the analysis on simulated \Fermi data where the GCE consists of purely DM-like emission, with different rows corresponding to five different simulated realizations. The left column shows the inferred source-count distribution posteriors for GCE-correlated (red) and disk-correlated (blue) PSs. Dashed vertical lines corresponding to the flux associated with 1 expected photon per source and the approximate 1-$\sigma$ threshold for detecting individual sources are shown for reference. Solid lines correspond to the inferred posterior median, and the lighter and darker bands represent the middle-68\% and 95\% posterior containments respectively, evaluated point-wise in flux $F$. The middle column shows the posteriors for the Poissonian templates. The right column shows the joins posterior on the flux fractions of DM-like and PS-like emission. The dotted lines (in the left two columns) and the stars (in the right column) correspond to the true simulated quantities. DM-like emission is successfully inferred in each case, with the other parameter posteriors corresponding faithfully to the true simulated values.} 
\label{fig:sim_sbi_dm}
\end{figure*}

\section{Tests on simulated data}
\label{sec:simulations}

We begin by validating our pipeline on simulated \Fermi data. We create simulated datasets with the parameters of interest in the forward model fixed to posterior medians obtained in a fit of the baseline model to real \Fermi data, and test the ability of our model to infer the presence of either DM-like or PS-like signals on top of the modeled astrophysical background.

Figure~\ref{fig:sim_sbi_dm} shows results of the analysis conditioning the trained baseline model on five simulated maps where the GCE consists of purely DM-like emission, drawing 20,000 representative samples in each case. The left column shows the median (solid lines) as well as middle-68/95\% containment (dark/light shaded regions) of the posteriors on the source-count distributions $F^2 \dd N/\dd F$ of GCE-correlated (red) and disk-correlated (blue) PS emission, evaluated point-wise in flux $F$. The dashed grey vertical lines correspond to the flux associated with a single expected photon count per source (below which Poissonian and PS-like emission is expected to be nearly degenerate) and the approximate 1-$\sigma$ threshold for detecting individual sources (below which the degeneracy is often observed in practice~\cite{Chang:2019ars,Buschmann:2020adf}). The middle column shows the posteriors on various modeled emission components, excluding emission from resolved 3FGL PSs as the posterior in that case is largely unconstrained owing to the fact that resolved PSs are masked out in the analysis. The right column shows the joint posterior on the fraction of DM- and PS-like emission in proportion to the total inferred flux in the ROI. The true underlying parameter values from which the data was generated are represented by dotted lines in the left and middle columns, and by star markers in the right column. We see that, in all cases shown, the pipeline successfully recovers the presence of DM-like emission, with little flux---$\lesssim 10\%$ of the total inferred GCE emission in all cases---attributed to PSs. 

\begin{figure*}[!htbp]
\centering
\includegraphics[width=0.95\textwidth]{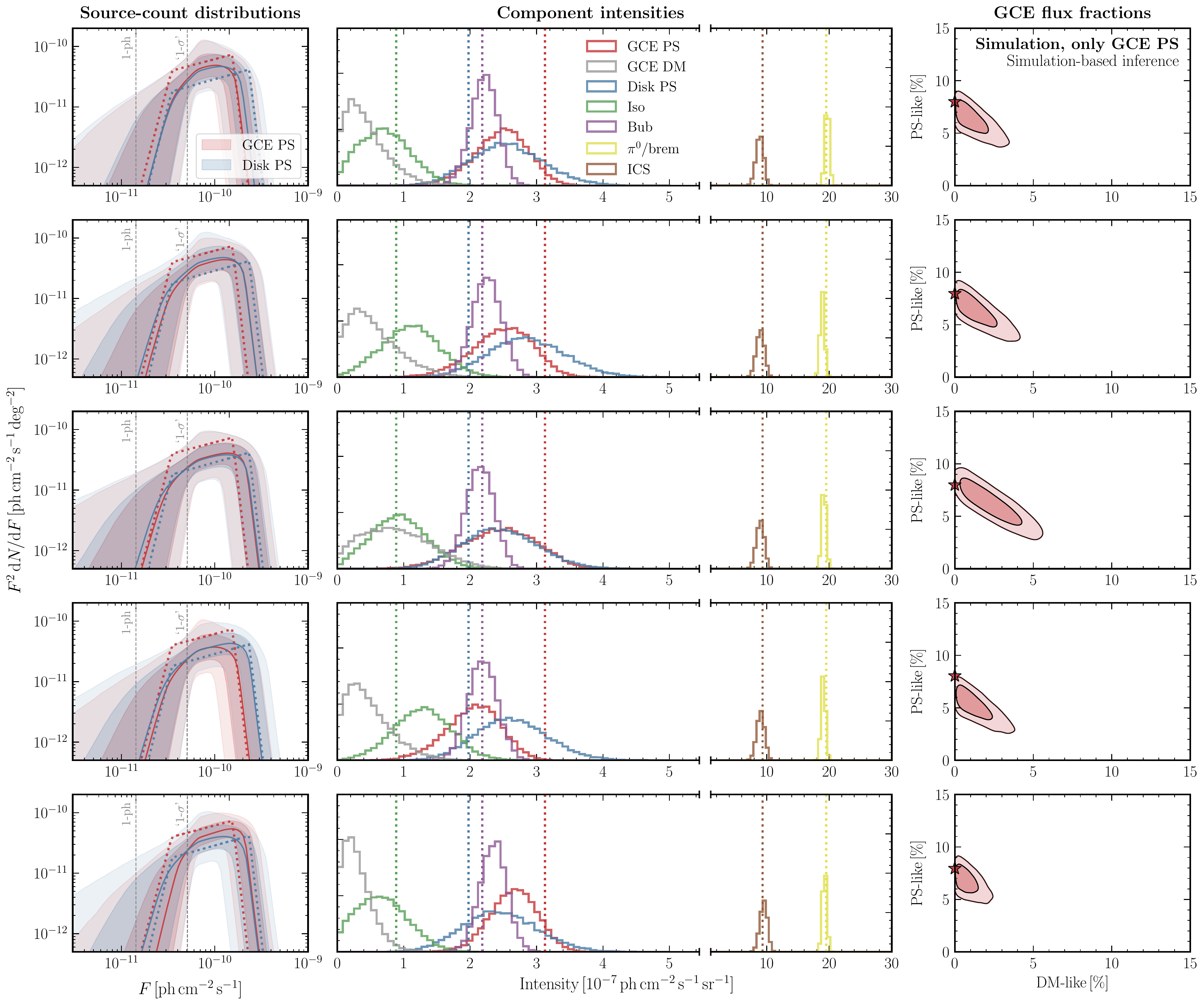}
\caption{Same as Fig.~\ref{fig:sim_sbi_dm}, but for five simulated realization of \Fermi data where the GCE consists of predominantly PS-like emission. PS-like emission is inferred in each case, with the other posteriors corresponding faithfully to their true simulated quantities. The GCE-correlated source-count distribution is also seen to be successfully recovered in the left panel. We note that, as detailed towards the end of Sec.~\ref{sec:datasets}, PS flux below $\sim 5$ photons is partially accounted for by the smooth DM-like component, which is responsible for the sharp turn-off in the modeled as well as inferred GCE-correlated SCD with decreasing flux.}
\label{fig:sim_sbi_ps}
\end{figure*}

Figure~\ref{fig:sim_sbi_ps} shows the corresponding results for simulated data containing PS-like emission correlated with the GCE. We see that PS-like emission is successfully inferred in each case, while at the same time exemplifying some degeneracy with the Poissonian component. Furthermore, as seen in the left column, the method is able to characterize the contribution of the two modeled PS components through the inferred source-count distribution. The inferred posteriors for the contribution of the DM-like component are seen to be compatible with zero. The overall flux of all modeled components, both PS and diffuse, is seen to be consistent with the true values used for the simulations in both sets of tests.

\section{Results on \Fermi data}
\label{sec:data}

We finally apply our neural simulation-based inference pipeline to the real \Fermi dataset. As a point of comparison, we also run the NPTF method described in Sec.~\ref{sec:likelihood-methods} on the data using the same spatial templates and prior assumptions as those used in the corresponding SBI analyses. A summary of the results for different analysis configurations, obtained by re-training the model using different assumptions about spatial templates or parameter priors, is shown in Table~\ref{tab:results}, including the fraction of overall emission attributed to the GCE, fraction of the GCE attributed to PS-like emission, flux corresponding to the highest break in the GCE broken power-law source count, fraction of the overall emission attributed to disk-correlated PSs, and flux corresponding to the highest break in the disk-correlated broken power-law source count. Medians as well as middle-68\% ranges on the respective posteriors are presented. In these analyses, we draw a larger number 50,000 of samples from the trained flow in order to reduce sample variance when quoting summary quantities of the inferred posteriors.

\subsection{Baseline analysis on \Fermi data}
\label{sec:baseline}

\begin{figure*}[!htbp]
\centering
\includegraphics[width=0.95\textwidth]{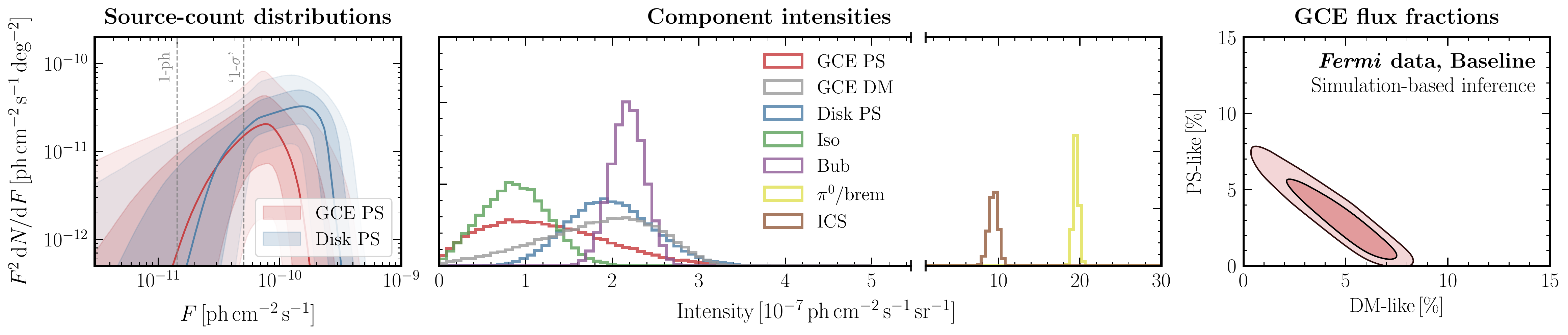}
\includegraphics[width=0.95\textwidth]{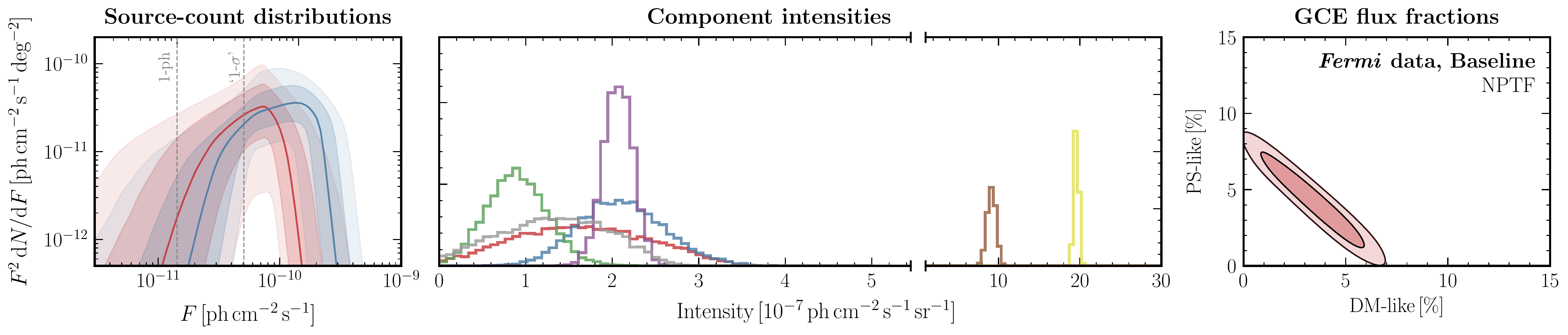}
\caption{Results of the baseline analysis on real \Fermi data. {(Top row)} Analysis using neural simulation-based inference with normalizing flows, and {(bottom row)} using the 1-point PDF likelihood implemented in the non-Poissonian template fitting (NPTF) framework. While moderate preference for a PS-like origin of the GCE is seen in the case of the NPTF analysis (bottom), the simulation-based inference analysis attributes a smaller fraction of the GCE to PS-like emission (top).}
\label{fig:fid_data}
\end{figure*}

Figure~\ref{fig:fid_data} shows posterior distributions for the baseline analysis on \Fermi data, with the top panel showing results for the SBI analysis and bottom panel corresponding to the NPTF analysis. Consistent with previous studies using a similar configuration, a significant fraction of the GCE---$55.0^{+8.8}_{-22.9}\%$---is attributed to PS-like emission within the NPTF framework. For the SBI analysis, although posteriors for the astrophysical background templates are seen to be broadly consistent with those inferred in the NPTF analysis, the preference for PSs is reduced, with $37.9^{+8.9}_{-19.2}\%$ of the GCE emission being PS-like. We also note that, in both cases, the inferred GCE-correlated source-count distribution sits at lower values than those inferred in previous NPTF analyses, which have generally found the bulk of PSs to lie just below the 3FGL PS detection threshold at $\sim2$--$3\times 10^{-10}$\,ph\,cm$^{-2}$\,s$^{-1}$~\cite{Lee:2015fea}. Given our doubly-broken power-law parameterization, the actual peak of the SCD is not physically meaningful---it can be driven simply by the position of the lowest break which marks a soft boundary between PS-like and smooth (but still possibly PS-driven) emission for accounting purposes. Instead, we use the upper break as a proxy for where the brightest unresolved sources are inferred to lie. In this baseline configuration, the SCD upper break is constrained to be $1.3^{+0.3}_{-0.4}\times 10^{-10}$\,ph\,cm$^{-2}$\,s$^{-1}$, corresponding to $\sim8$---$10$ photons. This `dimming' of the inferred SCD compared to previous NPTF analyses was also observes in Ref.~\cite{List:2021aer}, where it was found to be largely driven by the use of more up-to-date diffuse models. We note that even though the SBI analysis prefers a smaller GCE fraction in point sources, there is significant posterior overlap in the inferred joint PS and DM flux fraction posteriors between the SBI and NPTF analyses.

Besides the modeled GCE-correlated components, the fluxes associated with the other Poissonian astrophysical templates in our analysis are broadly consistent with previous studies of the Galactic Center. We note, in particular, that the relatively high ICS emission inferred in our analysis is consistent with previous studies of the Galactic Center region where this component was floated separately~\cite{Fermi-LAT:2015sau,List:2021aer,Calore:2014xka,Calore:2014nla}.

\begin{table*}[!t]
\footnotesize
\begin{center}
\begin{tabular}{C{2.3cm}C{2cm}|C{2cm}C{2cm}C{2.2cm}C{2cm}C{2.2cm}C{2cm}}
\toprule
\textbf{Configuration}  & \textbf{Method}  & $\dfrac{\textbf{GCE}}{\textbf{Total}}$	 & $\dfrac{\textbf{GCE PS}}{\textbf{GCE}}$  & $F_{\mathrm{b}, 1}^\mathrm{GCE}$	&   $\dfrac{\textbf{Disk PS}}{\textbf{Total}}$ &  $F_{\mathrm{b}, 1}^\mathrm{Disk}$	&  \textbf{Posteriors}\rule{0pt}{4.5ex}	\\[-2.3mm]
- & - & \scriptsize{[\%]} & \scriptsize{[\%]} & \scriptsize{[$10^{-10}$\,ph\,cm$^{-2}$\,s$^{-1}$]} & \scriptsize{[\%]} & \scriptsize{[$10^{-10}$\,ph\,cm$^{-2}$\,s$^{-1}$]}	& -\Tstrut\Bstrut \\
\Xhline{1\arrayrulewidth}
\multirow{2}{*}{Baseline} & SBI & $7.8^{+0.2}_{-0.6}$ & $37.9^{+8.9}_{-19.2}$ & $1.3^{+0.3}_{-0.4}$ & $5.0^{+0.5}_{-1.1}$ & $2.2^{+0.2}_{-0.5}$ & \multirow{2}{*}{Figure~\ref{fig:fid_data}}\Tstrut \\
& NPTF & $7.7^{+0.2}_{-0.6}$ & $55.0^{+8.8}_{-22.9}$ & $1.1^{+0.1}_{-0.2}$ & $5.4^{+0.5}_{-1.1}$ & $2.0^{+0.2}_{-0.5}$\Bstrut &\\ 
\hline
\multirow{2}{*}{Dif. {Model~A}} & SBI & $6.4^{+0.2}_{-0.6}$ & $57.3^{+9.9}_{-25.6}$ & $1.2^{+0.2}_{-0.3}$ & $4.9^{+0.6}_{-1.3}$ & $2.3^{+0.2}_{-0.5}$ & \multirow{2}{*}{Figure~\ref{fig:fid_data_modelA}}\Tstrut  \\ 
& NPTF & $6.7^{+0.2}_{-0.6}$ & $74.9^{+6.6}_{-22.5}$ & $1.1^{+0.1}_{-0.2}$ & $5.1^{+0.5}_{-1.3}$ & $2.2^{+0.2}_{-0.5}$\Bstrut &\\
\hline
\multirow{2}{*}{Dif. {Model~F}} & SBI & $5.0^{+0.2}_{-0.6}$ & $59.4^{+10.4}_{-26.3}$ & $1.4^{+0.3}_{-0.4}$ & $4.5^{+0.5}_{-1.1}$ & $3.3^{+0.3}_{-0.8}$
& \multirow{2}{*}{Figure~\ref{fig:fid_data_modelF}}\Tstrut \\
& NPTF & $5.2^{+0.2}_{-0.5}$ & $67.5^{+8.6}_{-26.7}$ & $1.1^{+0.2}_{-0.3}$ & $6.4^{+0.5}_{-1.1}$ & $2.0^{+0.2}_{-0.4}$\Bstrut &\\
\hline
\multirow{2}{*}{Thick disk} & SBI & $7.9^{+0.2}_{-0.6}$ & $42.2^{+9.6}_{-21.0}$ & $1.6^{+0.4}_{-0.6}$ & $3.5^{+0.6}_{-1.3}$ & $2.7^{+0.4}_{-0.8}$ & \multirow{2}{*}{Figure~\ref{fig:fid_data_thick_disk}}\Tstrut \\
& NPTF & $8.2^{+0.3}_{-0.7}$ & $75.0^{+7.1}_{-22.6}$ & $1.1^{+0.1}_{-0.2}$ & $2.3^{+0.7}_{-1.1}$ & $3.1^{+0.6}_{-1.2}$\Bstrut &\\
\hline
\multirow{2}{*}{Alt. priors} & SBI & $7.7^{+0.2}_{-0.6}$ & $54.2^{+11.9}_{-27.4}$ & $0.9^{+0.2}_{-0.4}$ & $5.9^{+0.5}_{-1.1}$ & $2.4^{+0.2}_{-0.4}$ & \multirow{2}{*}{Figure~\ref{fig:fid_data_alt_priors}}\Tstrut \\
& NPTF & $7.9^{+0.2}_{-0.6}$ & $77.7^{+6.5}_{-21.4}$ & $0.9^{+0.1}_{-0.3}$ & $5.9^{+0.5}_{-1.1}$ & $2.3^{+0.2}_{-0.4}$\Bstrut &\\
\botrule
\end{tabular}
\end{center}
\caption{Inferred values for the inferred GCE flux as a fraction of the total flux, the GCE PS-like flux as a fraction of the total GCE flux, the position of the upper source count flux break $F_{\mathrm{b}, 1}$ for the GCE and disk PS components, and the disk flux as a fraction of the total flux. For the baseline configuration as well as the various systematic variations explored, the median along with the 16th and 84th posterior percentile values are shown for the simulation-based inference (SBI) and NPTF analyses.}
\label{tab:results}
\end{table*}    

\subsection{Signal injection test on \Fermi data}
\label{sec:sig-injection}

A crucial test of self-consistency is the ability of the method to recover an artificial signal injected onto the real $\gamma$-ray data. As shown in Ref.~\cite{Leane:2019xiy}, initial applications of the NPTF to the GCE would generally fail this closure test, with implications for characterizing the nature of PSs in the Galactic Center explored in Refs.~\cite{Chang:2019ars,Buschmann:2020adf}. In particular, it was shown that this test can help diagnose underlying issues associated with mismodeling of the diffuse foreground emission, which have the potential to bias the characterization of PS populations. Recent NPTF analyses using improved descriptions of foreground modeling~\cite{Buschmann:2020adf} show consistent behavior under thist closure test. We perform a version of this test within our framework, testing the ability of our method to recover different mock signals injected onto the real \Fermi data.

\begin{figure*}[!htbp]
\centering
\includegraphics[width=0.95\textwidth]{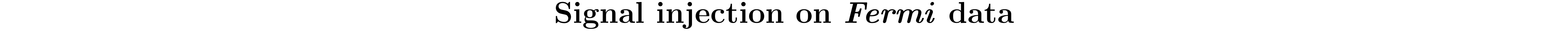}
\includegraphics[width=0.95\textwidth]{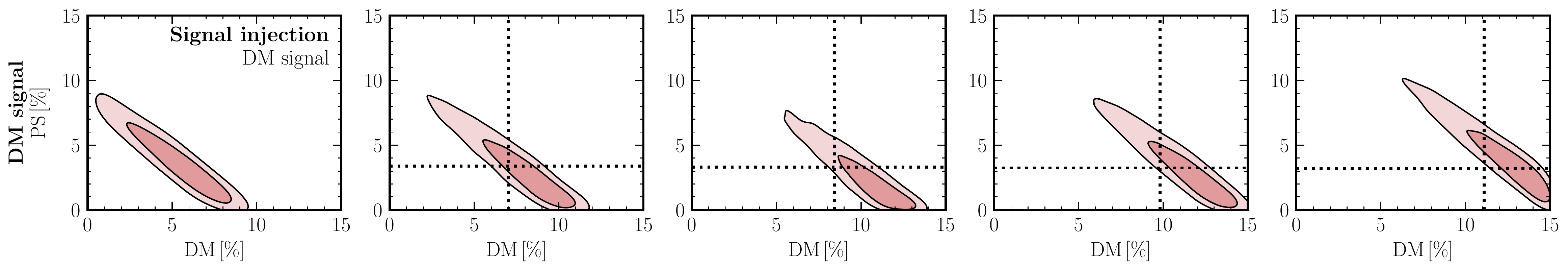}
\includegraphics[width=0.95\textwidth]{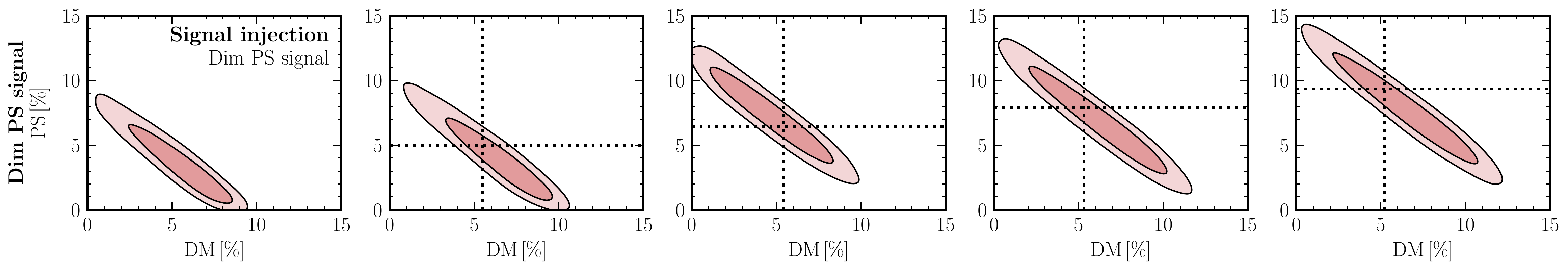}
\includegraphics[width=0.95\textwidth]{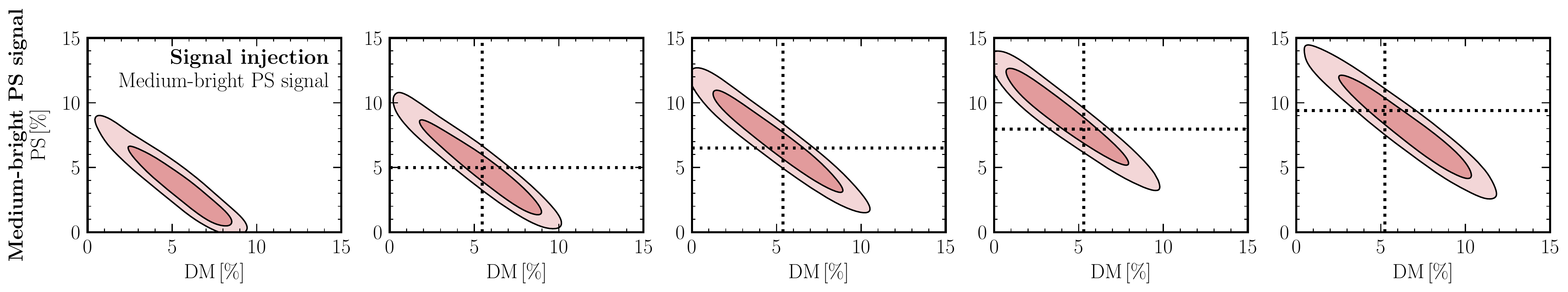}
\includegraphics[width=0.95\textwidth]{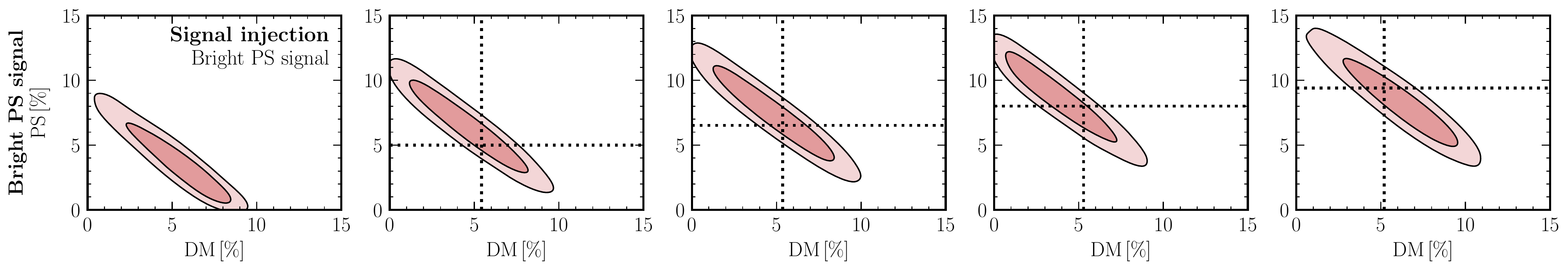}
\includegraphics[width=0.95\textwidth]{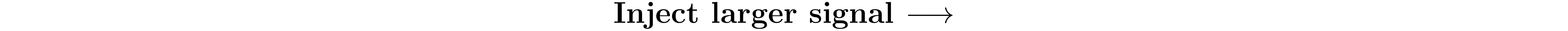}
\caption{Joint posterior for the flux fraction of PS-like and DM-like emission when an artificial DM signal is injected onto the real \Fermi data. The different rows correspond to different signal types, from top to bottom, purely DM, dim PSs (maximum of 5 expected counts per PS), moderately-bright PSs (maximum of 10 expected counts per PS), and bright PSs (maximum of 20 expected counts per PS). The leftmost panels shows the baseline analysis on \Fermi data, with subsequent panels showing results with progressively larger signals injected onto the data. The dotted lines show the expected total emissions including the injected signal and the median fluxes. The additional injected DM and PS signals are seen to correctly reconstructed within the respective posterior bounds in all cases.}
\label{fig:sig_inj_data}
\end{figure*}

Figure~\ref{fig:sig_inj_data} shows the results of this test, with the different rows corresponding to different signal configurations---purely DM, bright PSs, medium-bright PSs, and dim PSs. Bright, medium-bright, and dim PS configurations are taken have a maximum PS flux (given by the highest break in Eq.~\eqref{eq:scd_bpl}) at 20, 10, and 5 photon counts respectively, with other parameters set to median values inferred on real \Fermi data, except the lower break, which was set to 2 photon counts. The leftmost columns show the baseline analysis on \Fermi data, with subsequent columns showing signals of progressively larger sizes injected onto the data, up to approximately the size of the original GCE signal. The dotted horizontal and vertical lines show the total emissions including the injected signal and the median fluxes for the PS and DM components of the GCE inferred without any additional injected signal, respectively. 

The additional injected signal is seen to be reconstructed correctly within the inferred 95\% confidence interval in all four cases. For the DM signal (top row), the brighter inferred DM signals tend to be slightly overestimated. The injected PS signals (rows 2--4) are correctly reconstructed in all cases, with the dimmer PS signals showing a more prominent flat direction with Poissonian emission, as expected.

\subsection{Systematic variations on the analysis}
\label{sec:systematics}

\begin{figure*}[!htbp]
\centering
\includegraphics[width=0.95\textwidth]{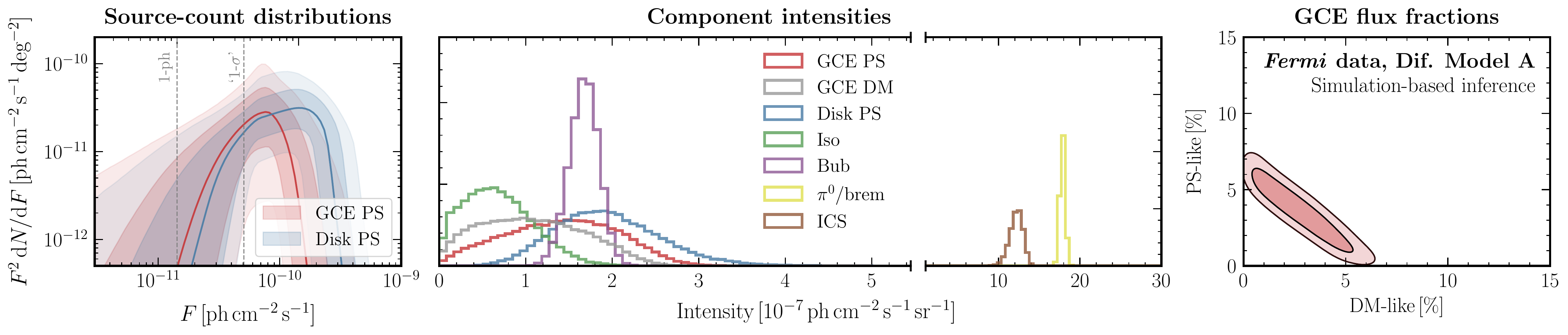}
\includegraphics[width=0.95\textwidth]{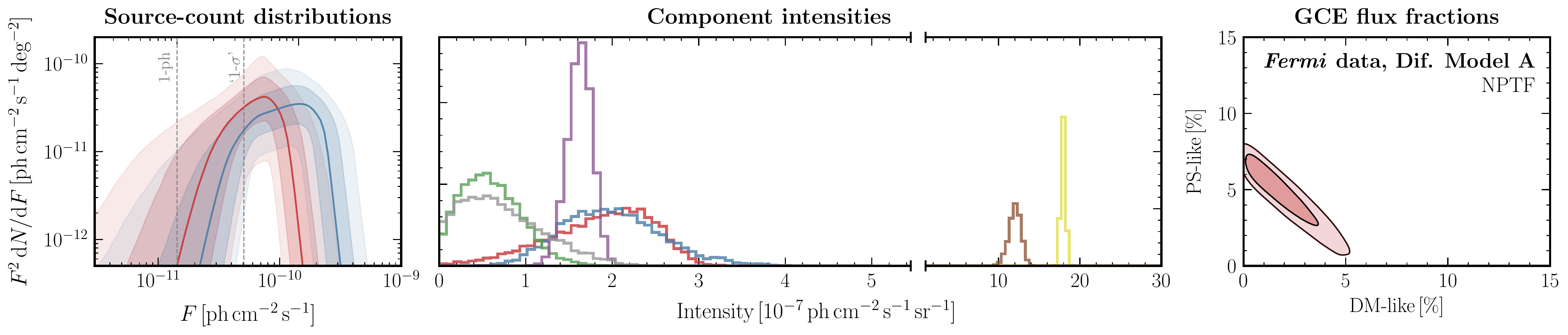}
\caption{Same as Fig.~\ref{fig:fid_data}, but with the diffuse foreground emission modeled using the alternative {Model~A}.}
\label{fig:fid_data_modelA}
\end{figure*}
\begin{figure*}[!htbp]
\centering
\includegraphics[width=0.95\textwidth]{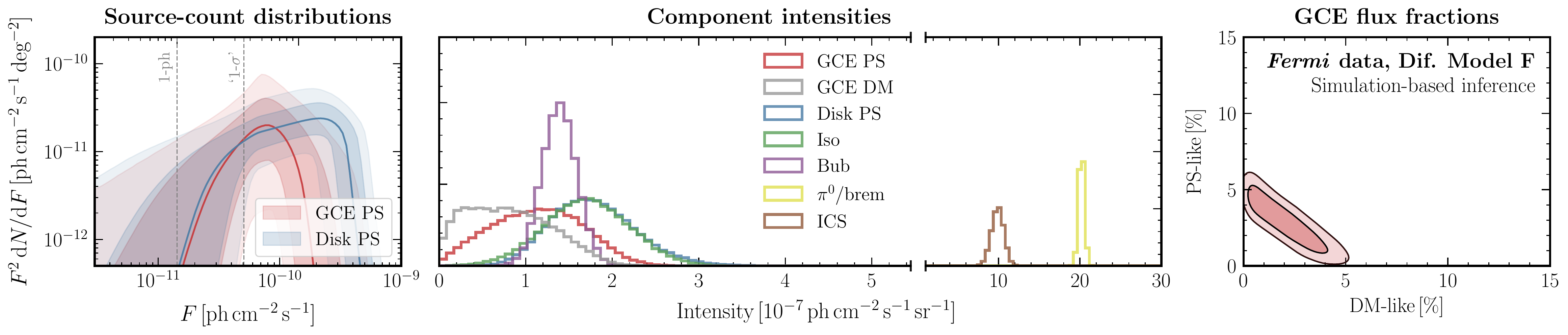}
\includegraphics[width=0.95\textwidth]{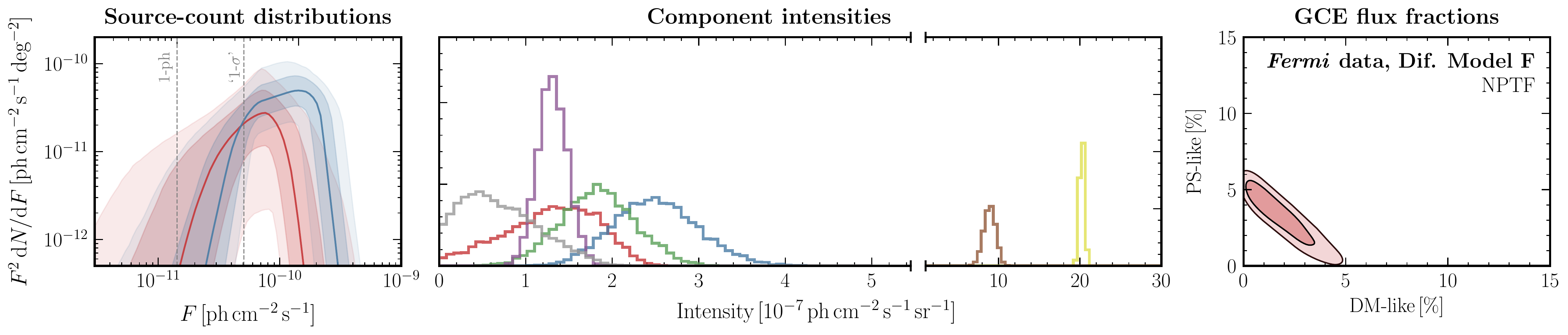}
\caption{Same as Fig.~\ref{fig:fid_data}, but with the diffuse foreground emission modeled using the alternative {Model~F}.}
\label{fig:fid_data_modelF}
\end{figure*}
\begin{figure*}[!htbp]
\centering
\includegraphics[width=0.95\textwidth]{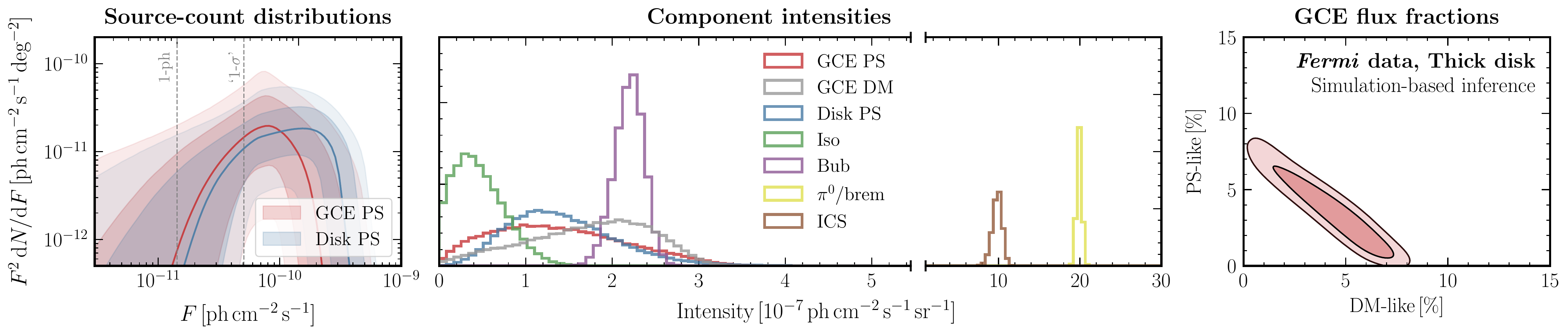}
\includegraphics[width=0.95\textwidth]{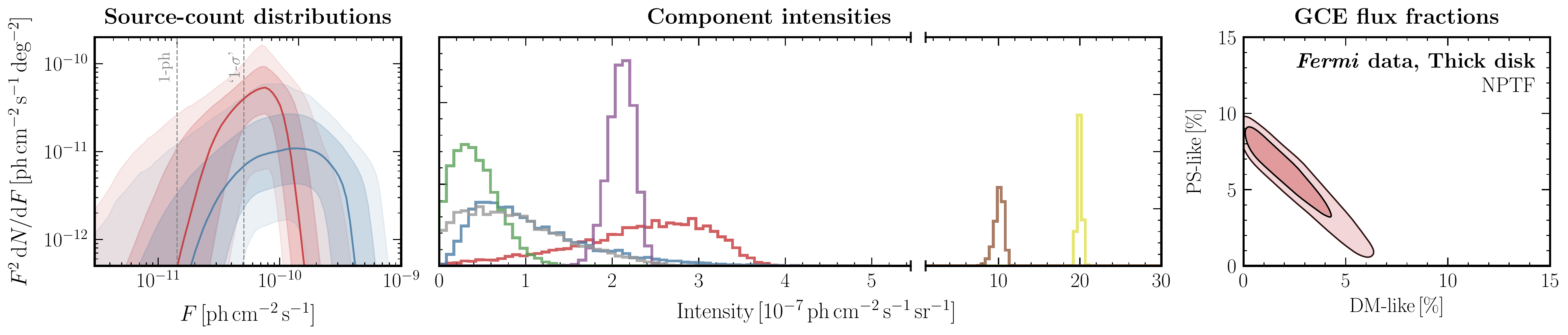}
\caption{Same as Fig.~\ref{fig:fid_data}, but with the spatial distribution of disk-correlated PSs modeled using a thick-disk template (scale factor $z_\mathrm{s}=1\,\mathrm{kpc}$ in Eq.~\eqref{eq:disk_spatial}) rather than the default thin-disk template ($z_\mathrm{s}=0.3\,\mathrm{kpc}$).}
\label{fig:fid_data_thick_disk}
\end{figure*}

We test the robustness of our results by exploring several systematic variations on the baseline analysis, using alternative descriptions for the diffuse foreground emission template, the spatial distribution of disk-correlated sources, and prior configuration. Here, the neural network is re-trained on a new set of simulations obtained using the alternative forward model before applying it to \Fermi data. Results of these analysis variations are summarized in Tab.~\ref{tab:results}. \\

\noindent
\textbf{Variation on the diffuse foreground model:}
In addition to diffuse {Model~O} considered in the baseline analysis, we consider the alternative Models A and F from Ref.~\cite{Calore:2014xka} to model the diffuse foreground emission, again including separate templates for gas-correlated emission and inverse Compton scattering. While shown to be a worse fit to the present dataset~\cite{Buschmann:2020adf}, these models have been previously used in the GCE literature~\cite{Buschmann:2020adf,Leane:2020pfc,Leane:2020nmi} and provide a useful comparison point.

Results for these variations are shown in Figs.~\ref{fig:fid_data_modelA} and~\ref{fig:fid_data_modelF}, respectively. In each case, results using the SBI pipeline are shown in the top row, with corresponding results using the NPTF pipeline in the bottom row. 
A somewhat larger fraction of the GCE, $57.3^{+9.9}_{-25.6}$, is attributed to PSs when using diffuse {Model~A} (Fig.~\ref{fig:fid_data_modelA}) compared to the baseline analysis using {Model~O}. The corresponding NPTF analysis finds a still larger fraction of $74.9^{+6.6}_{-22.5}\%$. Using {Model~F}, a similar $59.4^{+10.4}_{-26.3}$ of the GCE is attributed to PSs, with qualitatively similar results found by the NPTF analysis. The total emission absorbed by the GCE in this case is about $\sim 60\%$ of that found in the baseline scenario. This is consistent with the results of Ref.~\cite{Buschmann:2020adf}, which found that the total GCE flux could vary by up to a factor of $\sim 2$ between analyses using different diffuse models. \\

\noindent
\textbf{Variation on the disk template:}
The baseline scenario considered a disk-correlated PS population with a spatial distribution given by Eq.~\eqref{eq:disk_spatial}, setting the scale height scale height $z_\mathrm{s} = 0.3\,\mathrm{kpc}$ corresponding to the `thin-disk' scenario. Given uncertainties in the spatial distribution of the point source population (in particular, that of millisecond pulsars) associated with the Galactic disk, a `thick-disk' spatial distribution has been employed in the literature as an alternative model~\cite{Lee:2015fea,Leane:2019xiy,Buschmann:2020adf}, where the scale height is typically set to $z_\mathrm{s} = 1\,\mathrm{kpc}$. 

Results using a thick-disk template for the disk-correlated PS population are shown in Fig.~\ref{fig:fid_data_thick_disk}. For the SBI analysis, a slightly larger fraction $42.2^{+9.6}_{-21.0}$ of the GCE flux is attributed to a PS population in this case compared to the baseline scenario. Once again, the NPTF analysis estimates a higher relative fraction $75.0^{+7.1}_{-22.6}\%$ of the GCE in point sources. The emission attributed to disk-correlated PSs is reduced in this case compared to the baseline scenario, possibly indicating a redistribution of PS-like emission between the GCE- and disk-correlated components. \\

\noindent
\textbf{Alternative prior specification:}
In the baseline analysis, we have chosen to enforce a soft distinction between relatively-bright PSs emitting $\gtrsim 5$\,photons in expectation, and a combination of dimmer PSs and smooth emission following Poisson statistics taken together. This is done by placing a prior on the source-count slope below this chosen counts threshold that encourages a steeply-falling distribution with decreasing PS flux, allowing for a conservative interpretation of our results as a lower bound on the amount of PS emission. We also explore an alternative configuration where the lower break on the SCD is allowed to go down to a single photon, giving the PS component more overlap with the dim emission, and thus accounting for more emission in the PS-like component. The results of this analysis on data are summarized in the last row of Tab.~\ref{tab:results}.

Reassuringly, the total flux attributed to the GCE is consistent between the alternative and baseline prior choices. As expected, allowing the lower SCD break to go down to smaller expected counts increases the fraction of the GCE flux attributable to PS-like emission, for the SBI case increasing the median fraction by $\sim16\%$ relative to the baseline case. The NPTF analysis sees a larger increase in PS flux, with the median increasing by $\sim23\%$. The fact that the NPTF analysis is relatively more sensitive to the details of modeling close to the single-photon limit is not surprising---since mismodeling is most likely to affect this dimmer regime in the source-count distribution, this is also where the two methods can be expected to diverge more significantly. 
The position of the upper flux break, quantifying the inferred fluxes of the brightest sources in the PS population, is slightly reduced to $0.9^{+0.2}_{-0.4}\times 10^{-10}$\,ph\,cm$^{-2}$\,s$^{-1}$, corresponding to 5--7\,photons.
The posterior distributions for these cases, as well as the prior distribution on the source-count distribution corresponding to the two prior choices, are shown in App.~\ref{app:priors}. There, we also check that the inferred emission below 5 photons is consistently redistributed between the PS-like and DM-like components when using the two different prior configurations.

\section{Susceptibility to model misspecification}
\label{sec:mismodeling}

\begin{figure*}[!htbp]
\centering
\includegraphics[width=0.95\textwidth]{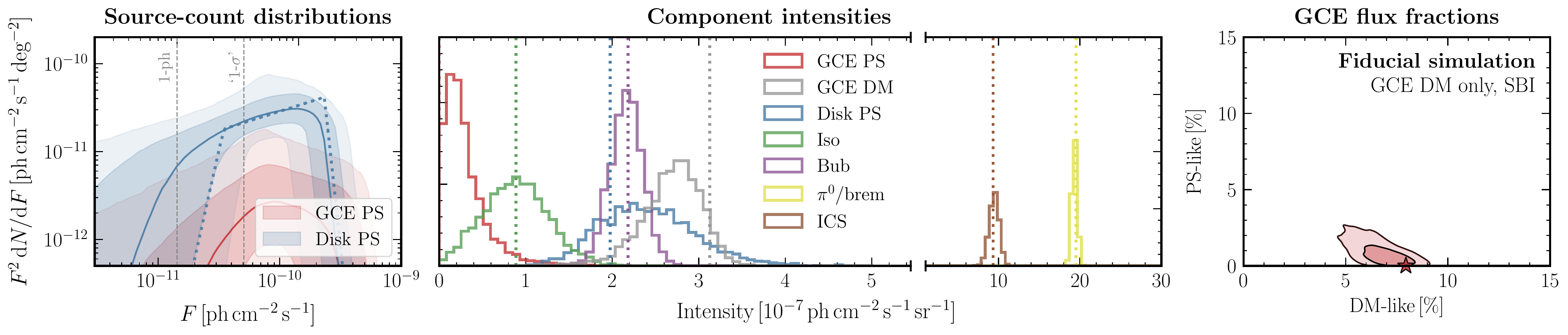}
\includegraphics[width=0.95\textwidth]{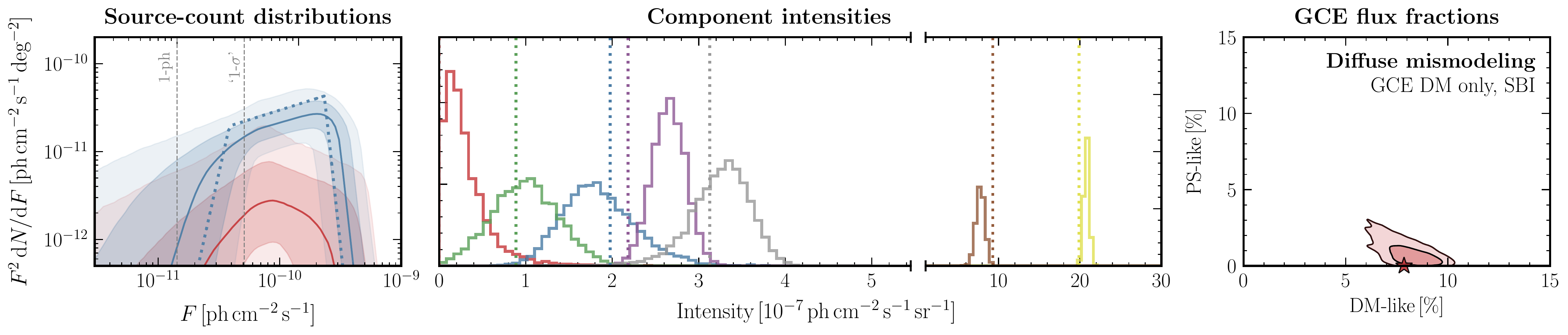}
\includegraphics[width=0.95\textwidth]{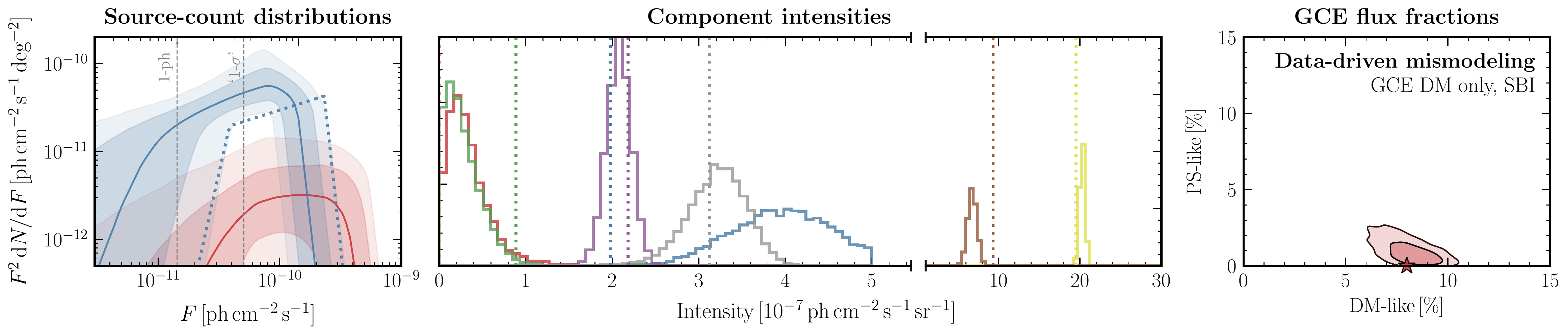}
\includegraphics[width=0.95\textwidth]{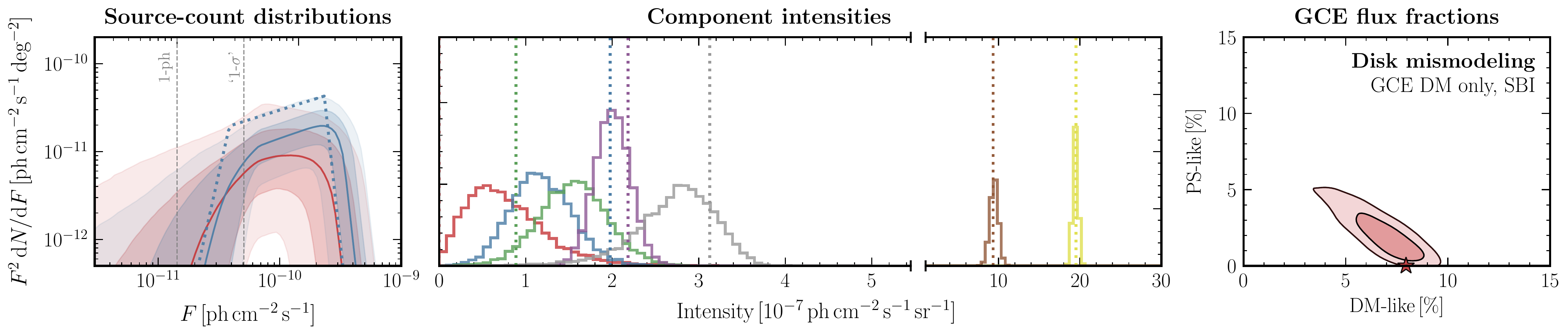}
\includegraphics[width=0.95\textwidth]{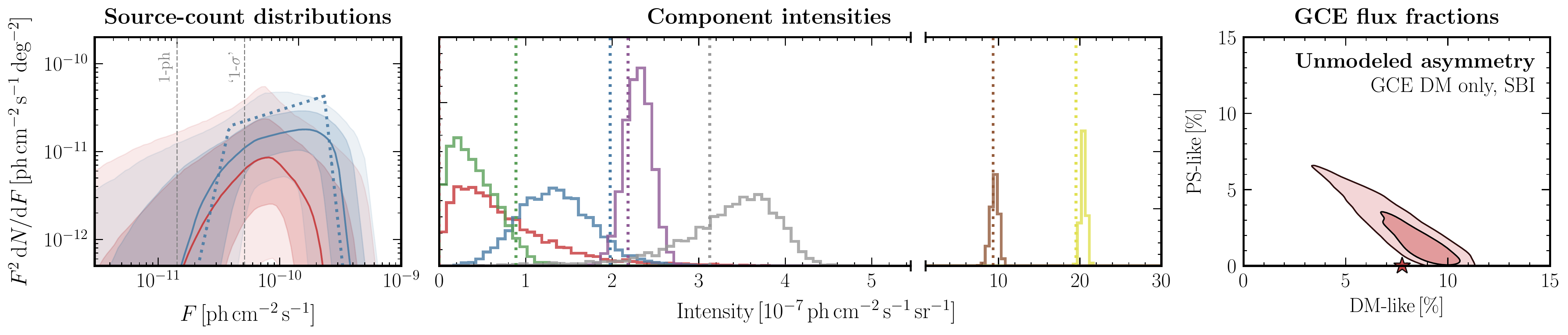}
\caption{Effect of mismodeling on a smooth GCE within our analysis framework. Each row shows aggregate posteriors collected over 10 simulated samples; row-wise from top to bottom: \emph{(i)}~No mismodeling; simulated data is constructed with the same templates as those used in the forward model for training. \emph{(ii)}~Mock data created with diffuse {Model~A}, showing a possible effect of diffuse mismodeling. \emph{(iii)}~Mock data where the diffuse template, described by {Model~O}, is modulated by draws from a Gaussian process modeling large-scale mismodeling inferred from the real \Fermi data. \emph{(iv)}~Mock data where the thick-disk template is used in lieu of the thin-disk template. \emph{(v)}~Mock data where the GCE signal in the Northern hemisphere is twice as large as that in the Southern hemisphere. While some PS-like emission is inferred, it is consistent with zero in all cases, and evidence for a smooth GCE is robust.}
\label{fig:sim_sbi_mismo}
\end{figure*}

Given the complex astrophysical environment in the Galactic Center, a key challenge in $\gamma$-ray analyses of the GCE is that associated with effects of mismodeled signal and background templates. As explored in detail in Refs.~\cite{Lee:2015fea,Leane:2020pfc,Leane:2020nmi,Buschmann:2020adf,Chang:2019ars} within the NPTF framework, mismodeling can hamper the characterization of an Inner Galaxy PS population and, if sufficiently severe, can result in the attribution of mismodeled residuals to a spurious PS population when the underlying emission is actually smooth in nature. 

In this section we assess the susceptibility of our simulation-based inference pipeline to several known sources of mismodeling. We do so by creating mock data with a smooth GCE signal and a background model that was perturbed compared to the model that was used to train the SBI pipeline, and analyzing it with our baseline neural network \emph{i.e.}, the one trained on the forward model described in Sec.~\ref{sec:datasets} and used in the baseline analysis on data in Sec.~\ref{sec:data}. The ability of our method to correctly characterize the injected signal is then indicative of the level of robustness that can be expected in the real data under corresponding circumstances. Results for the various tests performed are shown in Fig.~\ref{fig:sim_sbi_mismo}, and will be described below. 
In each case, we show posteriors obtained by combining 50,000 samples from analyses of 10 different mock datasets (thinned by a corresponding factor of 10) in order to characterize the `average' mismodeling associated with a given configuration. The first row of Fig.~\ref{fig:sim_sbi_mismo} shows the aggregate analysis without mismodeling \emph{i.e.}, conditioned on mock data created with the same forward model as that used for training the neural posterior estimator, as a point of comparison. \\

\noindent
\textbf{Test of diffuse mismodeling using an alternative diffuse emission template:}
We create mock data using diffuse {Model~A}, and analyze it using our baseline analysis pipeline trained with {Model~O}. The aggregated results over 10 different maps are shown in the second row of Fig.~\ref{fig:sim_sbi_mismo}. We see that even though some of the other diffuse component posteriors are shifted relative to their true values, the DM-like emission is faithfully recovered, and no additional PS-like flux is inferred. \\

\noindent
\textbf{A data-driven test of large-scale mismodeling:}
We construct a data-driven model of foreground mismodeling on large spatial scales (specifically, well above the scale of the instrumental PSF) and assess the ability of our method to recover a smooth DM-like signal in this case. Following Ref.~\cite{Mishra-Sharma:2020kjb}, we perform a Poissonian template analysis on the \Fermi dataset $x$, modulating the diffuse model template $T_{\mathrm{dif}}$, which describes the bremsstrahlung and neutral pion decay components of diffuse {Model~O}, by an (exponentiated) Gaussian process (GP) $f$:
\begin{equation}
x \sim \operatorname{Pois}\left(\sum_{i \neq \mathrm{dif}} A_{i} T_{i}+\exp \left(f\right) A_{\mathrm{dif}} T_{\mathrm{dif}}\right).
\end{equation}
The other Poissonian templates $T_{i}$, including a GCE DM template and the inverse Compton component of the diffuse foreground model, are treated as before using an overall normalization factor $A_{i}$. $f \sim \mathcal{N}(m, K)$ is the GP component with prior mean $m$ set to zero, and the covariance $K$ described using the Mat\'ern kernel with smoothness parameter $\nu = 5/2$. We refer to Ref.~\cite{Mishra-Sharma:2020kjb} for further details of the analysis, as well a validation of the GP-augmented template fitting pipeline on simulated data.

\begin{figure*}[t]
\centering
\includegraphics[width=1.\textwidth]{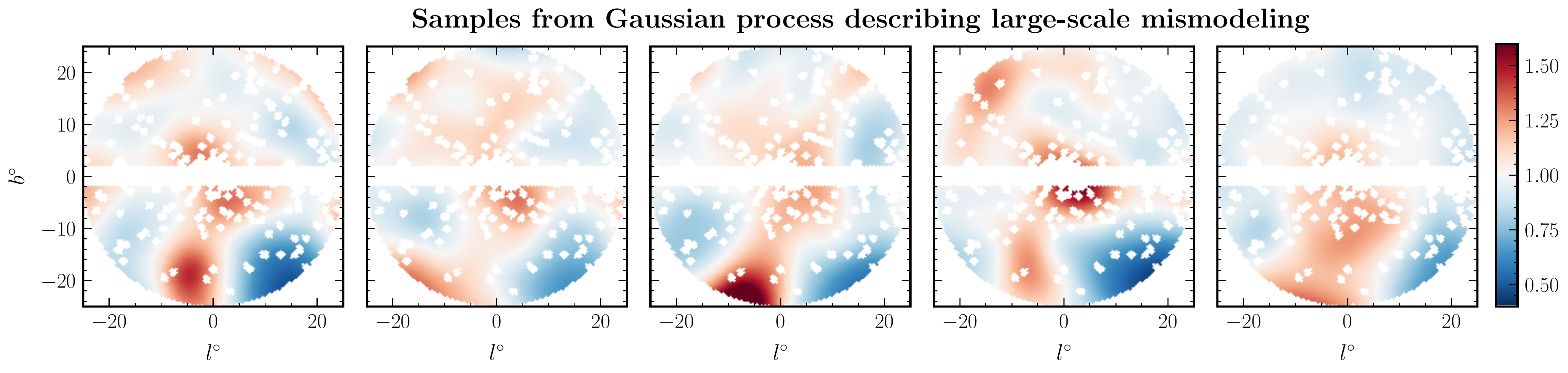}
\caption{Five random samples from the Gaussian process description of large-scale multiplicative mismodeling associated with the gas-correlated component of diffuse foreground {Model~O} when applied to the real \Fermi data.}
\label{fig:dd_mismo_map}
\end{figure*}

Five random samples from the Gaussian process describing multiplicative mismodeling relative to the real \Fermi data when using our baseline diffuse {Model~O} are shown in Fig.~\ref{fig:dd_mismo_map}. The largest mismodeling by magnitude in this case is inferred to be concentrated in the southern regions of the baseline ROI. We note that, when analyzing the real \Fermi data, the recovered GCE flux tends to be lower by up to 40\% when using the GP-modulated diffuse model compared to that obtain in a Poissonian fit without the GP, with the missing emission absorbed by the GP-modulated template. This is indicative of the fact that a component of the centrally concentrated emission could be better described by the modulated template rather than the generalized NFW template modeling DM annihilation. We leave a detailed study of implications of this fact for the morphology of the excess to future work. When creating simulated data containing DM-like emission in association with this modulated template, the fraction of DM-like flux in the simulation was correspondingly reduced by 40\%.

In order to test the effect of such mismodeling on recovery of a DM signal we modulate the bremsstrahlung and neutral pion decay-tracing components of {Model~O} using samples drawn from the inferred Gaussian process. These simulated samples are used as mock data that are then analyzed with our baseline model, where the unmodulated {Model~O} was used to create training samples.
The results of this test are shown in the third row of Fig.~\ref{fig:sim_sbi_mismo}. It can be seen that while large-scale mismodeling can distort the total flux attributed to individual modeled components, in particular causing the disk-correlated PS emission to be significantly overestimated, preference for a smooth origin of the signal remains robust. \\

\noindent
\textbf{Effect of mismodeling the disk spatial template:} 
We replace the thin-disk template, described by a scale height $z_\mathrm{s} = 0.3\,\mathrm{kpc}$ in Eq.~\eqref{eq:disk_spatial}, with a thick-disk template with $z_\mathrm{s} = 1\,\mathrm{kpc}$ in the simulated data. Results of then analyzing 10 mock maps using the thin-disk template used in the baseline configuration are shown in the fourth row of Fig.~\ref{fig:sim_sbi_mismo}. We see that the disk-correlated PS emission is underestimated, and a small amount of GCE PS-like emission is inferred while the marginal DM-like posterior is not significantly affected. This could be indicative of a reshuffling of the emission between disk- and GCE-correlated components. \\

\noindent
\textbf{Effect of an unmodeled asymmetry in the signal:}
Besides mismodeling associated with astrophysical background templates, another concern is that associated with mismodeling of the signal emission itself. In particular, as pointed out in Refs.~\cite{Leane:2020nmi,Leane:2020pfc}, a North-South asymmetry in a putative dark matter signal, if unaccounted for, could lead to spurious inference of a PS population associated with the purely smooth, asymmetric signal in the NPTF framework. Refs.~\cite{Leane:2020nmi,Leane:2020pfc} found preference for such a scenario in real \Fermi data, with the GCE signal in the Northern hemisphere a factor of $\sim2$ larger than that in the Southern hemisphere when the GCE template in the two regions is floated separately in a ROI defined by $r < 10^\circ$. In this case, for certain diffuse models, no preference for a PS-like GCE was found in contrast to the case when a single template was used to model the GCE. 

We test the impact of a North-South-asymmetric dark matter signal within our framework by running our baseline pipeline on simulated datasets where the dark matter-like signal in the Northern hemisphere of the ROI is 2 times larger than that in the Southern hemisphere, mimicking the preference in real data found in Refs.~\cite{Leane:2020nmi,Leane:2020pfc}. The result of this test over 10 such simulated realizations is shown in the last row of Fig.~\ref{fig:sim_sbi_mismo}. We see that even with the presence of a substantially asymmetric DM-like signal we retain preference for a predominantly smooth GCE. While some additional PS-like emission is inferred, the effect is small compared to that exhibited within the NPTF framework in analogous tests~\cite{Leane:2020nmi,Leane:2020pfc}. We attribute this to the fact that the \texttt{DeepSphere}-based convolutional neural network feature extractor can account for pixel-to-pixel correlations in the $\gamma$-ray counts map, and can thus be sensitive to \emph{local} PS-like structures. In contrast, the 1-point PDF-based NPTF framework, being agnostic to the ordering of the pixels, can notice spurious PS-like structures in the distribution of `residuals' associated with an asymmetric signal when analyzed with a symmetric template.
As done in Ref.~\cite{Buschmann:2020adf}, we emphasize that the presence of a substantial asymmetry in the GCE signal, if not attributed to diffuse mismodeling, would point towards astrophysical explanations of the GCE since a true dark matter signal would not be expected to be significantly asymmetric.

\bigskip

In all cases tested, while posteriors for certain templates can show systematic biases, preference for a smooth origin of the GCE remains robust and the fraction of inferred PS-like emission is compatible with zero. Finally, it is also interesting to similarly consider the effect of mismodeling on a PS-like GCE signal. We perform a subset of the tests described above on simulated GCE PS signals in App.~\ref{app:mismodeling_ps}, showing successful recovery of an overwhelmingly PS-like GCE in the face of mismodeling.

\section{Discussion and conclusions}
\label{sec:conclusion}

In this paper, we have leveraged recent advances in neural simulation-based inference in order to jointly characterize a putative DM-like signal and PS-like population associated with the observed \Fermi Galactic Center Excess. Consistent with Ref.~\cite{List:2020mzd} which used a Bayesian neural network and first leveraged a \texttt{DeepSphere}-based feature extraction architecture for analyzing $\gamma$-ray data in the Galactic Center region, our analysis based on conditional posterior density estimation with normalizing flows finds a reduced contribution associated with a potential population of unresolved PSs to the GCE compared to previous analyses based on the photon statistics of the $\gamma$-ray map. In particular, depending on the analysis configuration, we find a median value of $\sim40$--$60\%$ as the fraction of GCE emission that can be attributed to a PS population, with the brightest unresolved sources inferred to be at somewhat smaller fluxes $\sim 10^{-10}$\,ph\,cm$^{-2}$\,s$^{-1}$ compared to values found in previous analyses based on the non-Poissonian template fitting (NPTF) framework~\cite{Lee:2015fea}. The NPTF analyses performed in this work find a similarly dimmer source-count distribution, in all cases however attributing a larger fraction $\sim50$--$80\%$ of the GCE to a PS population as compared to the corresponding SBI analyses. Even though the SBI analyses presented here generically attributed a smaller fraction of the GCE flux to PSs, we note that there is significant overlap within posterior uncertainties between results returned by the two methods, as can be seen from Tab.~\ref{tab:results}. We note that, as detailed in Sec.~\ref{sec:analysis}, modulo details associated with the respective inference procedures (\emph{i.e.,} the posterior sampling algorithm in the NPTF case, and architecture specification and training in the SBI case), we expect differences in results between the two methods to be primarily driven by the distinct ways they respond to model misspecification and the existence of a finite point spread function.

It is interesting to ask whether the fraction of PS emission inferred in our analysis is consistent with physically-motivated models of astrophysical PS emission, \emph{e.g.} from a millisecond pulsar population, in the Galactic Center. While a detailed study is beyond the scope of this paper, Ref.~\cite{Dinsmore:2021nip} recently estimated this fraction for several models in the literature. They generally find an $\mathcal O(1)$ fraction of point sources to be resolvable (although with a wide range of variation given the uncertainties associated with modeling the astrophysical PS population), consistent with the results of our analysis. However, given the inherent degeneracy between a population of dim PSs and diffuse emission, a definitive statement about the origin of the smooth component cannot be made and more exotic contributors like dark matter annihilation cannot be ruled out.

The results of this paper are broadly consistent with and complementary to those obtained in Ref.~\cite{List:2021aer}, which used a \texttt{DeepSphere}-based architecture which was, in contrast to our parametric approach, combined with a novel neural network-based non-parameteric approach to infer the counts distributions associated to PS populations using histograms with modeled uncertainties~\cite{List2021}. Their approach does not explicitly distinguish between Poissonian and PS-like components, treating emission associated with the inferred counts PDF below some threshold as effectively Poissonian. While this makes a direct comparison to the results of their analysis challenging, the overall conclusions regarding the fraction of emission that can be attributed to PSs and the characteristics of the GCE and disk source-count distributions are qualitatively similar between the two studies. In particular, both papers find a dimmer GCE-correlated source-count distribution, with a smaller lower bound $\gtrsim \mathcal O(40\%)$ on the fraction of the total GCE emission associated to PSs compared to previous studies based on the NPTF.

Our qualitative conclusions are robust to the systematic variations we have explored, including different models for the diffuse foreground and spatial distribution of disk-correlated PS emission. We used a novel Gaussian process-based method to construct a data-driven model of large-scale spatial mismodeling, finding our method to be resilient to such effects when it comes to inferring the presence of DM-like emission. As in any Galactic Center $\gamma$-ray analysis, we caution of the potential of unknown systematics, such as mismodeling on the scale of the size of the \Fermi-LAT point-spread function, to bias the results and conclusions of our analysis. Although machine learning-based analyses can utilize more of the information encoded in the forward model, and in particular in the present case can take advantage of pixel-to-pixel correlations, this can also make them more susceptible to specific modeled features compared to traditional techniques based on data reduction to hand-crafted data summaries. We leave a more detailed investigation of the potential impact of these effects on our analysis to future work.

Several improvements to the framework presented here are possible. Although we have used a dataset restricted to the top quartile of photons by quality of PSF reconstruction, as shown in Ref.~\cite{Leane:2020pfc} the use of a larger data sample can provide improved sensitivity to a PS population while acting as a consistency check with results obtained on the smaller sample. Since our method does not rely on an approximate treatment of the PSF and can exploit pixel-to-pixel correlations in inferring the presence of PS populations, we expect that it should be able to better handle the presence of a larger PSF compared to the NPTF approach. For the same reason, widening the energy range employed below 2\,GeV may provide improved sensitivity since the GCE signal extends to lower energies. The inclusion of energy-binning information in the analysis can be implemented in a straightforward manner by splitting up the data and template maps into individual energy bins and feeding these as separate channels in the graph-convolutional feature extraction network. The use of more complex feature extraction architectures can additionally improve the robustness of our results. 
While we have considered a simulated-based inference framework based on posterior density estimation with normalizing flows, alternative frameworks based on likelihood-ratio estimation~\cite{Brehmer:2018eca,Brehmer:2018hga,Brehmer:2018kdj,Cranmer:2015bka, Hermans:2019ioj,Miller:2020hua,Miller:2021hys} or flow-based likelihood estimation~\cite{2019arXiv191200042W,pmlr-v89-papamakarios19a} can provide complementary ways to characterize the $\gamma$-ray PS population in the Galactic Center. Additionally, the use of sequential active-learning methods~\cite{pmlr-v89-papamakarios19a} and methods that make use of additional latent information from the simulator~\cite{Brehmer:2018eca,Brehmer:2018hga,Brehmer:2018kdj,Brehmer:2019xox,Stoye:2018ovl} can significantly improve the simulation sample efficiency and allow for extensions to more complex forward models, which can be important in particular for an energy-binned analysis and if including additional degrees of freedom for the astrophysical background models. 

Since diffuse mismodeling is the largest source of uncertainty in any analysis that aims to characterize the GCE, we also note the possibility of using adversarial learning methods~\cite{Louppe:2016ylz} or distance correlations~\cite{Kasieczka:2020yyl} to account for systematic differences between the modeled and real \Fermi data. Alternatively, generative modeling of the diffuse foreground either in a Gaussian process-based data-driven framework or using, \emph{e.g.}, autoencoders trained on an ensemble of plausible diffuse models, can provide a principled way to account for the large latent space associated with diffuse emission modeling. Motivated by quantitative variations in our results on \Fermi data when using different disk templates, self-consistently accounting for plausible variations in the spatial distribution of disk-correlated PSs can strengthen the results of our analysis when it comes to characterizing the PS population in the Galactic Center. 
These extensions can lead to a more robust characterization of an unresolved PS population in the Galactic Center region associated with the GCE, and we leave their study to future work.

The code used to obtain the results in this paper as well as a pre-trained neural network model associated with the baseline analysis presented here is available at \url{https://github.com/smsharma/fermi-gce-flows}.

\vspace{.2cm}

\begin{acknowledgments}

We thank Johann Brehmer and Tracy Slatyer for helpful conversations. We are grateful to Florian List and Nick Rodd for carefully reading an earlier version of this paper and for their many helpful comments.
SM would like to thank the Center for Computational Astrophysics at the Flatiron Institute for their hospitality while this work was being performed. 
This work was performed in part at the Aspen Center for Physics, which is supported by National Science Foundation grant PHY-1607611.
The participation of SM at the Aspen Center for Physics was supported by the Simons Foundation.
SM is supported by the NSF CAREER grant PHY-1554858, NSF grants PHY-1620727 and PHY-1915409, and the Simons Foundation. 
KC is partially supported by NSF awards ACI-1450310, OAC-1836650, and OAC-1841471, the NSF grant PHY-1505463, and the Moore-Sloan Data Science Environment at NYU. 
This work is supported by the National Science Foundation under Cooperative Agreement PHY-2019786 (The NSF AI Institute for Artificial Intelligence and Fundamental Interactions, \url{http://iaifi.org/}).
This material is based upon work supported by the U.S. Department of Energy, Office of Science, Office of High Energy Physics of U.S. Department of Energy under grant Contract Number DE-SC0012567.
We thank the \Fermi-LAT Collaboration for making publicly available the $\gamma$-ray data used in this work.
This work made use of the NYU IT High Performance Computing resources, services, and staff expertise. 
This research has made use of NASA's Astrophysics Data System. 
This research made use of the \texttt{astropy}~\cite{Price-Whelan:2018hus,Robitaille:2013mpa}, \texttt{dynesty}~\cite{Speagle_2020}, \texttt{getdist}~\cite{Lewis:2019xzd}, \texttt{IPython}~\cite{PER-GRA:2007}, \texttt{Jupyter}~\cite{Kluyver2016JupyterN}, \texttt{matplotlib}~\cite{Hunter:2007}, \texttt{MLflow}~\cite{10.1145/3399579.3399867}, \texttt{nflows}~\cite{nflows}, \texttt{NPTFit}~\cite{Mishra-Sharma:2016gis}, \texttt{NPTFit-Sim}~\cite{NPTFit-Sim}, \texttt{NumPy}~\cite{harris2020array}, \texttt{pandas}~\cite{pandas:2010}, \texttt{PyGSP}~\cite{michael_defferrard_2017_1003158}, \texttt{Pyro}~\cite{bingham2019pyro}, \texttt{PyTorch}~\cite{NEURIPS2019_9015}, \texttt{PyTorch Geometric}~\cite{Fey/Lenssen/2019}, \texttt{PyTorch Lightning}~\cite{william_falcon_2020_3828935}, \texttt{seaborn}~\cite{seaborn}, \texttt{sbi}~\cite{tejero-cantero2020sbi}, \texttt{scikit-learn}~\cite{JMLR:v12:pedregosa11a}, \texttt{SciPy}~\cite{2020SciPy-NMeth}, and \texttt{tqdm}~\cite{casper_da_costa_luis_2021_5517697} software packages. We acknowledge the use of data products and templates from the code repository associated with Ref.~\cite{List:2020mzd}.\footnote{\url{https://github.com/FloList/GCE_NN}} We acknowledge use of the \texttt{DeepSphere} graph-convolutional layer from the code repository associated with Ref.~\cite{DBLP:conf/iclr/DefferrardMGP20}.\footnote{\url{https://github.com/deepsphere/deepsphere-pytorch}}
\end{acknowledgments}

\vspace{.5cm}

\appendix
\section*{Appendix}

\section{Prior-predictive distributions and results for alternative priors}
\label{app:priors}

\begin{figure*}[!t]
\centering
\includegraphics[width=0.9\textwidth]{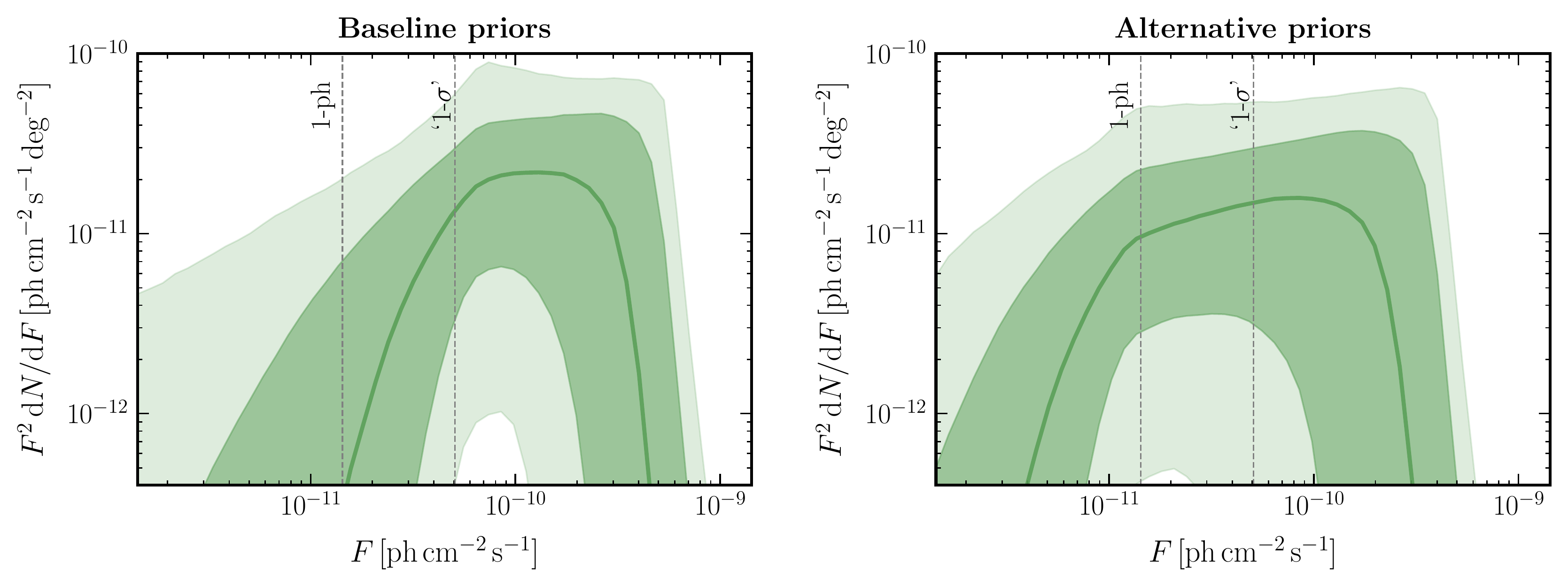}
\caption{Prior-predictive distribution on the source-count distribution for the baseline PS model priors (left) and alternative prior specification giving the PS component more overlap with emission close to the single-photon limit (right). The median (lines), middle-68\% containment (darker bands), and middle-95\% containment (lighter bands) regions are shown.}
\label{fig:pp_check}
\end{figure*}

Figure~\ref{fig:pp_check} shows the prior distribution induced on the source-count distribution for the baseline PS model with upper SCD break priors uniform in the interval $S_{\mathrm{b}, 1}  \in  [5, 40]$\,photons (left) and the alternative prior specification with $S_{\mathrm{b}, 1}  \in  [1, 30]$\,photons (right). The latter prescription gives the PS-like component more overlap with emission just above 1-photon, since the slope below the second break encourages the SCD to steeply drop. It can be seen that both prior choices still allow for significant PS-like emission below their respective counts soft thresholds.

\begin{figure*}[!t]
\centering
\includegraphics[width=1.\textwidth]{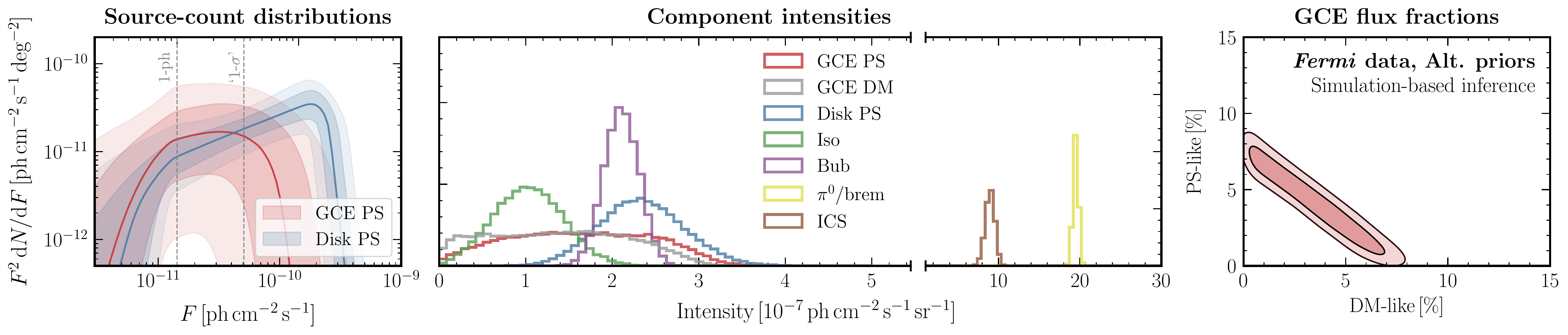}
\includegraphics[width=1.\textwidth]{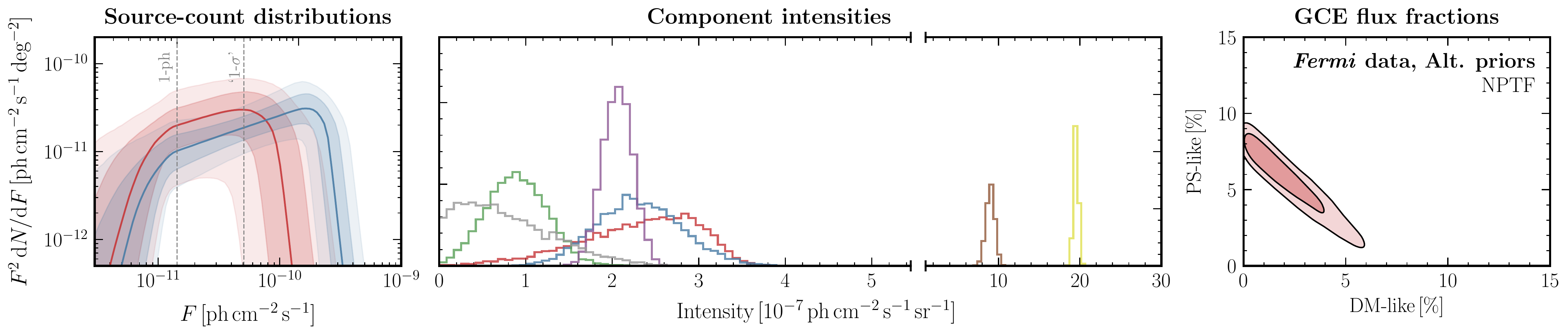}
\caption{Same as Fig.~\ref{fig:fid_data}, but using the alternative prior specification for the PS model, with the break on the lower SCD break uniform within $S_{\mathrm{b}, 1}  \in  [1, 30]$\,photons rather than $S_{\mathrm{b}, 1}  \in  [5, 40]$\,photons. This configuration gives the PS model more overlap with the dim PS-like emission.}
\label{fig:fid_data_alt_priors}
\end{figure*}

Figure~\ref{fig:fid_data_alt_priors} shows posterior distributions for the analysis using the alternative prior set. These results are summarized in the bottom row of Tab.~\ref{tab:results}. As expected, both the SBI (top row) and NPTF (bottom row) analyses show a larger inferred PS flux compared to the analysis using the baseline prior choice. Reassuringly, the total flux absorbed by both GCE components taken together remains consistent between the analyses with different prior choices.

\begin{figure*}[!htbp]
\floatbox[{\capbeside\thisfloatsetup{capbesideposition={right,center},capbesidewidth=0.32\textwidth}}]{figure}[\FBwidth]
{\caption{A heuristic check of the distribution of PS-like flux below 5 photons in the analyses with baseline and alternative priors. The excess dark matter flux (shown as counts per pixel $\langle S \rangle$) in the baseline prior configuration (topmost data point) is seen to be consistent with the cumulative excess flux below 5 photons in the alternative prior configuration compared to the baseline one (second data point from the top). When this excess flux is added to the total PS flux in the baseline configuration (middle data point), the combination (second data point from the bottom) is additionally seen to be consistent with the total PS flux in the alternative prior configuration (bottommost data point).}
\label{fig:consistency}}
{\includegraphics[width=0.6\textwidth]{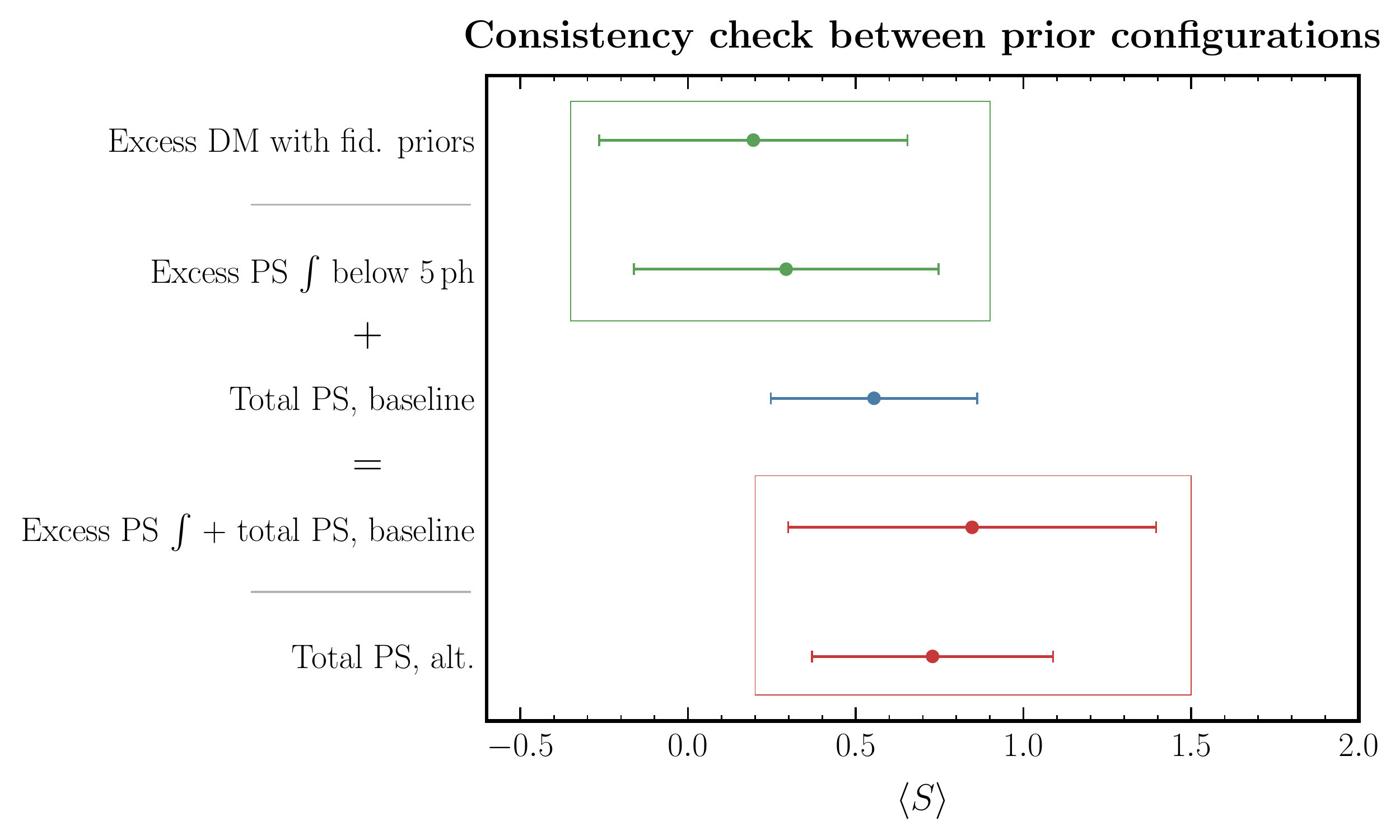}}
\end{figure*}

Finally, Fig.~\ref{fig:consistency} shows a check of how the partitioning of flux between PS-like and DM-like components varies between the two prior choices. 
The excess dark matter flux (shown as inferred counts per-pixel $\langle S \rangle$) in the baseline prior configuration (topmost data point) is seen to be consistent with the cumulative excess flux below 5 photons in the alternative prior configuration compared to the baseline one (second data point from the top). When this excess flux is added to the total PS flux in the baseline configuration (middle data point), the combination (second data point from the bottom) is additionally seen to be consistent with the total PS flux in the alternative prior configuration (bottommost data point). We note that this test is merely heuristic---in particular, since the posteriors for baseline and alternative prior analyses are described by independent samples, the component counts were combined or subtracted assuming uncorrelated errors (computed as standard deviations over the respective posteriors), which is certain to not be the case. However, this test is indicative of the fact that the inferred flux below the threshold we set for accounting purposes is redistributed between the PS and DM components, as would be expected if the two analyses were self-consistent.


\section{Mismodeling effects on a simulated GCE PS signal}
\label{app:mismodeling_ps}

\begin{figure*}[!htbp]
\centering
\includegraphics[width=1.\textwidth]{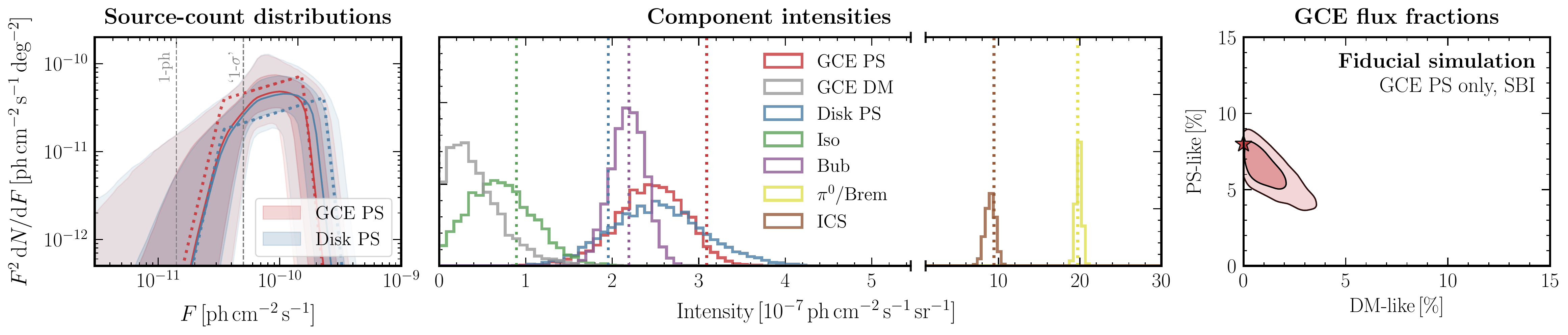}
\includegraphics[width=1.\textwidth]{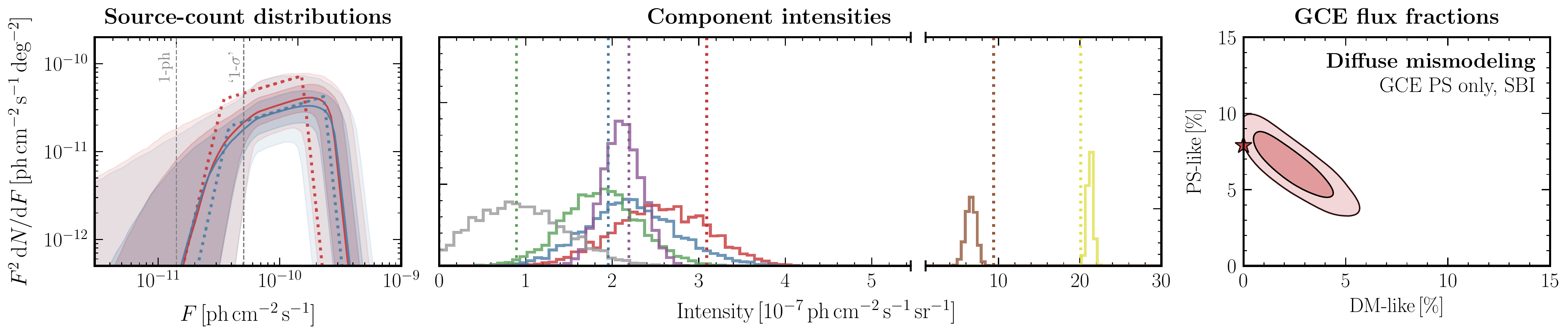}
\includegraphics[width=1.\textwidth]{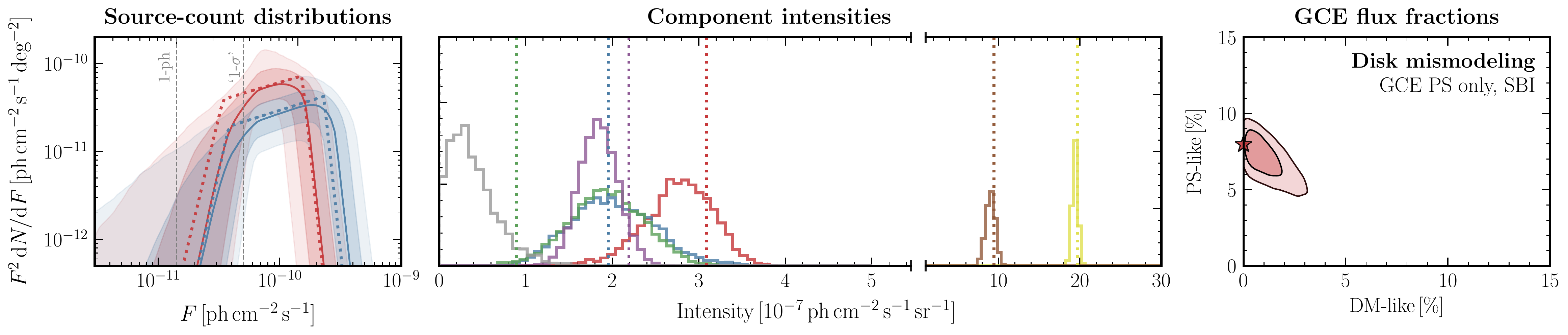}
\caption{Effect of mismodeling on a PS-like GCE within our analysis framework. Each row shows aggregate posteriors collected over 10 simulated samples; row-wise from top to bottom: \emph{(i)}~No mismodeling; simulated data is constructed with the same templates as those used in the forward model. \emph{(ii)}~Mock data created with diffuse foreground {Model~A}, showing a possible effect of diffuse mismodeling. \emph{(iii)}~Mock data where the thick-disk template is used in lieu of the thin-disk template. A substantially PS-like GCE is inferred, although a subdominant fraction of DM-like flux is inferred as well when considering diffuse foreground mismodeling.}
\label{fig:sim_sbi_mismo_ps}
\end{figure*}

Figure~\ref{fig:sim_sbi_mismo_ps} shows the analog of Fig.~\ref{fig:sim_sbi_mismo} where we test the effect of mismodeling on a PS-like rather than smooth GCE. As in the test on simulated maps with a smooth GCE described in Sec.~\ref{sec:mismodeling}, the data containing GCE-correlated PSs is created using a forward model that is different in a specific way from that used do train the neural network model: \emph{(i)}~No mismodeling; simulated data is constructed with the same templates as those in the forward model used for training the posterior estimator (top row). \emph{(ii)}~Mock data created with diffuse {Model~A}, showing the effect of diffuse mismodeling (middle row). \emph{(iii)}~Mock data where the thick-disk template is used in lieu of the thin-disk template (bottom row). 

In each case, the aggregate posterior described by 50,000 samples obtained over 10 simulations and then thinned by a factor of 10 is shown. Since the GP-modulated (smooth) diffuse template tends to absorb a substantial fraction of the GCE flux when applied to real \Fermi data, a test using this modulated template was not performed here as it would not yield self-consistent results.
It can be seen from the rightmost column of Fig.~\ref{fig:sim_sbi_mismo_ps} that a substantially PS-like GCE is recovered in both cases tested, although a small fraction of flux is attributed to DM-like emission when mismodeling of the diffuse emission template is considered.


\section{Effect of allowing a negative normalization for the DM template}
\label{app:negative_dm}

Ref.~\cite{Leane:2019xiy} showed that systematic mismodeling in GCE analyses can be manifest in terms of unphysical negative values of the normalizations of spatial templates, and in particular that a significantly negative Poissonian DM template normalization can be symptomatic of oversubstraction due to diffuse mismodeling. We perform a version of this test by re-training our our model using simulated samples where the prior on the normalization of a DM-like signal was allowed be negative---specifically, it was modified from $\langle S_{\rm GCE}^{\rm Poiss} \rangle \in [0, 2.5]\,\mathrm{ph/pix}$ in the baseline case to  $\langle S_{\rm GCE}^{\rm Poiss} \rangle \in [-1, 2.5]\,\mathrm{ph/pix}$. We note that since the amplitude of the GCE signal is much smaller than that of the Galactic diffuse emission, which is restricted to be positive, the total expected counts in a pixel are never negative even when allowing for a negative normalization for the DM template.

\begin{figure*}[!htbp]
\centering
\includegraphics[width=1.\textwidth]{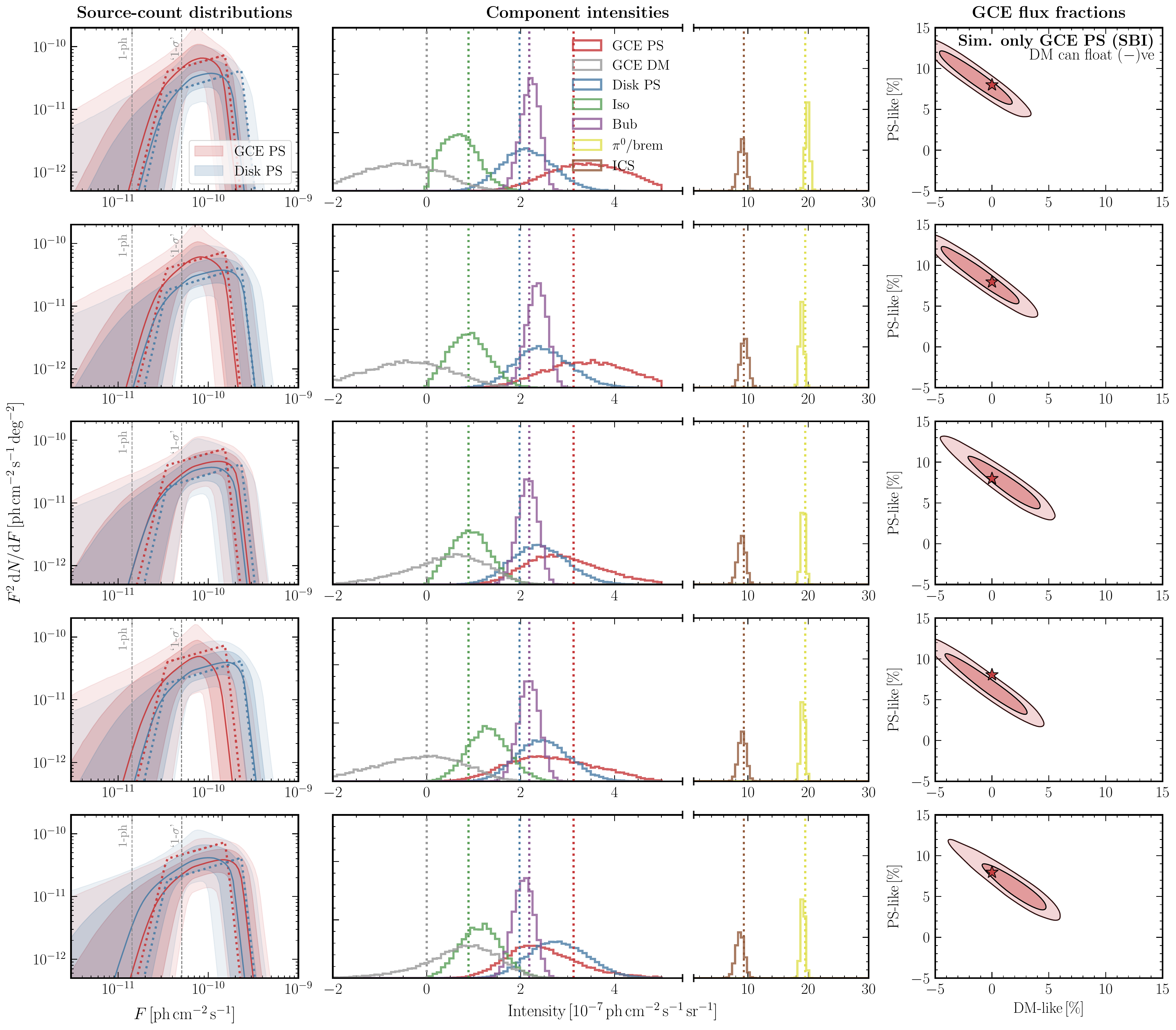}
\caption{Same as Fig.~\ref{fig:sim_sbi_ps} (SBI analysis on simulated \Fermi data with PSs only), but with the normalization of the DM-like template allowed to go negative. The posteriors for the DM-like flux now contain zero, as expected for a consistent analysis.}
\label{fig:sim_sbi_ps_neg}
\end{figure*}

Results with this model on simulations containing purely PS-like emission, in analogy with those shown in Fig.~\ref{fig:sim_sbi_ps} for the baseline case, are shown in Fig.~\ref{fig:sim_sbi_ps_neg}. The posterior on the contribution of DM-like emission now contains zero and, over the simulation ensemble, is roughly centered on it. This is as expected for a consistency when purely PS-like emission is present, further validating the analysis.

\begin{figure*}[!htbp]
\centering
\includegraphics[width=1.\textwidth]{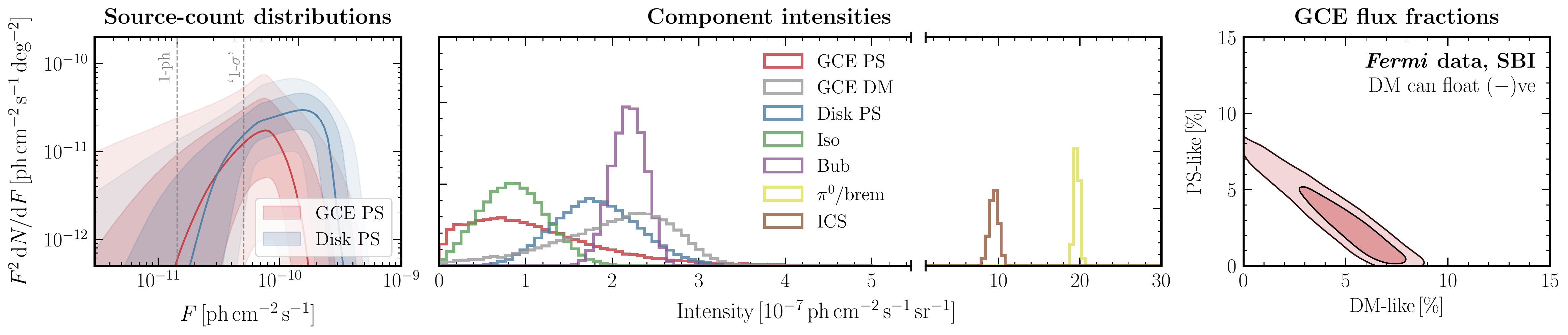}
\caption{Same as Fig.~\ref{fig:fid_data} (SBI analysis on real \Fermi data), but with the normalization of the DM-like template allowed to take on negative values. Results consistent with the baseline analysis are seen, and the DM-like flux does not take on unphysical negative values.}
\label{fig:fermi_sbi_neg}
\end{figure*}

We apply this model to \Fermi data, showing results in Fig.~\ref{fig:fermi_sbi_neg}. We see that the inferred posteriors are consistent with those obtained in the baseline analysis in Fig.~\ref{fig:fid_data} where the DM-like flux is restricted to positive values; in particular the DM-like flux does not tend to unphysical negative values.


\clearpage

\bibliographystyle{apsrev4-1}
\bibliography{fermi-gce-flows}

\begin{thebibliography}{128}%
\makeatletter
\providecommand \@ifxundefined [1]{%
 \@ifx{#1\undefined}
}%
\providecommand \@ifnum [1]{%
 \ifnum #1\expandafter \@firstoftwo
 \else \expandafter \@secondoftwo
 \fi
}%
\providecommand \@ifx [1]{%
 \ifx #1\expandafter \@firstoftwo
 \else \expandafter \@secondoftwo
 \fi
}%
\providecommand \natexlab [1]{#1}%
\providecommand \enquote  [1]{``#1''}%
\providecommand \bibnamefont  [1]{#1}%
\providecommand \bibfnamefont [1]{#1}%
\providecommand \citenamefont [1]{#1}%
\providecommand \href@noop [0]{\@secondoftwo}%
\providecommand \href [0]{\begingroup \@sanitize@url \@href}%
\providecommand \@href[1]{\@@startlink{#1}\@@href}%
\providecommand \@@href[1]{\endgroup#1\@@endlink}%
\providecommand \@sanitize@url [0]{\catcode `\\12\catcode `\$12\catcode
  `\&12\catcode `\#12\catcode `\^12\catcode `\_12\catcode `\%12\relax}%
\providecommand \@@startlink[1]{}%
\providecommand \@@endlink[0]{}%
\providecommand \url  [0]{\begingroup\@sanitize@url \@url }%
\providecommand \@url [1]{\endgroup\@href {#1}{\urlprefix }}%
\providecommand \urlprefix  [0]{URL }%
\providecommand \Eprint [0]{\href }%
\providecommand \doibase [0]{http://dx.doi.org/}%
\providecommand \selectlanguage [0]{\@gobble}%
\providecommand \bibinfo  [0]{\@secondoftwo}%
\providecommand \bibfield  [0]{\@secondoftwo}%
\providecommand \translation [1]{[#1]}%
\providecommand \BibitemOpen [0]{}%
\providecommand \bibitemStop [0]{}%
\providecommand \bibitemNoStop [0]{.\EOS\space}%
\providecommand \EOS [0]{\spacefactor3000\relax}%
\providecommand \BibitemShut  [1]{\csname bibitem#1\endcsname}%
\let\auto@bib@innerbib\@empty
\bibitem [{\citenamefont {Atwood}\ \emph {et~al.}(2009)\citenamefont {Atwood}
  \emph {et~al.}}]{Atwood:2009ez}%
  \BibitemOpen
  \bibfield  {author} {\bibinfo {author} {\bibfnamefont {W.~B.}\ \bibnamefont
  {Atwood}} \emph {et~al.} (\bibinfo {collaboration} {Fermi-LAT}),\ }\href
  {\doibase 10.1088/0004-637X/697/2/1071} {\bibfield  {journal} {\bibinfo
  {journal} {Astrophys. J.}\ }\textbf {\bibinfo {volume} {697}},\ \bibinfo
  {pages} {1071} (\bibinfo {year} {2009})},\ \Eprint
  {http://arxiv.org/abs/0902.1089} {arXiv:0902.1089 [astro-ph.IM]} \BibitemShut
  {NoStop}%
\bibitem [{\citenamefont {Goodenough}\ and\ \citenamefont
  {Hooper}(2009)}]{Goodenough:2009gk}%
  \BibitemOpen
  \bibfield  {author} {\bibinfo {author} {\bibfnamefont {L.}~\bibnamefont
  {Goodenough}}\ and\ \bibinfo {author} {\bibfnamefont {D.}~\bibnamefont
  {Hooper}},\ }\href@noop {} {\  (\bibinfo {year} {2009})},\ \Eprint
  {http://arxiv.org/abs/0910.2998} {arXiv:0910.2998 [hep-ph]} \BibitemShut
  {NoStop}%
\bibitem [{\citenamefont {Hooper}\ and\ \citenamefont
  {Goodenough}(2011)}]{Hooper:2010mq}%
  \BibitemOpen
  \bibfield  {author} {\bibinfo {author} {\bibfnamefont {D.}~\bibnamefont
  {Hooper}}\ and\ \bibinfo {author} {\bibfnamefont {L.}~\bibnamefont
  {Goodenough}},\ }\href {\doibase 10.1016/j.physletb.2011.02.029} {\bibfield
  {journal} {\bibinfo  {journal} {Phys. Lett. B}\ }\textbf {\bibinfo {volume}
  {697}},\ \bibinfo {pages} {412} (\bibinfo {year} {2011})},\ \Eprint
  {http://arxiv.org/abs/1010.2752} {arXiv:1010.2752 [hep-ph]} \BibitemShut
  {NoStop}%
\bibitem [{\citenamefont {Boyarsky}\ \emph {et~al.}(2011)\citenamefont
  {Boyarsky}, \citenamefont {Malyshev},\ and\ \citenamefont
  {Ruchayskiy}}]{Boyarsky:2010dr}%
  \BibitemOpen
  \bibfield  {author} {\bibinfo {author} {\bibfnamefont {A.}~\bibnamefont
  {Boyarsky}}, \bibinfo {author} {\bibfnamefont {D.}~\bibnamefont {Malyshev}},
  \ and\ \bibinfo {author} {\bibfnamefont {O.}~\bibnamefont {Ruchayskiy}},\
  }\href {\doibase 10.1016/j.physletb.2011.10.014} {\bibfield  {journal}
  {\bibinfo  {journal} {Phys. Lett. B}\ }\textbf {\bibinfo {volume} {705}},\
  \bibinfo {pages} {165} (\bibinfo {year} {2011})},\ \Eprint
  {http://arxiv.org/abs/1012.5839} {arXiv:1012.5839 [hep-ph]} \BibitemShut
  {NoStop}%
\bibitem [{\citenamefont {Hooper}\ and\ \citenamefont
  {Linden}(2011)}]{Hooper:2011ti}%
  \BibitemOpen
  \bibfield  {author} {\bibinfo {author} {\bibfnamefont {D.}~\bibnamefont
  {Hooper}}\ and\ \bibinfo {author} {\bibfnamefont {T.}~\bibnamefont
  {Linden}},\ }\href {\doibase 10.1103/PhysRevD.84.123005} {\bibfield
  {journal} {\bibinfo  {journal} {Phys. Rev. D}\ }\textbf {\bibinfo {volume}
  {84}},\ \bibinfo {pages} {123005} (\bibinfo {year} {2011})},\ \Eprint
  {http://arxiv.org/abs/1110.0006} {arXiv:1110.0006 [astro-ph.HE]} \BibitemShut
  {NoStop}%
\bibitem [{\citenamefont {Abazajian}\ and\ \citenamefont
  {Kaplinghat}(2012)}]{Abazajian:2012pn}%
  \BibitemOpen
  \bibfield  {author} {\bibinfo {author} {\bibfnamefont {K.~N.}\ \bibnamefont
  {Abazajian}}\ and\ \bibinfo {author} {\bibfnamefont {M.}~\bibnamefont
  {Kaplinghat}},\ }\href {\doibase 10.1103/PhysRevD.86.083511} {\bibfield
  {journal} {\bibinfo  {journal} {Phys. Rev. D}\ }\textbf {\bibinfo {volume}
  {86}},\ \bibinfo {pages} {083511} (\bibinfo {year} {2012})},\ \bibinfo {note}
  {[Erratum: Phys.Rev.D 87, 129902 (2013)]},\ \Eprint
  {http://arxiv.org/abs/1207.6047} {arXiv:1207.6047 [astro-ph.HE]} \BibitemShut
  {NoStop}%
\bibitem [{\citenamefont {Hooper}\ and\ \citenamefont
  {Slatyer}(2013)}]{Hooper:2013rwa}%
  \BibitemOpen
  \bibfield  {author} {\bibinfo {author} {\bibfnamefont {D.}~\bibnamefont
  {Hooper}}\ and\ \bibinfo {author} {\bibfnamefont {T.~R.}\ \bibnamefont
  {Slatyer}},\ }\href {\doibase 10.1016/j.dark.2013.06.003} {\bibfield
  {journal} {\bibinfo  {journal} {Phys. Dark Univ.}\ }\textbf {\bibinfo
  {volume} {2}},\ \bibinfo {pages} {118} (\bibinfo {year} {2013})},\ \Eprint
  {http://arxiv.org/abs/1302.6589} {arXiv:1302.6589 [astro-ph.HE]} \BibitemShut
  {NoStop}%
\bibitem [{\citenamefont {Gordon}\ and\ \citenamefont
  {Macias}(2013)}]{Gordon:2013vta}%
  \BibitemOpen
  \bibfield  {author} {\bibinfo {author} {\bibfnamefont {C.}~\bibnamefont
  {Gordon}}\ and\ \bibinfo {author} {\bibfnamefont {O.}~\bibnamefont
  {Macias}},\ }\href {\doibase 10.1103/PhysRevD.88.083521} {\bibfield
  {journal} {\bibinfo  {journal} {Phys. Rev. D}\ }\textbf {\bibinfo {volume}
  {88}},\ \bibinfo {pages} {083521} (\bibinfo {year} {2013})},\ \bibinfo {note}
  {[Erratum: Phys.Rev.D 89, 049901 (2014)]},\ \Eprint
  {http://arxiv.org/abs/1306.5725} {arXiv:1306.5725 [astro-ph.HE]} \BibitemShut
  {NoStop}%
\bibitem [{\citenamefont {Abazajian}\ \emph {et~al.}(2014)\citenamefont
  {Abazajian}, \citenamefont {Canac}, \citenamefont {Horiuchi},\ and\
  \citenamefont {Kaplinghat}}]{Abazajian:2014fta}%
  \BibitemOpen
  \bibfield  {author} {\bibinfo {author} {\bibfnamefont {K.~N.}\ \bibnamefont
  {Abazajian}}, \bibinfo {author} {\bibfnamefont {N.}~\bibnamefont {Canac}},
  \bibinfo {author} {\bibfnamefont {S.}~\bibnamefont {Horiuchi}}, \ and\
  \bibinfo {author} {\bibfnamefont {M.}~\bibnamefont {Kaplinghat}},\ }\href
  {\doibase 10.1103/PhysRevD.90.023526} {\bibfield  {journal} {\bibinfo
  {journal} {Phys. Rev. D}\ }\textbf {\bibinfo {volume} {90}},\ \bibinfo
  {pages} {023526} (\bibinfo {year} {2014})},\ \Eprint
  {http://arxiv.org/abs/1402.4090} {arXiv:1402.4090 [astro-ph.HE]} \BibitemShut
  {NoStop}%
\bibitem [{\citenamefont {Daylan}\ \emph {et~al.}(2016)\citenamefont {Daylan},
  \citenamefont {Finkbeiner}, \citenamefont {Hooper}, \citenamefont {Linden},
  \citenamefont {Portillo}, \citenamefont {Rodd},\ and\ \citenamefont
  {Slatyer}}]{Daylan:2014rsa}%
  \BibitemOpen
  \bibfield  {author} {\bibinfo {author} {\bibfnamefont {T.}~\bibnamefont
  {Daylan}}, \bibinfo {author} {\bibfnamefont {D.~P.}\ \bibnamefont
  {Finkbeiner}}, \bibinfo {author} {\bibfnamefont {D.}~\bibnamefont {Hooper}},
  \bibinfo {author} {\bibfnamefont {T.}~\bibnamefont {Linden}}, \bibinfo
  {author} {\bibfnamefont {S.~K.~N.}\ \bibnamefont {Portillo}}, \bibinfo
  {author} {\bibfnamefont {N.~L.}\ \bibnamefont {Rodd}}, \ and\ \bibinfo
  {author} {\bibfnamefont {T.~R.}\ \bibnamefont {Slatyer}},\ }\href {\doibase
  10.1016/j.dark.2015.12.005} {\bibfield  {journal} {\bibinfo  {journal} {Phys.
  Dark Univ.}\ }\textbf {\bibinfo {volume} {12}},\ \bibinfo {pages} {1}
  (\bibinfo {year} {2016})},\ \Eprint {http://arxiv.org/abs/1402.6703}
  {arXiv:1402.6703 [astro-ph.HE]} \BibitemShut {NoStop}%
\bibitem [{\citenamefont {Calore}\ \emph
  {et~al.}(2015{\natexlab{a}})\citenamefont {Calore}, \citenamefont {Cholis},\
  and\ \citenamefont {Weniger}}]{Calore:2014xka}%
  \BibitemOpen
  \bibfield  {author} {\bibinfo {author} {\bibfnamefont {F.}~\bibnamefont
  {Calore}}, \bibinfo {author} {\bibfnamefont {I.}~\bibnamefont {Cholis}}, \
  and\ \bibinfo {author} {\bibfnamefont {C.}~\bibnamefont {Weniger}},\ }\href
  {\doibase 10.1088/1475-7516/2015/03/038} {\bibfield  {journal} {\bibinfo
  {journal} {JCAP}\ }\textbf {\bibinfo {volume} {03}},\ \bibinfo {pages} {038}
  (\bibinfo {year} {2015}{\natexlab{a}})},\ \Eprint
  {http://arxiv.org/abs/1409.0042} {arXiv:1409.0042 [astro-ph.CO]} \BibitemShut
  {NoStop}%
\bibitem [{\citenamefont {Abazajian}\ \emph {et~al.}(2015)\citenamefont
  {Abazajian}, \citenamefont {Canac}, \citenamefont {Horiuchi}, \citenamefont
  {Kaplinghat},\ and\ \citenamefont {Kwa}}]{Abazajian:2014hsa}%
  \BibitemOpen
  \bibfield  {author} {\bibinfo {author} {\bibfnamefont {K.~N.}\ \bibnamefont
  {Abazajian}}, \bibinfo {author} {\bibfnamefont {N.}~\bibnamefont {Canac}},
  \bibinfo {author} {\bibfnamefont {S.}~\bibnamefont {Horiuchi}}, \bibinfo
  {author} {\bibfnamefont {M.}~\bibnamefont {Kaplinghat}}, \ and\ \bibinfo
  {author} {\bibfnamefont {A.}~\bibnamefont {Kwa}},\ }\href {\doibase
  10.1088/1475-7516/2015/07/013} {\bibfield  {journal} {\bibinfo  {journal}
  {JCAP}\ }\textbf {\bibinfo {volume} {07}},\ \bibinfo {pages} {013} (\bibinfo
  {year} {2015})},\ \Eprint {http://arxiv.org/abs/1410.6168} {arXiv:1410.6168
  [astro-ph.HE]} \BibitemShut {NoStop}%
\bibitem [{\citenamefont {Ajello}\ \emph {et~al.}(2016)\citenamefont {Ajello}
  \emph {et~al.}}]{Fermi-LAT:2015sau}%
  \BibitemOpen
  \bibfield  {author} {\bibinfo {author} {\bibfnamefont {M.}~\bibnamefont
  {Ajello}} \emph {et~al.} (\bibinfo {collaboration} {Fermi-LAT}),\ }\href
  {\doibase 10.3847/0004-637X/819/1/44} {\bibfield  {journal} {\bibinfo
  {journal} {Astrophys. J.}\ }\textbf {\bibinfo {volume} {819}},\ \bibinfo
  {pages} {44} (\bibinfo {year} {2016})},\ \Eprint
  {http://arxiv.org/abs/1511.02938} {arXiv:1511.02938 [astro-ph.HE]}
  \BibitemShut {NoStop}%
\bibitem [{\citenamefont {Linden}\ \emph {et~al.}(2016)\citenamefont {Linden},
  \citenamefont {Rodd}, \citenamefont {Safdi},\ and\ \citenamefont
  {Slatyer}}]{Linden:2016rcf}%
  \BibitemOpen
  \bibfield  {author} {\bibinfo {author} {\bibfnamefont {T.}~\bibnamefont
  {Linden}}, \bibinfo {author} {\bibfnamefont {N.~L.}\ \bibnamefont {Rodd}},
  \bibinfo {author} {\bibfnamefont {B.~R.}\ \bibnamefont {Safdi}}, \ and\
  \bibinfo {author} {\bibfnamefont {T.~R.}\ \bibnamefont {Slatyer}},\ }\href
  {\doibase 10.1103/PhysRevD.94.103013} {\bibfield  {journal} {\bibinfo
  {journal} {Phys. Rev. D}\ }\textbf {\bibinfo {volume} {94}},\ \bibinfo
  {pages} {103013} (\bibinfo {year} {2016})},\ \Eprint
  {http://arxiv.org/abs/1604.01026} {arXiv:1604.01026 [astro-ph.HE]}
  \BibitemShut {NoStop}%
\bibitem [{\citenamefont {Macias}\ \emph {et~al.}(2018)\citenamefont {Macias},
  \citenamefont {Gordon}, \citenamefont {Crocker}, \citenamefont {Coleman},
  \citenamefont {Paterson}, \citenamefont {Horiuchi},\ and\ \citenamefont
  {Pohl}}]{Macias:2016nev}%
  \BibitemOpen
  \bibfield  {author} {\bibinfo {author} {\bibfnamefont {O.}~\bibnamefont
  {Macias}}, \bibinfo {author} {\bibfnamefont {C.}~\bibnamefont {Gordon}},
  \bibinfo {author} {\bibfnamefont {R.~M.}\ \bibnamefont {Crocker}}, \bibinfo
  {author} {\bibfnamefont {B.}~\bibnamefont {Coleman}}, \bibinfo {author}
  {\bibfnamefont {D.}~\bibnamefont {Paterson}}, \bibinfo {author}
  {\bibfnamefont {S.}~\bibnamefont {Horiuchi}}, \ and\ \bibinfo {author}
  {\bibfnamefont {M.}~\bibnamefont {Pohl}},\ }\href {\doibase
  10.1038/s41550-018-0414-3} {\bibfield  {journal} {\bibinfo  {journal} {Nature
  Astron.}\ }\textbf {\bibinfo {volume} {2}},\ \bibinfo {pages} {387} (\bibinfo
  {year} {2018})},\ \Eprint {http://arxiv.org/abs/1611.06644} {arXiv:1611.06644
  [astro-ph.HE]} \BibitemShut {NoStop}%
\bibitem [{\citenamefont {Clark}\ \emph {et~al.}(2018)\citenamefont {Clark},
  \citenamefont {Scott}, \citenamefont {Trotta},\ and\ \citenamefont
  {Lewis}}]{Clark:2016mbb}%
  \BibitemOpen
  \bibfield  {author} {\bibinfo {author} {\bibfnamefont {H.~A.}\ \bibnamefont
  {Clark}}, \bibinfo {author} {\bibfnamefont {P.}~\bibnamefont {Scott}},
  \bibinfo {author} {\bibfnamefont {R.}~\bibnamefont {Trotta}}, \ and\ \bibinfo
  {author} {\bibfnamefont {G.~F.}\ \bibnamefont {Lewis}},\ }\href {\doibase
  10.1088/1475-7516/2018/07/060} {\bibfield  {journal} {\bibinfo  {journal}
  {JCAP}\ }\textbf {\bibinfo {volume} {07}},\ \bibinfo {pages} {060} (\bibinfo
  {year} {2018})},\ \Eprint {http://arxiv.org/abs/1612.01539} {arXiv:1612.01539
  [astro-ph.HE]} \BibitemShut {NoStop}%
\bibitem [{\citenamefont {Abazajian}(2011)}]{Abazajian:2010zy}%
  \BibitemOpen
  \bibfield  {author} {\bibinfo {author} {\bibfnamefont {K.~N.}\ \bibnamefont
  {Abazajian}},\ }\href {\doibase 10.1088/1475-7516/2011/03/010} {\bibfield
  {journal} {\bibinfo  {journal} {JCAP}\ }\textbf {\bibinfo {volume} {03}},\
  \bibinfo {pages} {010} (\bibinfo {year} {2011})},\ \Eprint
  {http://arxiv.org/abs/1011.4275} {arXiv:1011.4275 [astro-ph.HE]} \BibitemShut
  {NoStop}%
\bibitem [{\citenamefont {Hooper}\ \emph {et~al.}(2013)\citenamefont {Hooper},
  \citenamefont {Cholis}, \citenamefont {Linden}, \citenamefont
  {Siegal-Gaskins},\ and\ \citenamefont {Slatyer}}]{Hooper:2013nhl}%
  \BibitemOpen
  \bibfield  {author} {\bibinfo {author} {\bibfnamefont {D.}~\bibnamefont
  {Hooper}}, \bibinfo {author} {\bibfnamefont {I.}~\bibnamefont {Cholis}},
  \bibinfo {author} {\bibfnamefont {T.}~\bibnamefont {Linden}}, \bibinfo
  {author} {\bibfnamefont {J.}~\bibnamefont {Siegal-Gaskins}}, \ and\ \bibinfo
  {author} {\bibfnamefont {T.}~\bibnamefont {Slatyer}},\ }\href {\doibase
  10.1103/PhysRevD.88.083009} {\bibfield  {journal} {\bibinfo  {journal} {Phys.
  Rev. D}\ }\textbf {\bibinfo {volume} {88}},\ \bibinfo {pages} {083009}
  (\bibinfo {year} {2013})},\ \Eprint {http://arxiv.org/abs/1305.0830}
  {arXiv:1305.0830 [astro-ph.HE]} \BibitemShut {NoStop}%
\bibitem [{\citenamefont {Calore}\ \emph {et~al.}(2014)\citenamefont {Calore},
  \citenamefont {Di~Mauro},\ and\ \citenamefont {Donato}}]{Calore:2014oga}%
  \BibitemOpen
  \bibfield  {author} {\bibinfo {author} {\bibfnamefont {F.}~\bibnamefont
  {Calore}}, \bibinfo {author} {\bibfnamefont {M.}~\bibnamefont {Di~Mauro}}, \
  and\ \bibinfo {author} {\bibfnamefont {F.}~\bibnamefont {Donato}},\ }\href
  {\doibase 10.1088/0004-637X/796/1/14} {\bibfield  {journal} {\bibinfo
  {journal} {Astrophys. J.}\ }\textbf {\bibinfo {volume} {796}},\ \bibinfo
  {pages} {1} (\bibinfo {year} {2014})},\ \Eprint
  {http://arxiv.org/abs/1406.2706} {arXiv:1406.2706 [astro-ph.HE]} \BibitemShut
  {NoStop}%
\bibitem [{\citenamefont {Cholis}\ \emph {et~al.}(2015)\citenamefont {Cholis},
  \citenamefont {Hooper},\ and\ \citenamefont {Linden}}]{Cholis:2014lta}%
  \BibitemOpen
  \bibfield  {author} {\bibinfo {author} {\bibfnamefont {I.}~\bibnamefont
  {Cholis}}, \bibinfo {author} {\bibfnamefont {D.}~\bibnamefont {Hooper}}, \
  and\ \bibinfo {author} {\bibfnamefont {T.}~\bibnamefont {Linden}},\ }\href
  {\doibase 10.1088/1475-7516/2015/06/043} {\bibfield  {journal} {\bibinfo
  {journal} {JCAP}\ }\textbf {\bibinfo {volume} {06}},\ \bibinfo {pages} {043}
  (\bibinfo {year} {2015})},\ \Eprint {http://arxiv.org/abs/1407.5625}
  {arXiv:1407.5625 [astro-ph.HE]} \BibitemShut {NoStop}%
\bibitem [{\citenamefont {Petrovi\'c}\ \emph {et~al.}(2015)\citenamefont
  {Petrovi\'c}, \citenamefont {Serpico},\ and\ \citenamefont
  {Zaharijas}}]{Petrovic:2014xra}%
  \BibitemOpen
  \bibfield  {author} {\bibinfo {author} {\bibfnamefont {J.}~\bibnamefont
  {Petrovi\'c}}, \bibinfo {author} {\bibfnamefont {P.~D.}\ \bibnamefont
  {Serpico}}, \ and\ \bibinfo {author} {\bibfnamefont {G.}~\bibnamefont
  {Zaharijas}},\ }\href {\doibase 10.1088/1475-7516/2015/02/023} {\bibfield
  {journal} {\bibinfo  {journal} {JCAP}\ }\textbf {\bibinfo {volume} {02}},\
  \bibinfo {pages} {023} (\bibinfo {year} {2015})},\ \Eprint
  {http://arxiv.org/abs/1411.2980} {arXiv:1411.2980 [astro-ph.HE]} \BibitemShut
  {NoStop}%
\bibitem [{\citenamefont {Yuan}\ and\ \citenamefont
  {Ioka}(2015)}]{Yuan:2014yda}%
  \BibitemOpen
  \bibfield  {author} {\bibinfo {author} {\bibfnamefont {Q.}~\bibnamefont
  {Yuan}}\ and\ \bibinfo {author} {\bibfnamefont {K.}~\bibnamefont {Ioka}},\
  }\href {\doibase 10.1088/0004-637X/802/2/124} {\bibfield  {journal} {\bibinfo
   {journal} {Astrophys. J.}\ }\textbf {\bibinfo {volume} {802}},\ \bibinfo
  {pages} {124} (\bibinfo {year} {2015})},\ \Eprint
  {http://arxiv.org/abs/1411.4363} {arXiv:1411.4363 [astro-ph.HE]} \BibitemShut
  {NoStop}%
\bibitem [{\citenamefont {Brandt}\ and\ \citenamefont
  {Kocsis}(2015)}]{Brandt:2015ula}%
  \BibitemOpen
  \bibfield  {author} {\bibinfo {author} {\bibfnamefont {T.~D.}\ \bibnamefont
  {Brandt}}\ and\ \bibinfo {author} {\bibfnamefont {B.}~\bibnamefont
  {Kocsis}},\ }\href {\doibase 10.1088/0004-637X/812/1/15} {\bibfield
  {journal} {\bibinfo  {journal} {Astrophys. J.}\ }\textbf {\bibinfo {volume}
  {812}},\ \bibinfo {pages} {15} (\bibinfo {year} {2015})},\ \Eprint
  {http://arxiv.org/abs/1507.05616} {arXiv:1507.05616 [astro-ph.HE]}
  \BibitemShut {NoStop}%
\bibitem [{\citenamefont {Gautam}\ \emph {et~al.}(2021)\citenamefont {Gautam},
  \citenamefont {Crocker}, \citenamefont {Ferrario}, \citenamefont {Ruiter},
  \citenamefont {Ploeg}, \citenamefont {Gordon},\ and\ \citenamefont
  {Macias}}]{Gautam:2021wqn}%
  \BibitemOpen
  \bibfield  {author} {\bibinfo {author} {\bibfnamefont {A.}~\bibnamefont
  {Gautam}}, \bibinfo {author} {\bibfnamefont {R.~M.}\ \bibnamefont {Crocker}},
  \bibinfo {author} {\bibfnamefont {L.}~\bibnamefont {Ferrario}}, \bibinfo
  {author} {\bibfnamefont {A.~J.}\ \bibnamefont {Ruiter}}, \bibinfo {author}
  {\bibfnamefont {H.}~\bibnamefont {Ploeg}}, \bibinfo {author} {\bibfnamefont
  {C.}~\bibnamefont {Gordon}}, \ and\ \bibinfo {author} {\bibfnamefont
  {O.}~\bibnamefont {Macias}},\ }\href@noop {} {\  (\bibinfo {year} {2021})},\
  \Eprint {http://arxiv.org/abs/2106.00222} {arXiv:2106.00222 [astro-ph.HE]}
  \BibitemShut {NoStop}%
\bibitem [{\citenamefont {Ploeg}\ \emph {et~al.}(2020)\citenamefont {Ploeg},
  \citenamefont {Gordon}, \citenamefont {Crocker},\ and\ \citenamefont
  {Macias}}]{Ploeg:2020jeh}%
  \BibitemOpen
  \bibfield  {author} {\bibinfo {author} {\bibfnamefont {H.}~\bibnamefont
  {Ploeg}}, \bibinfo {author} {\bibfnamefont {C.}~\bibnamefont {Gordon}},
  \bibinfo {author} {\bibfnamefont {R.}~\bibnamefont {Crocker}}, \ and\
  \bibinfo {author} {\bibfnamefont {O.}~\bibnamefont {Macias}},\ }\href
  {\doibase 10.1088/1475-7516/2020/12/035} {\bibfield  {journal} {\bibinfo
  {journal} {JCAP}\ }\textbf {\bibinfo {volume} {12}},\ \bibinfo {pages} {035}
  (\bibinfo {year} {2020})},\ \Eprint {http://arxiv.org/abs/2008.10821}
  {arXiv:2008.10821 [astro-ph.HE]} \BibitemShut {NoStop}%
\bibitem [{\citenamefont {Macias}\ \emph {et~al.}(2019)\citenamefont {Macias},
  \citenamefont {Horiuchi}, \citenamefont {Kaplinghat}, \citenamefont {Gordon},
  \citenamefont {Crocker},\ and\ \citenamefont {Nataf}}]{Macias:2019omb}%
  \BibitemOpen
  \bibfield  {author} {\bibinfo {author} {\bibfnamefont {O.}~\bibnamefont
  {Macias}}, \bibinfo {author} {\bibfnamefont {S.}~\bibnamefont {Horiuchi}},
  \bibinfo {author} {\bibfnamefont {M.}~\bibnamefont {Kaplinghat}}, \bibinfo
  {author} {\bibfnamefont {C.}~\bibnamefont {Gordon}}, \bibinfo {author}
  {\bibfnamefont {R.~M.}\ \bibnamefont {Crocker}}, \ and\ \bibinfo {author}
  {\bibfnamefont {D.~M.}\ \bibnamefont {Nataf}},\ }\href {\doibase
  10.1088/1475-7516/2019/09/042} {\bibfield  {journal} {\bibinfo  {journal}
  {JCAP}\ }\textbf {\bibinfo {volume} {09}},\ \bibinfo {pages} {042} (\bibinfo
  {year} {2019})},\ \Eprint {http://arxiv.org/abs/1901.03822} {arXiv:1901.03822
  [astro-ph.HE]} \BibitemShut {NoStop}%
\bibitem [{\citenamefont {Bartels}\ \emph
  {et~al.}(2018{\natexlab{a}})\citenamefont {Bartels}, \citenamefont {Storm},
  \citenamefont {Weniger},\ and\ \citenamefont {Calore}}]{Bartels:2017vsx}%
  \BibitemOpen
  \bibfield  {author} {\bibinfo {author} {\bibfnamefont {R.}~\bibnamefont
  {Bartels}}, \bibinfo {author} {\bibfnamefont {E.}~\bibnamefont {Storm}},
  \bibinfo {author} {\bibfnamefont {C.}~\bibnamefont {Weniger}}, \ and\
  \bibinfo {author} {\bibfnamefont {F.}~\bibnamefont {Calore}},\ }\href
  {\doibase 10.1038/s41550-018-0531-z} {\bibfield  {journal} {\bibinfo
  {journal} {Nature Astron.}\ }\textbf {\bibinfo {volume} {2}},\ \bibinfo
  {pages} {819} (\bibinfo {year} {2018}{\natexlab{a}})},\ \Eprint
  {http://arxiv.org/abs/1711.04778} {arXiv:1711.04778 [astro-ph.HE]}
  \BibitemShut {NoStop}%
\bibitem [{\citenamefont {Di~Mauro}(2020)}]{DiMauro:2020rcr}%
  \BibitemOpen
  \bibfield  {author} {\bibinfo {author} {\bibfnamefont {M.}~\bibnamefont
  {Di~Mauro}},\ }\href {\doibase 10.1103/PhysRevD.102.103013} {\bibfield
  {journal} {\bibinfo  {journal} {Phys. Rev. D}\ }\textbf {\bibinfo {volume}
  {102}},\ \bibinfo {pages} {103013} (\bibinfo {year} {2020})},\ \Eprint
  {http://arxiv.org/abs/2010.02231} {arXiv:2010.02231 [astro-ph.HE]}
  \BibitemShut {NoStop}%
\bibitem [{\citenamefont {Di~Mauro}(2021)}]{DiMauro:2021raz}%
  \BibitemOpen
  \bibfield  {author} {\bibinfo {author} {\bibfnamefont {M.}~\bibnamefont
  {Di~Mauro}},\ }\href {\doibase 10.1103/PhysRevD.103.063029} {\bibfield
  {journal} {\bibinfo  {journal} {Phys. Rev. D}\ }\textbf {\bibinfo {volume}
  {103}},\ \bibinfo {pages} {063029} (\bibinfo {year} {2021})},\ \Eprint
  {http://arxiv.org/abs/2101.04694} {arXiv:2101.04694 [astro-ph.HE]}
  \BibitemShut {NoStop}%
\bibitem [{\citenamefont {Lee}\ \emph {et~al.}(2016)\citenamefont {Lee},
  \citenamefont {Lisanti}, \citenamefont {Safdi}, \citenamefont {Slatyer},\
  and\ \citenamefont {Xue}}]{Lee:2015fea}%
  \BibitemOpen
  \bibfield  {author} {\bibinfo {author} {\bibfnamefont {S.~K.}\ \bibnamefont
  {Lee}}, \bibinfo {author} {\bibfnamefont {M.}~\bibnamefont {Lisanti}},
  \bibinfo {author} {\bibfnamefont {B.~R.}\ \bibnamefont {Safdi}}, \bibinfo
  {author} {\bibfnamefont {T.~R.}\ \bibnamefont {Slatyer}}, \ and\ \bibinfo
  {author} {\bibfnamefont {W.}~\bibnamefont {Xue}},\ }\href {\doibase
  10.1103/PhysRevLett.116.051103} {\bibfield  {journal} {\bibinfo  {journal}
  {Phys. Rev. Lett.}\ }\textbf {\bibinfo {volume} {116}},\ \bibinfo {pages}
  {051103} (\bibinfo {year} {2016})},\ \Eprint
  {http://arxiv.org/abs/1506.05124} {arXiv:1506.05124 [astro-ph.HE]}
  \BibitemShut {NoStop}%
\bibitem [{\citenamefont {Bartels}\ \emph {et~al.}(2016)\citenamefont
  {Bartels}, \citenamefont {Krishnamurthy},\ and\ \citenamefont
  {Weniger}}]{Bartels:2015aea}%
  \BibitemOpen
  \bibfield  {author} {\bibinfo {author} {\bibfnamefont {R.}~\bibnamefont
  {Bartels}}, \bibinfo {author} {\bibfnamefont {S.}~\bibnamefont
  {Krishnamurthy}}, \ and\ \bibinfo {author} {\bibfnamefont {C.}~\bibnamefont
  {Weniger}},\ }\href {\doibase 10.1103/PhysRevLett.116.051102} {\bibfield
  {journal} {\bibinfo  {journal} {Phys. Rev. Lett.}\ }\textbf {\bibinfo
  {volume} {116}},\ \bibinfo {pages} {051102} (\bibinfo {year} {2016})},\
  \Eprint {http://arxiv.org/abs/1506.05104} {arXiv:1506.05104 [astro-ph.HE]}
  \BibitemShut {NoStop}%
\bibitem [{\citenamefont {Buschmann}\ \emph {et~al.}(2020)\citenamefont
  {Buschmann}, \citenamefont {Rodd}, \citenamefont {Safdi}, \citenamefont
  {Chang}, \citenamefont {Mishra-Sharma}, \citenamefont {Lisanti},\ and\
  \citenamefont {Macias}}]{Buschmann:2020adf}%
  \BibitemOpen
  \bibfield  {author} {\bibinfo {author} {\bibfnamefont {M.}~\bibnamefont
  {Buschmann}}, \bibinfo {author} {\bibfnamefont {N.~L.}\ \bibnamefont {Rodd}},
  \bibinfo {author} {\bibfnamefont {B.~R.}\ \bibnamefont {Safdi}}, \bibinfo
  {author} {\bibfnamefont {L.~J.}\ \bibnamefont {Chang}}, \bibinfo {author}
  {\bibfnamefont {S.}~\bibnamefont {Mishra-Sharma}}, \bibinfo {author}
  {\bibfnamefont {M.}~\bibnamefont {Lisanti}}, \ and\ \bibinfo {author}
  {\bibfnamefont {O.}~\bibnamefont {Macias}},\ }\href {\doibase
  10.1103/PhysRevD.102.023023} {\bibfield  {journal} {\bibinfo  {journal}
  {Phys. Rev. D}\ }\textbf {\bibinfo {volume} {102}},\ \bibinfo {pages}
  {023023} (\bibinfo {year} {2020})},\ \Eprint
  {http://arxiv.org/abs/2002.12373} {arXiv:2002.12373 [astro-ph.HE]}
  \BibitemShut {NoStop}%
\bibitem [{\citenamefont {Chang}\ \emph {et~al.}(2020)\citenamefont {Chang},
  \citenamefont {Mishra-Sharma}, \citenamefont {Lisanti}, \citenamefont
  {Buschmann}, \citenamefont {Rodd},\ and\ \citenamefont
  {Safdi}}]{Chang:2019ars}%
  \BibitemOpen
  \bibfield  {author} {\bibinfo {author} {\bibfnamefont {L.~J.}\ \bibnamefont
  {Chang}}, \bibinfo {author} {\bibfnamefont {S.}~\bibnamefont
  {Mishra-Sharma}}, \bibinfo {author} {\bibfnamefont {M.}~\bibnamefont
  {Lisanti}}, \bibinfo {author} {\bibfnamefont {M.}~\bibnamefont {Buschmann}},
  \bibinfo {author} {\bibfnamefont {N.~L.}\ \bibnamefont {Rodd}}, \ and\
  \bibinfo {author} {\bibfnamefont {B.~R.}\ \bibnamefont {Safdi}},\ }\href
  {\doibase 10.1103/PhysRevD.101.023014} {\bibfield  {journal} {\bibinfo
  {journal} {Phys. Rev. D}\ }\textbf {\bibinfo {volume} {101}},\ \bibinfo
  {pages} {023014} (\bibinfo {year} {2020})},\ \Eprint
  {http://arxiv.org/abs/1908.10874} {arXiv:1908.10874 [astro-ph.CO]}
  \BibitemShut {NoStop}%
\bibitem [{\citenamefont {Leane}\ and\ \citenamefont
  {Slatyer}(2019)}]{Leane:2019xiy}%
  \BibitemOpen
  \bibfield  {author} {\bibinfo {author} {\bibfnamefont {R.~K.}\ \bibnamefont
  {Leane}}\ and\ \bibinfo {author} {\bibfnamefont {T.~R.}\ \bibnamefont
  {Slatyer}},\ }\href {\doibase 10.1103/PhysRevLett.123.241101} {\bibfield
  {journal} {\bibinfo  {journal} {Phys. Rev. Lett.}\ }\textbf {\bibinfo
  {volume} {123}},\ \bibinfo {pages} {241101} (\bibinfo {year} {2019})},\
  \Eprint {http://arxiv.org/abs/1904.08430} {arXiv:1904.08430 [astro-ph.HE]}
  \BibitemShut {NoStop}%
\bibitem [{\citenamefont {Malyshev}\ and\ \citenamefont
  {Hogg}(2011)}]{Malyshev:2011zi}%
  \BibitemOpen
  \bibfield  {author} {\bibinfo {author} {\bibfnamefont {D.}~\bibnamefont
  {Malyshev}}\ and\ \bibinfo {author} {\bibfnamefont {D.~W.}\ \bibnamefont
  {Hogg}},\ }\href {\doibase 10.1088/0004-637X/738/2/181} {\bibfield  {journal}
  {\bibinfo  {journal} {Astrophys. J.}\ }\textbf {\bibinfo {volume} {738}},\
  \bibinfo {pages} {181} (\bibinfo {year} {2011})},\ \Eprint
  {http://arxiv.org/abs/1104.0010} {arXiv:1104.0010 [astro-ph.CO]} \BibitemShut
  {NoStop}%
\bibitem [{\citenamefont {Lee}\ \emph {et~al.}(2015)\citenamefont {Lee},
  \citenamefont {Lisanti},\ and\ \citenamefont {Safdi}}]{Lee:2014mza}%
  \BibitemOpen
  \bibfield  {author} {\bibinfo {author} {\bibfnamefont {S.~K.}\ \bibnamefont
  {Lee}}, \bibinfo {author} {\bibfnamefont {M.}~\bibnamefont {Lisanti}}, \ and\
  \bibinfo {author} {\bibfnamefont {B.~R.}\ \bibnamefont {Safdi}},\ }\href
  {\doibase 10.1088/1475-7516/2015/05/056} {\bibfield  {journal} {\bibinfo
  {journal} {JCAP}\ }\textbf {\bibinfo {volume} {05}},\ \bibinfo {pages} {056}
  (\bibinfo {year} {2015})},\ \Eprint {http://arxiv.org/abs/1412.6099}
  {arXiv:1412.6099 [astro-ph.CO]} \BibitemShut {NoStop}%
\bibitem [{\citenamefont {Balaji}\ \emph {et~al.}(2018)\citenamefont {Balaji},
  \citenamefont {Cholis}, \citenamefont {Fox},\ and\ \citenamefont
  {McDermott}}]{Balaji:2018rwz}%
  \BibitemOpen
  \bibfield  {author} {\bibinfo {author} {\bibfnamefont {B.}~\bibnamefont
  {Balaji}}, \bibinfo {author} {\bibfnamefont {I.}~\bibnamefont {Cholis}},
  \bibinfo {author} {\bibfnamefont {P.~J.}\ \bibnamefont {Fox}}, \ and\
  \bibinfo {author} {\bibfnamefont {S.~D.}\ \bibnamefont {McDermott}},\ }\href
  {\doibase 10.1103/PhysRevD.98.043009} {\bibfield  {journal} {\bibinfo
  {journal} {Phys. Rev. D}\ }\textbf {\bibinfo {volume} {98}},\ \bibinfo
  {pages} {043009} (\bibinfo {year} {2018})},\ \Eprint
  {http://arxiv.org/abs/1803.01952} {arXiv:1803.01952 [astro-ph.HE]}
  \BibitemShut {NoStop}%
\bibitem [{\citenamefont {McDermott}\ \emph {et~al.}(2016)\citenamefont
  {McDermott}, \citenamefont {Fox}, \citenamefont {Cholis},\ and\ \citenamefont
  {Lee}}]{McDermott:2015ydv}%
  \BibitemOpen
  \bibfield  {author} {\bibinfo {author} {\bibfnamefont {S.~D.}\ \bibnamefont
  {McDermott}}, \bibinfo {author} {\bibfnamefont {P.~J.}\ \bibnamefont {Fox}},
  \bibinfo {author} {\bibfnamefont {I.}~\bibnamefont {Cholis}}, \ and\ \bibinfo
  {author} {\bibfnamefont {S.~K.}\ \bibnamefont {Lee}},\ }\href {\doibase
  10.1088/1475-7516/2016/07/045} {\bibfield  {journal} {\bibinfo  {journal}
  {JCAP}\ }\textbf {\bibinfo {volume} {07}},\ \bibinfo {pages} {045} (\bibinfo
  {year} {2016})},\ \Eprint {http://arxiv.org/abs/1512.00012} {arXiv:1512.00012
  [astro-ph.HE]} \BibitemShut {NoStop}%
\bibitem [{\citenamefont {Zhong}\ \emph {et~al.}(2020)\citenamefont {Zhong},
  \citenamefont {McDermott}, \citenamefont {Cholis},\ and\ \citenamefont
  {Fox}}]{Zhong:2019ycb}%
  \BibitemOpen
  \bibfield  {author} {\bibinfo {author} {\bibfnamefont {Y.-M.}\ \bibnamefont
  {Zhong}}, \bibinfo {author} {\bibfnamefont {S.~D.}\ \bibnamefont
  {McDermott}}, \bibinfo {author} {\bibfnamefont {I.}~\bibnamefont {Cholis}}, \
  and\ \bibinfo {author} {\bibfnamefont {P.~J.}\ \bibnamefont {Fox}},\ }\href
  {\doibase 10.1103/PhysRevLett.124.231103} {\bibfield  {journal} {\bibinfo
  {journal} {Phys. Rev. Lett.}\ }\textbf {\bibinfo {volume} {124}},\ \bibinfo
  {pages} {231103} (\bibinfo {year} {2020})},\ \Eprint
  {http://arxiv.org/abs/1911.12369} {arXiv:1911.12369 [astro-ph.HE]}
  \BibitemShut {NoStop}%
\bibitem [{\citenamefont {Caron}\ \emph {et~al.}(2021)\citenamefont {Caron}
  \emph {et~al.}}]{Caron:2021map}%
  \BibitemOpen
  \bibfield  {author} {\bibinfo {author} {\bibfnamefont {S.}~\bibnamefont
  {Caron}} \emph {et~al.},\ }\href@noop {} {\  (\bibinfo {year} {2021})},\
  \Eprint {http://arxiv.org/abs/2103.11068} {arXiv:2103.11068 [astro-ph.HE]}
  \BibitemShut {NoStop}%
\bibitem [{\citenamefont {List}\ \emph {et~al.}(2020)\citenamefont {List},
  \citenamefont {Rodd}, \citenamefont {Lewis},\ and\ \citenamefont
  {Bhat}}]{List:2020mzd}%
  \BibitemOpen
  \bibfield  {author} {\bibinfo {author} {\bibfnamefont {F.}~\bibnamefont
  {List}}, \bibinfo {author} {\bibfnamefont {N.~L.}\ \bibnamefont {Rodd}},
  \bibinfo {author} {\bibfnamefont {G.~F.}\ \bibnamefont {Lewis}}, \ and\
  \bibinfo {author} {\bibfnamefont {I.}~\bibnamefont {Bhat}},\ }\href {\doibase
  10.1103/PhysRevLett.125.241102} {\bibfield  {journal} {\bibinfo  {journal}
  {Phys. Rev. Lett.}\ }\textbf {\bibinfo {volume} {125}},\ \bibinfo {pages}
  {241102} (\bibinfo {year} {2020})},\ \Eprint
  {http://arxiv.org/abs/2006.12504} {arXiv:2006.12504 [astro-ph.HE]}
  \BibitemShut {NoStop}%
\bibitem [{\citenamefont {List}\ \emph {et~al.}(2021)\citenamefont {List},
  \citenamefont {Rodd},\ and\ \citenamefont {Lewis}}]{List:2021aer}%
  \BibitemOpen
  \bibfield  {author} {\bibinfo {author} {\bibfnamefont {F.}~\bibnamefont
  {List}}, \bibinfo {author} {\bibfnamefont {N.~L.}\ \bibnamefont {Rodd}}, \
  and\ \bibinfo {author} {\bibfnamefont {G.~F.}\ \bibnamefont {Lewis}},\
  }\href@noop {} {\  (\bibinfo {year} {2021})},\ \Eprint
  {http://arxiv.org/abs/2107.09070} {arXiv:2107.09070 [astro-ph.HE]}
  \BibitemShut {NoStop}%
\bibitem [{\citenamefont {Caron}\ \emph {et~al.}(2018)\citenamefont {Caron},
  \citenamefont {G\'omez-Vargas}, \citenamefont {Hendriks},\ and\ \citenamefont
  {Ruiz~de Austri}}]{Caron:2017udl}%
  \BibitemOpen
  \bibfield  {author} {\bibinfo {author} {\bibfnamefont {S.}~\bibnamefont
  {Caron}}, \bibinfo {author} {\bibfnamefont {G.~A.}\ \bibnamefont
  {G\'omez-Vargas}}, \bibinfo {author} {\bibfnamefont {L.}~\bibnamefont
  {Hendriks}}, \ and\ \bibinfo {author} {\bibfnamefont {R.}~\bibnamefont
  {Ruiz~de Austri}},\ }\href {\doibase 10.1088/1475-7516/2018/05/058}
  {\bibfield  {journal} {\bibinfo  {journal} {JCAP}\ }\textbf {\bibinfo
  {volume} {05}},\ \bibinfo {pages} {058} (\bibinfo {year} {2018})},\ \Eprint
  {http://arxiv.org/abs/1708.06706} {arXiv:1708.06706 [astro-ph.HE]}
  \BibitemShut {NoStop}%
\bibitem [{\citenamefont {List}(2021)}]{List2021}%
  \BibitemOpen
  \bibfield  {author} {\bibinfo {author} {\bibfnamefont {F.}~\bibnamefont
  {List}},\ }in\ \href {https://arxiv.org/pdf/2106.02051.pdf} {\emph {\bibinfo
  {booktitle} {Proceedings of the 38th International Conference on Machine
  Learning}}}\ (\bibinfo {year} {2021})\BibitemShut {NoStop}%
\bibitem [{\citenamefont {Cranmer}\ \emph {et~al.}(2020)\citenamefont
  {Cranmer}, \citenamefont {Brehmer},\ and\ \citenamefont
  {Louppe}}]{Cranmer:2019eaq}%
  \BibitemOpen
  \bibfield  {author} {\bibinfo {author} {\bibfnamefont {K.}~\bibnamefont
  {Cranmer}}, \bibinfo {author} {\bibfnamefont {J.}~\bibnamefont {Brehmer}}, \
  and\ \bibinfo {author} {\bibfnamefont {G.}~\bibnamefont {Louppe}},\ }\href
  {\doibase 10.1073/pnas.1912789117} {\bibfield  {journal} {\bibinfo  {journal}
  {Proc. Nat. Acad. Sci.}\ }\textbf {\bibinfo {volume} {117}},\ \bibinfo
  {pages} {30055} (\bibinfo {year} {2020})},\ \Eprint
  {http://arxiv.org/abs/1911.01429} {arXiv:1911.01429 [stat.ML]} \BibitemShut
  {NoStop}%
\bibitem [{\citenamefont {Papamakarios}\ \emph
  {et~al.}(2019{\natexlab{a}})\citenamefont {Papamakarios}, \citenamefont
  {Nalisnick}, \citenamefont {Rezende}, \citenamefont {Mohamed},\ and\
  \citenamefont {Lakshminarayanan}}]{papamakarios2019normalizing}%
  \BibitemOpen
  \bibfield  {author} {\bibinfo {author} {\bibfnamefont {G.}~\bibnamefont
  {Papamakarios}}, \bibinfo {author} {\bibfnamefont {E.}~\bibnamefont
  {Nalisnick}}, \bibinfo {author} {\bibfnamefont {D.~J.}\ \bibnamefont
  {Rezende}}, \bibinfo {author} {\bibfnamefont {S.}~\bibnamefont {Mohamed}}, \
  and\ \bibinfo {author} {\bibfnamefont {B.}~\bibnamefont {Lakshminarayanan}},\
  }\href@noop {} {\  (\bibinfo {year} {2019}{\natexlab{a}})},\ \Eprint
  {http://arxiv.org/abs/1912.02762} {arXiv:1912.02762 [cs.LG]} \BibitemShut
  {NoStop}%
\bibitem [{\citenamefont {Rezende}\ and\ \citenamefont
  {Mohamed}(2015)}]{DBLP:conf/icml/RezendeM15}%
  \BibitemOpen
  \bibfield  {author} {\bibinfo {author} {\bibfnamefont {D.~J.}\ \bibnamefont
  {Rezende}}\ and\ \bibinfo {author} {\bibfnamefont {S.}~\bibnamefont
  {Mohamed}},\ }in\ \href {http://proceedings.mlr.press/v37/rezende15.html}
  {\emph {\bibinfo {booktitle} {Proceedings of the 32nd International
  Conference on Machine Learning, {ICML} 2015, Lille, France, 6-11 July
  2015}}},\ \bibinfo {series} {{JMLR} Workshop and Conference Proceedings},
  Vol.~\bibinfo {volume} {37},\ \bibinfo {editor} {edited by\ \bibinfo {editor}
  {\bibfnamefont {F.~R.}\ \bibnamefont {Bach}}\ and\ \bibinfo {editor}
  {\bibfnamefont {D.~M.}\ \bibnamefont {Blei}}}\ (\bibinfo {year} {2015})\ pp.\
  \bibinfo {pages} {1530--1538}\BibitemShut {NoStop}%
\bibitem [{\citenamefont {Mishra-Sharma}\ \emph {et~al.}(2016)\citenamefont
  {Mishra-Sharma}, \citenamefont {Rodd},\ and\ \citenamefont
  {Safdi}}]{rodd_nicholas_safdi_siddharth_2016}%
  \BibitemOpen
  \bibfield  {author} {\bibinfo {author} {\bibfnamefont {S.}~\bibnamefont
  {Mishra-Sharma}}, \bibinfo {author} {\bibfnamefont {N.~L.}\ \bibnamefont
  {Rodd}}, \ and\ \bibinfo {author} {\bibfnamefont {B.~R.}\ \bibnamefont
  {Safdi}},\ }\href {https://dspace.mit.edu/handle/1721.1/105492} {\enquote
  {\bibinfo {title} {{Supplementary material for NPTFit}},}\ } (\bibinfo {year}
  {2016})\BibitemShut {NoStop}%
\bibitem [{\citenamefont {Mishra-Sharma}\ \emph {et~al.}(2017)\citenamefont
  {Mishra-Sharma}, \citenamefont {Rodd},\ and\ \citenamefont
  {Safdi}}]{Mishra-Sharma:2016gis}%
  \BibitemOpen
  \bibfield  {author} {\bibinfo {author} {\bibfnamefont {S.}~\bibnamefont
  {Mishra-Sharma}}, \bibinfo {author} {\bibfnamefont {N.~L.}\ \bibnamefont
  {Rodd}}, \ and\ \bibinfo {author} {\bibfnamefont {B.~R.}\ \bibnamefont
  {Safdi}},\ }\href {\doibase 10.3847/1538-3881/aa6d5f} {\bibfield  {journal}
  {\bibinfo  {journal} {Astron. J.}\ }\textbf {\bibinfo {volume} {153}},\
  \bibinfo {pages} {253} (\bibinfo {year} {2017})},\ \Eprint
  {http://arxiv.org/abs/1612.03173} {arXiv:1612.03173 [astro-ph.HE]}
  \BibitemShut {NoStop}%
\bibitem [{\citenamefont {Gorski}\ \emph {et~al.}(2005)\citenamefont {Gorski},
  \citenamefont {Hivon}, \citenamefont {Banday}, \citenamefont {Wandelt},
  \citenamefont {Hansen}, \citenamefont {Reinecke},\ and\ \citenamefont
  {Bartelman}}]{Gorski:2004by}%
  \BibitemOpen
  \bibfield  {author} {\bibinfo {author} {\bibfnamefont {K.~M.}\ \bibnamefont
  {Gorski}}, \bibinfo {author} {\bibfnamefont {E.}~\bibnamefont {Hivon}},
  \bibinfo {author} {\bibfnamefont {A.~J.}\ \bibnamefont {Banday}}, \bibinfo
  {author} {\bibfnamefont {B.~D.}\ \bibnamefont {Wandelt}}, \bibinfo {author}
  {\bibfnamefont {F.~K.}\ \bibnamefont {Hansen}}, \bibinfo {author}
  {\bibfnamefont {M.}~\bibnamefont {Reinecke}}, \ and\ \bibinfo {author}
  {\bibfnamefont {M.}~\bibnamefont {Bartelman}},\ }\href {\doibase
  10.1086/427976} {\bibfield  {journal} {\bibinfo  {journal} {Astrophys. J.}\
  }\textbf {\bibinfo {volume} {622}},\ \bibinfo {pages} {759} (\bibinfo {year}
  {2005})},\ \Eprint {http://arxiv.org/abs/astro-ph/0409513}
  {arXiv:astro-ph/0409513} \BibitemShut {NoStop}%
\bibitem [{\citenamefont {Chang}\ \emph {et~al.}(2018)\citenamefont {Chang},
  \citenamefont {Lisanti},\ and\ \citenamefont
  {Mishra-Sharma}}]{Chang:2018bpt}%
  \BibitemOpen
  \bibfield  {author} {\bibinfo {author} {\bibfnamefont {L.~J.}\ \bibnamefont
  {Chang}}, \bibinfo {author} {\bibfnamefont {M.}~\bibnamefont {Lisanti}}, \
  and\ \bibinfo {author} {\bibfnamefont {S.}~\bibnamefont {Mishra-Sharma}},\
  }\href {\doibase 10.1103/PhysRevD.98.123004} {\bibfield  {journal} {\bibinfo
  {journal} {Phys. Rev. D}\ }\textbf {\bibinfo {volume} {98}},\ \bibinfo
  {pages} {123004} (\bibinfo {year} {2018})},\ \Eprint
  {http://arxiv.org/abs/1804.04132} {arXiv:1804.04132 [astro-ph.CO]}
  \BibitemShut {NoStop}%
\bibitem [{\citenamefont {Acero}\ \emph {et~al.}(2015)\citenamefont {Acero}
  \emph {et~al.}}]{Fermi-LAT:2015bhf}%
  \BibitemOpen
  \bibfield  {author} {\bibinfo {author} {\bibfnamefont {F.}~\bibnamefont
  {Acero}} \emph {et~al.} (\bibinfo {collaboration} {Fermi-LAT}),\ }\href
  {\doibase 10.1088/0067-0049/218/2/23} {\bibfield  {journal} {\bibinfo
  {journal} {Astrophys. J. Suppl.}\ }\textbf {\bibinfo {volume} {218}},\
  \bibinfo {pages} {23} (\bibinfo {year} {2015})},\ \Eprint
  {http://arxiv.org/abs/1501.02003} {arXiv:1501.02003 [astro-ph.HE]}
  \BibitemShut {NoStop}%
\bibitem [{\citenamefont {Su}\ \emph {et~al.}(2010)\citenamefont {Su},
  \citenamefont {Slatyer},\ and\ \citenamefont {Finkbeiner}}]{Su:2010qj}%
  \BibitemOpen
  \bibfield  {author} {\bibinfo {author} {\bibfnamefont {M.}~\bibnamefont
  {Su}}, \bibinfo {author} {\bibfnamefont {T.~R.}\ \bibnamefont {Slatyer}}, \
  and\ \bibinfo {author} {\bibfnamefont {D.~P.}\ \bibnamefont {Finkbeiner}},\
  }\href {\doibase 10.1088/0004-637X/724/2/1044} {\bibfield  {journal}
  {\bibinfo  {journal} {Astrophys. J.}\ }\textbf {\bibinfo {volume} {724}},\
  \bibinfo {pages} {1044} (\bibinfo {year} {2010})},\ \Eprint
  {http://arxiv.org/abs/1005.5480} {arXiv:1005.5480 [astro-ph.HE]} \BibitemShut
  {NoStop}%
\bibitem [{\citenamefont {Navarro}\ \emph {et~al.}(1996)\citenamefont
  {Navarro}, \citenamefont {Frenk},\ and\ \citenamefont
  {White}}]{Navarro:1995iw}%
  \BibitemOpen
  \bibfield  {author} {\bibinfo {author} {\bibfnamefont {J.~F.}\ \bibnamefont
  {Navarro}}, \bibinfo {author} {\bibfnamefont {C.~S.}\ \bibnamefont {Frenk}},
  \ and\ \bibinfo {author} {\bibfnamefont {S.~D.~M.}\ \bibnamefont {White}},\
  }\href {\doibase 10.1086/177173} {\bibfield  {journal} {\bibinfo  {journal}
  {Astrophys. J.}\ }\textbf {\bibinfo {volume} {462}},\ \bibinfo {pages} {563}
  (\bibinfo {year} {1996})},\ \Eprint {http://arxiv.org/abs/astro-ph/9508025}
  {arXiv:astro-ph/9508025 [astro-ph]} \BibitemShut {NoStop}%
\bibitem [{\citenamefont {Navarro}\ \emph {et~al.}(1997)\citenamefont
  {Navarro}, \citenamefont {Frenk},\ and\ \citenamefont
  {White}}]{Navarro:1996gj}%
  \BibitemOpen
  \bibfield  {author} {\bibinfo {author} {\bibfnamefont {J.~F.}\ \bibnamefont
  {Navarro}}, \bibinfo {author} {\bibfnamefont {C.~S.}\ \bibnamefont {Frenk}},
  \ and\ \bibinfo {author} {\bibfnamefont {S.~D.}\ \bibnamefont {White}},\
  }\href {\doibase 10.1086/304888} {\bibfield  {journal} {\bibinfo  {journal}
  {Astrophys.J.}\ }\textbf {\bibinfo {volume} {490}},\ \bibinfo {pages} {493}
  (\bibinfo {year} {1997})},\ \Eprint {http://arxiv.org/abs/astro-ph/9611107}
  {arXiv:astro-ph/9611107 [astro-ph]} \BibitemShut {NoStop}%
\bibitem [{\citenamefont {Zhou}\ \emph {et~al.}(2015)\citenamefont {Zhou},
  \citenamefont {Liang}, \citenamefont {Huang}, \citenamefont {Li},
  \citenamefont {Fan}, \citenamefont {Feng},\ and\ \citenamefont
  {Chang}}]{Zhou:2014lva}%
  \BibitemOpen
  \bibfield  {author} {\bibinfo {author} {\bibfnamefont {B.}~\bibnamefont
  {Zhou}}, \bibinfo {author} {\bibfnamefont {Y.-F.}\ \bibnamefont {Liang}},
  \bibinfo {author} {\bibfnamefont {X.}~\bibnamefont {Huang}}, \bibinfo
  {author} {\bibfnamefont {X.}~\bibnamefont {Li}}, \bibinfo {author}
  {\bibfnamefont {Y.-Z.}\ \bibnamefont {Fan}}, \bibinfo {author} {\bibfnamefont
  {L.}~\bibnamefont {Feng}}, \ and\ \bibinfo {author} {\bibfnamefont
  {J.}~\bibnamefont {Chang}},\ }\href {\doibase 10.1103/PhysRevD.91.123010}
  {\bibfield  {journal} {\bibinfo  {journal} {Phys. Rev. D}\ }\textbf {\bibinfo
  {volume} {91}},\ \bibinfo {pages} {123010} (\bibinfo {year} {2015})},\
  \Eprint {http://arxiv.org/abs/1406.6948} {arXiv:1406.6948 [astro-ph.HE]}
  \BibitemShut {NoStop}%
\bibitem [{\citenamefont {{Bovy}}(2020)}]{2020arXiv201202169B}%
  \BibitemOpen
  \bibfield  {author} {\bibinfo {author} {\bibfnamefont {J.}~\bibnamefont
  {{Bovy}}},\ }\href@noop {} {\bibfield  {journal} {\bibinfo  {journal} {arXiv
  e-prints}\ ,\ \bibinfo {eid} {arXiv:2012.02169}} (\bibinfo {year} {2020})},\
  \Eprint {http://arxiv.org/abs/2012.02169} {arXiv:2012.02169 [astro-ph.GA]}
  \BibitemShut {NoStop}%
\bibitem [{\citenamefont {{Gravity
  Collaboration}}(2019)}]{2019A&A...625L..10G}%
  \BibitemOpen
  \bibfield  {author} {\bibinfo {author} {\bibnamefont {{Gravity
  Collaboration}}},\ }\href {\doibase 10.1051/0004-6361/201935656} {\bibfield
  {journal} {\bibinfo  {journal} {Astron. Astrophys.}\ }\textbf {\bibinfo
  {volume} {625}},\ \bibinfo {eid} {L10} (\bibinfo {year} {2019})},\ \Eprint
  {http://arxiv.org/abs/1904.05721} {arXiv:1904.05721 [astro-ph.GA]}
  \BibitemShut {NoStop}%
\bibitem [{\citenamefont {Rodd}\ and\ \citenamefont
  {Toomey}(2017)}]{NPTFit-Sim}%
  \BibitemOpen
  \bibfield  {author} {\bibinfo {author} {\bibfnamefont {N.~L.}\ \bibnamefont
  {Rodd}}\ and\ \bibinfo {author} {\bibfnamefont {M.~W.}\ \bibnamefont
  {Toomey}},\ }\href {https://github.com/nickrodd/NPTFit-Sim} {\emph {\bibinfo
  {title} {{NPTFit-Sim}}}} (\bibinfo {year} {2017})\BibitemShut {NoStop}%
\bibitem [{\citenamefont {Lorimer}\ \emph {et~al.}(2006)\citenamefont {Lorimer}
  \emph {et~al.}}]{Lorimer:2006qs}%
  \BibitemOpen
  \bibfield  {author} {\bibinfo {author} {\bibfnamefont {D.~R.}\ \bibnamefont
  {Lorimer}} \emph {et~al.},\ }\href {\doibase
  10.1111/j.1365-2966.2006.10887.x} {\bibfield  {journal} {\bibinfo  {journal}
  {Mon. Not. Roy. Astron. Soc.}\ }\textbf {\bibinfo {volume} {372}},\ \bibinfo
  {pages} {777} (\bibinfo {year} {2006})},\ \Eprint
  {http://arxiv.org/abs/astro-ph/0607640} {arXiv:astro-ph/0607640 [astro-ph]}
  \BibitemShut {NoStop}%
\bibitem [{\citenamefont {Bartels}\ \emph
  {et~al.}(2018{\natexlab{b}})\citenamefont {Bartels}, \citenamefont
  {Edwards},\ and\ \citenamefont {Weniger}}]{Bartels:2018xom}%
  \BibitemOpen
  \bibfield  {author} {\bibinfo {author} {\bibfnamefont {R.~T.}\ \bibnamefont
  {Bartels}}, \bibinfo {author} {\bibfnamefont {T.~D.~P.}\ \bibnamefont
  {Edwards}}, \ and\ \bibinfo {author} {\bibfnamefont {C.}~\bibnamefont
  {Weniger}},\ }\href {\doibase 10.1093/mnras/sty2529} {\bibfield  {journal}
  {\bibinfo  {journal} {Mon. Not. Roy. Astron. Soc.}\ }\textbf {\bibinfo
  {volume} {481}},\ \bibinfo {pages} {3966} (\bibinfo {year}
  {2018}{\natexlab{b}})},\ \Eprint {http://arxiv.org/abs/1805.11097}
  {arXiv:1805.11097 [astro-ph.HE]} \BibitemShut {NoStop}%
\bibitem [{\citenamefont {Collin}\ \emph {et~al.}(2021)\citenamefont {Collin},
  \citenamefont {Rodd}, \citenamefont {Erjavec},\ and\ \citenamefont
  {Perez}}]{Collin:2021ufc}%
  \BibitemOpen
  \bibfield  {author} {\bibinfo {author} {\bibfnamefont {G.~H.}\ \bibnamefont
  {Collin}}, \bibinfo {author} {\bibfnamefont {N.~L.}\ \bibnamefont {Rodd}},
  \bibinfo {author} {\bibfnamefont {T.}~\bibnamefont {Erjavec}}, \ and\
  \bibinfo {author} {\bibfnamefont {K.}~\bibnamefont {Perez}},\ }\href@noop {}
  {\  (\bibinfo {year} {2021})},\ \Eprint {http://arxiv.org/abs/2104.04529}
  {arXiv:2104.04529 [astro-ph.IM]} \BibitemShut {NoStop}%
\bibitem [{\citenamefont {{Brewer}}\ \emph {et~al.}(2013)\citenamefont
  {{Brewer}}, \citenamefont {{Foreman-Mackey}},\ and\ \citenamefont
  {{Hogg}}}]{2013AJ....146....7B}%
  \BibitemOpen
  \bibfield  {author} {\bibinfo {author} {\bibfnamefont {B.~J.}\ \bibnamefont
  {{Brewer}}}, \bibinfo {author} {\bibfnamefont {D.}~\bibnamefont
  {{Foreman-Mackey}}}, \ and\ \bibinfo {author} {\bibfnamefont {D.~W.}\
  \bibnamefont {{Hogg}}},\ }\href {\doibase 10.1088/0004-6256/146/1/7}
  {\bibfield  {journal} {\bibinfo  {journal} {Astron. J.}\ }\textbf {\bibinfo
  {volume} {146}},\ \bibinfo {eid} {7} (\bibinfo {year} {2013})},\ \Eprint
  {http://arxiv.org/abs/1211.5805} {arXiv:1211.5805 [astro-ph.IM]} \BibitemShut
  {NoStop}%
\bibitem [{\citenamefont {{Liu}}\ \emph {et~al.}(2021)\citenamefont {{Liu}},
  \citenamefont {{McAuliffe}},\ and\ \citenamefont
  {{Regier}}}]{2021arXiv210202409L}%
  \BibitemOpen
  \bibfield  {author} {\bibinfo {author} {\bibfnamefont {R.}~\bibnamefont
  {{Liu}}}, \bibinfo {author} {\bibfnamefont {J.~D.}\ \bibnamefont
  {{McAuliffe}}}, \ and\ \bibinfo {author} {\bibfnamefont {J.}~\bibnamefont
  {{Regier}}},\ }\href@noop {} {\  (\bibinfo {year} {2021})},\ \Eprint
  {http://arxiv.org/abs/2102.02409} {arXiv:2102.02409 [astro-ph.IM]}
  \BibitemShut {NoStop}%
\bibitem [{\citenamefont {Daylan}\ \emph {et~al.}(2017)\citenamefont {Daylan},
  \citenamefont {Portillo},\ and\ \citenamefont {Finkbeiner}}]{Daylan:2016tia}%
  \BibitemOpen
  \bibfield  {author} {\bibinfo {author} {\bibfnamefont {T.}~\bibnamefont
  {Daylan}}, \bibinfo {author} {\bibfnamefont {S.~K.~N.}\ \bibnamefont
  {Portillo}}, \ and\ \bibinfo {author} {\bibfnamefont {D.~P.}\ \bibnamefont
  {Finkbeiner}},\ }\href {\doibase 10.3847/1538-4357/aa679e} {\bibfield
  {journal} {\bibinfo  {journal} {Astrophys. J.}\ }\textbf {\bibinfo {volume}
  {839}},\ \bibinfo {pages} {4} (\bibinfo {year} {2017})},\ \Eprint
  {http://arxiv.org/abs/1607.04637} {arXiv:1607.04637 [astro-ph.IM]}
  \BibitemShut {NoStop}%
\bibitem [{\citenamefont {Feroz}\ \emph {et~al.}(2013)\citenamefont {Feroz},
  \citenamefont {Hobson}, \citenamefont {Cameron},\ and\ \citenamefont
  {Pettitt}}]{Feroz:2013hea}%
  \BibitemOpen
  \bibfield  {author} {\bibinfo {author} {\bibfnamefont {F.}~\bibnamefont
  {Feroz}}, \bibinfo {author} {\bibfnamefont {M.~P.}\ \bibnamefont {Hobson}},
  \bibinfo {author} {\bibfnamefont {E.}~\bibnamefont {Cameron}}, \ and\
  \bibinfo {author} {\bibfnamefont {A.~N.}\ \bibnamefont {Pettitt}},\
  }\href@noop {} {\  (\bibinfo {year} {2013})},\ \Eprint
  {http://arxiv.org/abs/1306.2144} {arXiv:1306.2144 [astro-ph.IM]} \BibitemShut
  {NoStop}%
\bibitem [{\citenamefont {Skilling}(2006)}]{skilling2006}%
  \BibitemOpen
  \bibfield  {author} {\bibinfo {author} {\bibfnamefont {J.}~\bibnamefont
  {Skilling}},\ }\href {\doibase 10.1214/06-BA127} {\bibfield  {journal}
  {\bibinfo  {journal} {Bayesian Anal.}\ }\textbf {\bibinfo {volume} {1}},\
  \bibinfo {pages} {833} (\bibinfo {year} {2006})}\BibitemShut {NoStop}%
\bibitem [{\citenamefont {Speagle}(2020)}]{Speagle_2020}%
  \BibitemOpen
  \bibfield  {author} {\bibinfo {author} {\bibfnamefont {J.~S.}\ \bibnamefont
  {Speagle}},\ }\href {\doibase 10.1093/mnras/staa278} {\bibfield  {journal}
  {\bibinfo  {journal} {Monthly Notices of the Royal Astronomical Society}\
  }\textbf {\bibinfo {volume} {493}},\ \bibinfo {pages} {3132} (\bibinfo {year}
  {2020})}\BibitemShut {NoStop}%
\bibitem [{\citenamefont {Leane}\ and\ \citenamefont
  {Slatyer}(2020{\natexlab{a}})}]{Leane:2020pfc}%
  \BibitemOpen
  \bibfield  {author} {\bibinfo {author} {\bibfnamefont {R.~K.}\ \bibnamefont
  {Leane}}\ and\ \bibinfo {author} {\bibfnamefont {T.~R.}\ \bibnamefont
  {Slatyer}},\ }\href {\doibase 10.1103/PhysRevD.102.063019} {\bibfield
  {journal} {\bibinfo  {journal} {Phys. Rev. D}\ }\textbf {\bibinfo {volume}
  {102}},\ \bibinfo {pages} {063019} (\bibinfo {year} {2020}{\natexlab{a}})},\
  \Eprint {http://arxiv.org/abs/2002.12371} {arXiv:2002.12371 [astro-ph.HE]}
  \BibitemShut {NoStop}%
\bibitem [{\citenamefont {Leane}\ and\ \citenamefont
  {Slatyer}(2020{\natexlab{b}})}]{Leane:2020nmi}%
  \BibitemOpen
  \bibfield  {author} {\bibinfo {author} {\bibfnamefont {R.~K.}\ \bibnamefont
  {Leane}}\ and\ \bibinfo {author} {\bibfnamefont {T.~R.}\ \bibnamefont
  {Slatyer}},\ }\href {\doibase 10.1103/PhysRevLett.125.121105} {\bibfield
  {journal} {\bibinfo  {journal} {Phys. Rev. Lett.}\ }\textbf {\bibinfo
  {volume} {125}},\ \bibinfo {pages} {121105} (\bibinfo {year}
  {2020}{\natexlab{b}})},\ \Eprint {http://arxiv.org/abs/2002.12370}
  {arXiv:2002.12370 [astro-ph.HE]} \BibitemShut {NoStop}%
\bibitem [{\citenamefont {Calore}\ \emph {et~al.}(2021)\citenamefont {Calore},
  \citenamefont {Donato},\ and\ \citenamefont {Manconi}}]{Calore:2021bty}%
  \BibitemOpen
  \bibfield  {author} {\bibinfo {author} {\bibfnamefont {F.}~\bibnamefont
  {Calore}}, \bibinfo {author} {\bibfnamefont {F.}~\bibnamefont {Donato}}, \
  and\ \bibinfo {author} {\bibfnamefont {S.}~\bibnamefont {Manconi}},\
  }\href@noop {} {\  (\bibinfo {year} {2021})},\ \Eprint
  {http://arxiv.org/abs/2102.12497} {arXiv:2102.12497 [astro-ph.HE]}
  \BibitemShut {NoStop}%
\bibitem [{\citenamefont {Lisanti}\ \emph {et~al.}(2016)\citenamefont
  {Lisanti}, \citenamefont {Mishra-Sharma}, \citenamefont {Necib},\ and\
  \citenamefont {Safdi}}]{Lisanti:2016jub}%
  \BibitemOpen
  \bibfield  {author} {\bibinfo {author} {\bibfnamefont {M.}~\bibnamefont
  {Lisanti}}, \bibinfo {author} {\bibfnamefont {S.}~\bibnamefont
  {Mishra-Sharma}}, \bibinfo {author} {\bibfnamefont {L.}~\bibnamefont
  {Necib}}, \ and\ \bibinfo {author} {\bibfnamefont {B.~R.}\ \bibnamefont
  {Safdi}},\ }\href {\doibase 10.3847/0004-637X/832/2/117} {\bibfield
  {journal} {\bibinfo  {journal} {Astrophys. J.}\ }\textbf {\bibinfo {volume}
  {832}},\ \bibinfo {pages} {117} (\bibinfo {year} {2016})},\ \Eprint
  {http://arxiv.org/abs/1606.04101} {arXiv:1606.04101 [astro-ph.HE]}
  \BibitemShut {NoStop}%
\bibitem [{\citenamefont {Zechlin}\ \emph
  {et~al.}(2016{\natexlab{a}})\citenamefont {Zechlin}, \citenamefont {Cuoco},
  \citenamefont {Donato}, \citenamefont {Fornengo},\ and\ \citenamefont
  {Regis}}]{Zechlin:2016pme}%
  \BibitemOpen
  \bibfield  {author} {\bibinfo {author} {\bibfnamefont {H.-S.}\ \bibnamefont
  {Zechlin}}, \bibinfo {author} {\bibfnamefont {A.}~\bibnamefont {Cuoco}},
  \bibinfo {author} {\bibfnamefont {F.}~\bibnamefont {Donato}}, \bibinfo
  {author} {\bibfnamefont {N.}~\bibnamefont {Fornengo}}, \ and\ \bibinfo
  {author} {\bibfnamefont {M.}~\bibnamefont {Regis}},\ }\href {\doibase
  10.3847/2041-8205/826/2/L31} {\bibfield  {journal} {\bibinfo  {journal}
  {Astrophys. J. Lett.}\ }\textbf {\bibinfo {volume} {826}},\ \bibinfo {pages}
  {L31} (\bibinfo {year} {2016}{\natexlab{a}})},\ \Eprint
  {http://arxiv.org/abs/1605.04256} {arXiv:1605.04256 [astro-ph.HE]}
  \BibitemShut {NoStop}%
\bibitem [{\citenamefont {Zechlin}\ \emph
  {et~al.}(2016{\natexlab{b}})\citenamefont {Zechlin}, \citenamefont {Cuoco},
  \citenamefont {Donato}, \citenamefont {Fornengo},\ and\ \citenamefont
  {Vittino}}]{Zechlin:2015wdz}%
  \BibitemOpen
  \bibfield  {author} {\bibinfo {author} {\bibfnamefont {H.-S.}\ \bibnamefont
  {Zechlin}}, \bibinfo {author} {\bibfnamefont {A.}~\bibnamefont {Cuoco}},
  \bibinfo {author} {\bibfnamefont {F.}~\bibnamefont {Donato}}, \bibinfo
  {author} {\bibfnamefont {N.}~\bibnamefont {Fornengo}}, \ and\ \bibinfo
  {author} {\bibfnamefont {A.}~\bibnamefont {Vittino}},\ }\href {\doibase
  10.3847/0067-0049/225/2/18} {\bibfield  {journal} {\bibinfo  {journal}
  {Astrophys. J. Suppl.}\ }\textbf {\bibinfo {volume} {225}},\ \bibinfo {pages}
  {18} (\bibinfo {year} {2016}{\natexlab{b}})},\ \Eprint
  {http://arxiv.org/abs/1512.07190} {arXiv:1512.07190 [astro-ph.HE]}
  \BibitemShut {NoStop}%
\bibitem [{\citenamefont {Somalwar}\ \emph {et~al.}(2021)\citenamefont
  {Somalwar}, \citenamefont {Chang}, \citenamefont {Mishra-Sharma},\ and\
  \citenamefont {Lisanti}}]{Somalwar:2020awt}%
  \BibitemOpen
  \bibfield  {author} {\bibinfo {author} {\bibfnamefont {J.~J.}\ \bibnamefont
  {Somalwar}}, \bibinfo {author} {\bibfnamefont {L.~J.}\ \bibnamefont {Chang}},
  \bibinfo {author} {\bibfnamefont {S.}~\bibnamefont {Mishra-Sharma}}, \ and\
  \bibinfo {author} {\bibfnamefont {M.}~\bibnamefont {Lisanti}},\ }\href
  {\doibase 10.3847/1538-4357/abc87d} {\bibfield  {journal} {\bibinfo
  {journal} {Astrophys. J.}\ }\textbf {\bibinfo {volume} {906}},\ \bibinfo
  {pages} {57} (\bibinfo {year} {2021})},\ \Eprint
  {http://arxiv.org/abs/2009.00021} {arXiv:2009.00021 [astro-ph.CO]}
  \BibitemShut {NoStop}%
\bibitem [{\citenamefont {Rubin}(1984)}]{10.1214/aos/1176346785}%
  \BibitemOpen
  \bibfield  {author} {\bibinfo {author} {\bibfnamefont {D.~B.}\ \bibnamefont
  {Rubin}},\ }\href {\doibase 10.1214/aos/1176346785} {\bibfield  {journal}
  {\bibinfo  {journal} {The Annals of Statistics}\ }\textbf {\bibinfo {volume}
  {12}},\ \bibinfo {pages} {1151 } (\bibinfo {year} {1984})}\BibitemShut
  {NoStop}%
\bibitem [{\citenamefont {Papamakarios}\ and\ \citenamefont
  {Murray}(2016)}]{10.5555/3157096.3157212}%
  \BibitemOpen
  \bibfield  {author} {\bibinfo {author} {\bibfnamefont {G.}~\bibnamefont
  {Papamakarios}}\ and\ \bibinfo {author} {\bibfnamefont {I.}~\bibnamefont
  {Murray}},\ }in\ \href
  {https://proceedings.neurips.cc/paper/2016/hash/6aca97005c68f1206823815f66102863-Abstract.html}
  {\emph {\bibinfo {booktitle} {Proceedings of the 30th International
  Conference on Neural Information Processing Systems}}},\ \bibinfo {series and
  number} {NIPS'16}\ (\bibinfo  {publisher} {Curran Associates Inc.},\ \bibinfo
  {address} {Red Hook, NY, USA},\ \bibinfo {year} {2016})\ pp.\ \bibinfo
  {pages} {1036--1044},\ \Eprint {http://arxiv.org/abs/1605.06376}
  {arXiv:1605.06376 [stat.ML]} \BibitemShut {NoStop}%
\bibitem [{\citenamefont {Cranmer}\ and\ \citenamefont
  {Louppe}(2016)}]{cranmer_kyle_2016_198541}%
  \BibitemOpen
  \bibfield  {author} {\bibinfo {author} {\bibfnamefont {K.}~\bibnamefont
  {Cranmer}}\ and\ \bibinfo {author} {\bibfnamefont {G.}~\bibnamefont
  {Louppe}},\ }\href {\doibase 10.5281/zenodo.198541} {\bibfield  {journal}
  {\bibinfo  {journal} {J. Brief Ideas}\ } (\bibinfo {year} {2016}),\
  10.5281/zenodo.198541}\BibitemShut {NoStop}%
\bibitem [{\citenamefont {Papamakarios}\ \emph {et~al.}(2017)\citenamefont
  {Papamakarios}, \citenamefont {Pavlakou},\ and\ \citenamefont
  {Murray}}]{10.5555/3294771.3294994}%
  \BibitemOpen
  \bibfield  {author} {\bibinfo {author} {\bibfnamefont {G.}~\bibnamefont
  {Papamakarios}}, \bibinfo {author} {\bibfnamefont {T.}~\bibnamefont
  {Pavlakou}}, \ and\ \bibinfo {author} {\bibfnamefont {I.}~\bibnamefont
  {Murray}},\ }in\ \href
  {https://papers.nips.cc/paper/2017/hash/6c1da886822c67822bcf3679d04369fa-Abstract.html}
  {\emph {\bibinfo {booktitle} {Proceedings of the 31st International
  Conference on Neural Information Processing Systems}}},\ \bibinfo {series and
  number} {NIPS'17}\ (\bibinfo  {publisher} {Curran Associates Inc.},\ \bibinfo
  {address} {Red Hook, NY, USA},\ \bibinfo {year} {2017})\ pp.\ \bibinfo
  {pages} {2335--2344}\BibitemShut {NoStop}%
\bibitem [{\citenamefont {Kingma}\ \emph {et~al.}(2016)\citenamefont {Kingma},
  \citenamefont {Salimans}, \citenamefont {Jozefowicz}, \citenamefont {Chen},
  \citenamefont {Sutskever},\ and\ \citenamefont
  {Welling}}]{10.5555/3157382.3157627}%
  \BibitemOpen
  \bibfield  {author} {\bibinfo {author} {\bibfnamefont {D.~P.}\ \bibnamefont
  {Kingma}}, \bibinfo {author} {\bibfnamefont {T.}~\bibnamefont {Salimans}},
  \bibinfo {author} {\bibfnamefont {R.}~\bibnamefont {Jozefowicz}}, \bibinfo
  {author} {\bibfnamefont {X.}~\bibnamefont {Chen}}, \bibinfo {author}
  {\bibfnamefont {I.}~\bibnamefont {Sutskever}}, \ and\ \bibinfo {author}
  {\bibfnamefont {M.}~\bibnamefont {Welling}},\ }in\ \href
  {https://papers.nips.cc/paper/2016/hash/ddeebdeefdb7e7e7a697e1c3e3d8ef54-Abstract.html}
  {\emph {\bibinfo {booktitle} {Proceedings of the 30th International
  Conference on Neural Information Processing Systems}}},\ \bibinfo {series and
  number} {NIPS'16}\ (\bibinfo  {publisher} {Curran Associates Inc.},\ \bibinfo
  {address} {Red Hook, NY, USA},\ \bibinfo {year} {2016})\ pp.\ \bibinfo
  {pages} {4743--4751}\BibitemShut {NoStop}%
\bibitem [{\citenamefont {Dinh}\ \emph {et~al.}(2017)\citenamefont {Dinh},
  \citenamefont {Sohl{-}Dickstein},\ and\ \citenamefont
  {Bengio}}]{DBLP:conf/iclr/DinhSB17}%
  \BibitemOpen
  \bibfield  {author} {\bibinfo {author} {\bibfnamefont {L.}~\bibnamefont
  {Dinh}}, \bibinfo {author} {\bibfnamefont {J.}~\bibnamefont
  {Sohl{-}Dickstein}}, \ and\ \bibinfo {author} {\bibfnamefont
  {S.}~\bibnamefont {Bengio}},\ }in\ \href
  {https://openreview.net/forum?id=HkpbnH9lx} {\emph {\bibinfo {booktitle} {5th
  International Conference on Learning Representations, {ICLR} 2017, Toulon,
  France, April 24-26, 2017, Conference Track Proceedings}}}\ (\bibinfo {year}
  {2017})\BibitemShut {NoStop}%
\bibitem [{\citenamefont {Dinh}\ \emph {et~al.}(2015)\citenamefont {Dinh},
  \citenamefont {Krueger},\ and\ \citenamefont
  {Bengio}}]{DBLP:journals/corr/DinhKB14}%
  \BibitemOpen
  \bibfield  {author} {\bibinfo {author} {\bibfnamefont {L.}~\bibnamefont
  {Dinh}}, \bibinfo {author} {\bibfnamefont {D.}~\bibnamefont {Krueger}}, \
  and\ \bibinfo {author} {\bibfnamefont {Y.}~\bibnamefont {Bengio}},\ }in\
  \href {http://arxiv.org/abs/1410.8516} {\emph {\bibinfo {booktitle} {3rd
  International Conference on Learning Representations, {ICLR} 2015, San Diego,
  CA, USA, May 7-9, 2015, Workshop Track Proceedings}}},\ \bibinfo {editor}
  {edited by\ \bibinfo {editor} {\bibfnamefont {Y.}~\bibnamefont {Bengio}}\
  and\ \bibinfo {editor} {\bibfnamefont {Y.}~\bibnamefont {LeCun}}}\ (\bibinfo
  {year} {2015})\BibitemShut {NoStop}%
\bibitem [{\citenamefont {Durkan}\ \emph
  {et~al.}(2019{\natexlab{a}})\citenamefont {Durkan}, \citenamefont {Bekasov},
  \citenamefont {Murray},\ and\ \citenamefont
  {Papamakarios}}]{DBLP:conf/nips/DurkanB0P19}%
  \BibitemOpen
  \bibfield  {author} {\bibinfo {author} {\bibfnamefont {C.}~\bibnamefont
  {Durkan}}, \bibinfo {author} {\bibfnamefont {A.}~\bibnamefont {Bekasov}},
  \bibinfo {author} {\bibfnamefont {I.}~\bibnamefont {Murray}}, \ and\ \bibinfo
  {author} {\bibfnamefont {G.}~\bibnamefont {Papamakarios}},\ }in\ \href
  {https://proceedings.neurips.cc/paper/2019/hash/7ac71d433f282034e088473244df8c02-Abstract.html}
  {\emph {\bibinfo {booktitle} {Advances in Neural Information Processing
  Systems 32: Annual Conference on Neural Information Processing Systems 2019,
  NeurIPS 2019, December 8-14, 2019, Vancouver, BC, Canada}}},\ \bibinfo
  {editor} {edited by\ \bibinfo {editor} {\bibfnamefont {H.~M.}\ \bibnamefont
  {Wallach}}, \bibinfo {editor} {\bibfnamefont {H.}~\bibnamefont {Larochelle}},
  \bibinfo {editor} {\bibfnamefont {A.}~\bibnamefont {Beygelzimer}}, \bibinfo
  {editor} {\bibfnamefont {F.}~\bibnamefont {d'Alch{\'{e}}{-}Buc}}, \bibinfo
  {editor} {\bibfnamefont {E.~B.}\ \bibnamefont {Fox}}, \ and\ \bibinfo
  {editor} {\bibfnamefont {R.}~\bibnamefont {Garnett}}}\ (\bibinfo {year}
  {2019})\ pp.\ \bibinfo {pages} {7509--7520}\BibitemShut {NoStop}%
\bibitem [{\citenamefont {Durkan}\ \emph
  {et~al.}(2019{\natexlab{b}})\citenamefont {Durkan}, \citenamefont {Bekasov},
  \citenamefont {Murray},\ and\ \citenamefont
  {Papamakarios}}]{durkan2019cubic}%
  \BibitemOpen
  \bibfield  {author} {\bibinfo {author} {\bibfnamefont {C.}~\bibnamefont
  {Durkan}}, \bibinfo {author} {\bibfnamefont {A.}~\bibnamefont {Bekasov}},
  \bibinfo {author} {\bibfnamefont {I.}~\bibnamefont {Murray}}, \ and\ \bibinfo
  {author} {\bibfnamefont {G.}~\bibnamefont {Papamakarios}},\ }in\ \href
  {https://arxiv.org/abs/1906.02145} {\emph {\bibinfo {booktitle} {1st Workshop
  on Invertible Neural Networks and Normalizing Flows at ICML 2019}}}\
  (\bibinfo {year} {2019})\ \Eprint {http://arxiv.org/abs/1906.02145}
  {arXiv:1906.02145} \BibitemShut {NoStop}%
\bibitem [{\citenamefont {Grathwohl}\ \emph {et~al.}(2019)\citenamefont
  {Grathwohl}, \citenamefont {Chen}, \citenamefont {Bettencourt}, \citenamefont
  {Sutskever},\ and\ \citenamefont
  {Duvenaud}}]{DBLP:conf/iclr/GrathwohlCBSD19}%
  \BibitemOpen
  \bibfield  {author} {\bibinfo {author} {\bibfnamefont {W.}~\bibnamefont
  {Grathwohl}}, \bibinfo {author} {\bibfnamefont {R.~T.~Q.}\ \bibnamefont
  {Chen}}, \bibinfo {author} {\bibfnamefont {J.}~\bibnamefont {Bettencourt}},
  \bibinfo {author} {\bibfnamefont {I.}~\bibnamefont {Sutskever}}, \ and\
  \bibinfo {author} {\bibfnamefont {D.}~\bibnamefont {Duvenaud}},\ }in\ \href
  {https://openreview.net/forum?id=rJxgknCcK7} {\emph {\bibinfo {booktitle}
  {7th International Conference on Learning Representations, {ICLR} 2019, New
  Orleans, LA, USA, May 6-9, 2019}}}\ (\bibinfo {year} {2019})\ \Eprint
  {http://arxiv.org/abs/1810.01367} {arXiv:1810.01367 [cs.LG]} \BibitemShut
  {NoStop}%
\bibitem [{\citenamefont {Germain}\ \emph {et~al.}(2015)\citenamefont
  {Germain}, \citenamefont {Gregor}, \citenamefont {Murray},\ and\
  \citenamefont {Larochelle}}]{DBLP:conf/icml/GermainGML15}%
  \BibitemOpen
  \bibfield  {author} {\bibinfo {author} {\bibfnamefont {M.}~\bibnamefont
  {Germain}}, \bibinfo {author} {\bibfnamefont {K.}~\bibnamefont {Gregor}},
  \bibinfo {author} {\bibfnamefont {I.}~\bibnamefont {Murray}}, \ and\ \bibinfo
  {author} {\bibfnamefont {H.}~\bibnamefont {Larochelle}},\ }in\ \href
  {http://proceedings.mlr.press/v37/germain15.html} {\emph {\bibinfo
  {booktitle} {Proceedings of the 32nd International Conference on Machine
  Learning, {ICML} 2015, Lille, France, 6-11 July 2015}}},\ \bibinfo {series}
  {{JMLR} Workshop and Conference Proceedings}, Vol.~\bibinfo {volume} {37},\
  \bibinfo {editor} {edited by\ \bibinfo {editor} {\bibfnamefont {F.~R.}\
  \bibnamefont {Bach}}\ and\ \bibinfo {editor} {\bibfnamefont {D.~M.}\
  \bibnamefont {Blei}}}\ (\bibinfo {year} {2015})\ pp.\ \bibinfo {pages}
  {881--889}\BibitemShut {NoStop}%
\bibitem [{\citenamefont {Defferrard}\ \emph {et~al.}(2020)\citenamefont
  {Defferrard}, \citenamefont {Milani}, \citenamefont {Gusset},\ and\
  \citenamefont {Perraudin}}]{DBLP:conf/iclr/DefferrardMGP20}%
  \BibitemOpen
  \bibfield  {author} {\bibinfo {author} {\bibfnamefont {M.}~\bibnamefont
  {Defferrard}}, \bibinfo {author} {\bibfnamefont {M.}~\bibnamefont {Milani}},
  \bibinfo {author} {\bibfnamefont {F.}~\bibnamefont {Gusset}}, \ and\ \bibinfo
  {author} {\bibfnamefont {N.}~\bibnamefont {Perraudin}},\ }in\ \href
  {https://openreview.net/forum?id=B1e3OlStPB} {\emph {\bibinfo {booktitle}
  {8th International Conference on Learning Representations, {ICLR} 2020, Addis
  Ababa, Ethiopia, April 26-30, 2020}}}\ (\bibinfo {year} {2020})\BibitemShut
  {NoStop}%
\bibitem [{\citenamefont {Perraudin}\ \emph {et~al.}(2019)\citenamefont
  {Perraudin}, \citenamefont {Defferrard}, \citenamefont {Kacprzak},\ and\
  \citenamefont {Sgier}}]{Perraudin:2018rbt}%
  \BibitemOpen
  \bibfield  {author} {\bibinfo {author} {\bibfnamefont {N.}~\bibnamefont
  {Perraudin}}, \bibinfo {author} {\bibfnamefont {M.}~\bibnamefont
  {Defferrard}}, \bibinfo {author} {\bibfnamefont {T.}~\bibnamefont
  {Kacprzak}}, \ and\ \bibinfo {author} {\bibfnamefont {R.}~\bibnamefont
  {Sgier}},\ }\href {\doibase 10.1016/j.ascom.2019.03.004} {\bibfield
  {journal} {\bibinfo  {journal} {Astron. Comput.}\ }\textbf {\bibinfo {volume}
  {27}},\ \bibinfo {pages} {130} (\bibinfo {year} {2019})},\ \Eprint
  {http://arxiv.org/abs/1810.12186} {arXiv:1810.12186 [astro-ph.CO]}
  \BibitemShut {NoStop}%
\bibitem [{\citenamefont {Defferrard}\ \emph {et~al.}(2019)\citenamefont
  {Defferrard}, \citenamefont {Perraudin}, \citenamefont {Kacprzak},\ and\
  \citenamefont {Sgier}}]{deepsphere_rlgm}%
  \BibitemOpen
  \bibfield  {author} {\bibinfo {author} {\bibfnamefont {M.}~\bibnamefont
  {Defferrard}}, \bibinfo {author} {\bibfnamefont {N.}~\bibnamefont
  {Perraudin}}, \bibinfo {author} {\bibfnamefont {T.}~\bibnamefont {Kacprzak}},
  \ and\ \bibinfo {author} {\bibfnamefont {R.}~\bibnamefont {Sgier}},\ }in\
  \href@noop {} {\emph {\bibinfo {booktitle} {ICLR Workshop on Representation
  Learning on Graphs and Manifolds}}}\ (\bibinfo {year} {2019})\ \Eprint
  {http://arxiv.org/abs/1904.05146} {arXiv:1904.05146 [cs.LG]} \BibitemShut
  {NoStop}%
\bibitem [{\citenamefont {Defferrard}\ \emph {et~al.}(2016)\citenamefont
  {Defferrard}, \citenamefont {Bresson},\ and\ \citenamefont
  {Vandergheynst}}]{DBLP:conf/nips/DefferrardBV16}%
  \BibitemOpen
  \bibfield  {author} {\bibinfo {author} {\bibfnamefont {M.}~\bibnamefont
  {Defferrard}}, \bibinfo {author} {\bibfnamefont {X.}~\bibnamefont {Bresson}},
  \ and\ \bibinfo {author} {\bibfnamefont {P.}~\bibnamefont {Vandergheynst}},\
  }in\ \href
  {https://proceedings.neurips.cc/paper/2016/hash/04df4d434d481c5bb723be1b6df1ee65-Abstract.html}
  {\emph {\bibinfo {booktitle} {Advances in Neural Information Processing
  Systems 29: Annual Conference on Neural Information Processing Systems 2016,
  December 5-10, 2016, Barcelona, Spain}}},\ \bibinfo {editor} {edited by\
  \bibinfo {editor} {\bibfnamefont {D.~D.}\ \bibnamefont {Lee}}, \bibinfo
  {editor} {\bibfnamefont {M.}~\bibnamefont {Sugiyama}}, \bibinfo {editor}
  {\bibfnamefont {U.}~\bibnamefont {von Luxburg}}, \bibinfo {editor}
  {\bibfnamefont {I.}~\bibnamefont {Guyon}}, \ and\ \bibinfo {editor}
  {\bibfnamefont {R.}~\bibnamefont {Garnett}}}\ (\bibinfo {year} {2016})\ pp.\
  \bibinfo {pages} {3837--3845}\BibitemShut {NoStop}%
\bibitem [{\citenamefont {Kingma}\ and\ \citenamefont
  {Ba}(2015)}]{DBLP:journals/corr/KingmaB14}%
  \BibitemOpen
  \bibfield  {author} {\bibinfo {author} {\bibfnamefont {D.~P.}\ \bibnamefont
  {Kingma}}\ and\ \bibinfo {author} {\bibfnamefont {J.}~\bibnamefont {Ba}},\
  }in\ \href {http://arxiv.org/abs/1412.6980} {\emph {\bibinfo {booktitle} {3rd
  International Conference on Learning Representations, {ICLR} 2015, San Diego,
  CA, USA, May 7-9, 2015, Conference Track Proceedings}}},\ \bibinfo {editor}
  {edited by\ \bibinfo {editor} {\bibfnamefont {Y.}~\bibnamefont {Bengio}}\
  and\ \bibinfo {editor} {\bibfnamefont {Y.}~\bibnamefont {LeCun}}}\ (\bibinfo
  {year} {2015})\BibitemShut {NoStop}%
\bibitem [{\citenamefont {Loshchilov}\ and\ \citenamefont
  {Hutter}(2019)}]{DBLP:conf/iclr/LoshchilovH19}%
  \BibitemOpen
  \bibfield  {author} {\bibinfo {author} {\bibfnamefont {I.}~\bibnamefont
  {Loshchilov}}\ and\ \bibinfo {author} {\bibfnamefont {F.}~\bibnamefont
  {Hutter}},\ }in\ \href {https://openreview.net/forum?id=Bkg6RiCqY7} {\emph
  {\bibinfo {booktitle} {7th International Conference on Learning
  Representations, {ICLR} 2019, New Orleans, LA, USA, May 6-9, 2019}}}\
  (\bibinfo {year} {2019})\BibitemShut {NoStop}%
\bibitem [{\citenamefont {Calore}\ \emph
  {et~al.}(2015{\natexlab{b}})\citenamefont {Calore}, \citenamefont {Cholis},
  \citenamefont {McCabe},\ and\ \citenamefont {Weniger}}]{Calore:2014nla}%
  \BibitemOpen
  \bibfield  {author} {\bibinfo {author} {\bibfnamefont {F.}~\bibnamefont
  {Calore}}, \bibinfo {author} {\bibfnamefont {I.}~\bibnamefont {Cholis}},
  \bibinfo {author} {\bibfnamefont {C.}~\bibnamefont {McCabe}}, \ and\ \bibinfo
  {author} {\bibfnamefont {C.}~\bibnamefont {Weniger}},\ }\href {\doibase
  10.1103/PhysRevD.91.063003} {\bibfield  {journal} {\bibinfo  {journal} {Phys.
  Rev. D}\ }\textbf {\bibinfo {volume} {91}},\ \bibinfo {pages} {063003}
  (\bibinfo {year} {2015}{\natexlab{b}})},\ \Eprint
  {http://arxiv.org/abs/1411.4647} {arXiv:1411.4647 [hep-ph]} \BibitemShut
  {NoStop}%
\bibitem [{\citenamefont {Mishra-Sharma}\ and\ \citenamefont
  {Cranmer}(2020)}]{Mishra-Sharma:2020kjb}%
  \BibitemOpen
  \bibfield  {author} {\bibinfo {author} {\bibfnamefont {S.}~\bibnamefont
  {Mishra-Sharma}}\ and\ \bibinfo {author} {\bibfnamefont {K.}~\bibnamefont
  {Cranmer}},\ }in\ \href@noop {} {\emph {\bibinfo {booktitle} {{Machine
  Learning and the Physical Sciences Workshop at the 34th Conference on Neural
  Information Processing Systems (NeurIPS)}}}}\ (\bibinfo {year} {2020})\
  \Eprint {http://arxiv.org/abs/2010.10450} {arXiv:2010.10450 [astro-ph.HE]}
  \BibitemShut {NoStop}%
\bibitem [{\citenamefont {Dinsmore}\ and\ \citenamefont
  {Slatyer}(2021)}]{Dinsmore:2021nip}%
  \BibitemOpen
  \bibfield  {author} {\bibinfo {author} {\bibfnamefont {J.~T.}\ \bibnamefont
  {Dinsmore}}\ and\ \bibinfo {author} {\bibfnamefont {T.~R.}\ \bibnamefont
  {Slatyer}},\ }\href@noop {} {\  (\bibinfo {year} {2021})},\ \Eprint
  {http://arxiv.org/abs/2112.09699} {arXiv:2112.09699 [astro-ph.HE]}
  \BibitemShut {NoStop}%
\bibitem [{\citenamefont {Brehmer}\ \emph
  {et~al.}(2018{\natexlab{a}})\citenamefont {Brehmer}, \citenamefont {Cranmer},
  \citenamefont {Louppe},\ and\ \citenamefont {Pavez}}]{Brehmer:2018eca}%
  \BibitemOpen
  \bibfield  {author} {\bibinfo {author} {\bibfnamefont {J.}~\bibnamefont
  {Brehmer}}, \bibinfo {author} {\bibfnamefont {K.}~\bibnamefont {Cranmer}},
  \bibinfo {author} {\bibfnamefont {G.}~\bibnamefont {Louppe}}, \ and\ \bibinfo
  {author} {\bibfnamefont {J.}~\bibnamefont {Pavez}},\ }\href {\doibase
  10.1103/PhysRevD.98.052004} {\bibfield  {journal} {\bibinfo  {journal} {Phys.
  Rev. D}\ }\textbf {\bibinfo {volume} {98}},\ \bibinfo {pages} {052004}
  (\bibinfo {year} {2018}{\natexlab{a}})},\ \Eprint
  {http://arxiv.org/abs/1805.00020} {arXiv:1805.00020 [hep-ph]} \BibitemShut
  {NoStop}%
\bibitem [{\citenamefont {Brehmer}\ \emph
  {et~al.}(2020{\natexlab{a}})\citenamefont {Brehmer}, \citenamefont {Louppe},
  \citenamefont {Pavez},\ and\ \citenamefont {Cranmer}}]{Brehmer:2018hga}%
  \BibitemOpen
  \bibfield  {author} {\bibinfo {author} {\bibfnamefont {J.}~\bibnamefont
  {Brehmer}}, \bibinfo {author} {\bibfnamefont {G.}~\bibnamefont {Louppe}},
  \bibinfo {author} {\bibfnamefont {J.}~\bibnamefont {Pavez}}, \ and\ \bibinfo
  {author} {\bibfnamefont {K.}~\bibnamefont {Cranmer}},\ }\href {\doibase
  10.1073/pnas.1915980117} {\bibfield  {journal} {\bibinfo  {journal} {Proc.
  Nat. Acad. Sci.}\ }\textbf {\bibinfo {volume} {117}},\ \bibinfo {pages}
  {5242} (\bibinfo {year} {2020}{\natexlab{a}})},\ \Eprint
  {http://arxiv.org/abs/1805.12244} {arXiv:1805.12244 [stat.ML]} \BibitemShut
  {NoStop}%
\bibitem [{\citenamefont {Brehmer}\ \emph
  {et~al.}(2018{\natexlab{b}})\citenamefont {Brehmer}, \citenamefont {Cranmer},
  \citenamefont {Louppe},\ and\ \citenamefont {Pavez}}]{Brehmer:2018kdj}%
  \BibitemOpen
  \bibfield  {author} {\bibinfo {author} {\bibfnamefont {J.}~\bibnamefont
  {Brehmer}}, \bibinfo {author} {\bibfnamefont {K.}~\bibnamefont {Cranmer}},
  \bibinfo {author} {\bibfnamefont {G.}~\bibnamefont {Louppe}}, \ and\ \bibinfo
  {author} {\bibfnamefont {J.}~\bibnamefont {Pavez}},\ }\href {\doibase
  10.1103/PhysRevLett.121.111801} {\bibfield  {journal} {\bibinfo  {journal}
  {Phys. Rev. Lett.}\ }\textbf {\bibinfo {volume} {121}},\ \bibinfo {pages}
  {111801} (\bibinfo {year} {2018}{\natexlab{b}})},\ \Eprint
  {http://arxiv.org/abs/1805.00013} {arXiv:1805.00013 [hep-ph]} \BibitemShut
  {NoStop}%
\bibitem [{\citenamefont {Cranmer}\ \emph {et~al.}(2015)\citenamefont
  {Cranmer}, \citenamefont {Pavez},\ and\ \citenamefont
  {Louppe}}]{Cranmer:2015bka}%
  \BibitemOpen
  \bibfield  {author} {\bibinfo {author} {\bibfnamefont {K.}~\bibnamefont
  {Cranmer}}, \bibinfo {author} {\bibfnamefont {J.}~\bibnamefont {Pavez}}, \
  and\ \bibinfo {author} {\bibfnamefont {G.}~\bibnamefont {Louppe}},\
  }\href@noop {} {\  (\bibinfo {year} {2015})},\ \Eprint
  {http://arxiv.org/abs/1506.02169} {arXiv:1506.02169 [stat.AP]} \BibitemShut
  {NoStop}%
\bibitem [{\citenamefont {Hermans}\ \emph {et~al.}(2020)\citenamefont
  {Hermans}, \citenamefont {Begy},\ and\ \citenamefont
  {Louppe}}]{Hermans:2019ioj}%
  \BibitemOpen
  \bibfield  {author} {\bibinfo {author} {\bibfnamefont {J.}~\bibnamefont
  {Hermans}}, \bibinfo {author} {\bibfnamefont {V.}~\bibnamefont {Begy}}, \
  and\ \bibinfo {author} {\bibfnamefont {G.}~\bibnamefont {Louppe}},\ }\href
  {http://arxiv.org/abs/1903.04057} {\bibfield  {journal} {\bibinfo  {journal}
  {arXiv:1903.04057 [cs, stat]}\ } (\bibinfo {year} {2020})},\ \bibinfo {note}
  {arXiv: 1903.04057}\BibitemShut {NoStop}%
\bibitem [{\citenamefont {Miller}\ \emph {et~al.}(2020)\citenamefont {Miller},
  \citenamefont {Cole}, \citenamefont {Louppe},\ and\ \citenamefont
  {Weniger}}]{Miller:2020hua}%
  \BibitemOpen
  \bibfield  {author} {\bibinfo {author} {\bibfnamefont {B.~K.}\ \bibnamefont
  {Miller}}, \bibinfo {author} {\bibfnamefont {A.}~\bibnamefont {Cole}},
  \bibinfo {author} {\bibfnamefont {G.}~\bibnamefont {Louppe}}, \ and\ \bibinfo
  {author} {\bibfnamefont {C.}~\bibnamefont {Weniger}},\ }\href@noop {} {\
  (\bibinfo {year} {2020})},\ \Eprint {http://arxiv.org/abs/2011.13951}
  {arXiv:2011.13951 [astro-ph.IM]} \BibitemShut {NoStop}%
\bibitem [{\citenamefont {Miller}\ \emph {et~al.}(2021)\citenamefont {Miller},
  \citenamefont {Cole}, \citenamefont {Forr\'e}, \citenamefont {Louppe},\ and\
  \citenamefont {Weniger}}]{Miller:2021hys}%
  \BibitemOpen
  \bibfield  {author} {\bibinfo {author} {\bibfnamefont {B.~K.}\ \bibnamefont
  {Miller}}, \bibinfo {author} {\bibfnamefont {A.}~\bibnamefont {Cole}},
  \bibinfo {author} {\bibfnamefont {P.}~\bibnamefont {Forr\'e}}, \bibinfo
  {author} {\bibfnamefont {G.}~\bibnamefont {Louppe}}, \ and\ \bibinfo {author}
  {\bibfnamefont {C.}~\bibnamefont {Weniger}},\ }\href {\doibase
  10.5281/zenodo.5043707} {\  (\bibinfo {year} {2021}),\
  10.5281/zenodo.5043707},\ \Eprint {http://arxiv.org/abs/2107.01214}
  {arXiv:2107.01214 [stat.ML]} \BibitemShut {NoStop}%
\bibitem [{\citenamefont {{Winkler}}\ \emph {et~al.}(2019)\citenamefont
  {{Winkler}}, \citenamefont {{Worrall}}, \citenamefont {{Hoogeboom}},\ and\
  \citenamefont {{Welling}}}]{2019arXiv191200042W}%
  \BibitemOpen
  \bibfield  {author} {\bibinfo {author} {\bibfnamefont {C.}~\bibnamefont
  {{Winkler}}}, \bibinfo {author} {\bibfnamefont {D.}~\bibnamefont
  {{Worrall}}}, \bibinfo {author} {\bibfnamefont {E.}~\bibnamefont
  {{Hoogeboom}}}, \ and\ \bibinfo {author} {\bibfnamefont {M.}~\bibnamefont
  {{Welling}}},\ }\href@noop {} {\  (\bibinfo {year} {2019})},\ \Eprint
  {http://arxiv.org/abs/1912.00042} {arXiv:1912.00042 [cs.LG]} \BibitemShut
  {NoStop}%
\bibitem [{\citenamefont {Papamakarios}\ \emph
  {et~al.}(2019{\natexlab{b}})\citenamefont {Papamakarios}, \citenamefont
  {Sterratt},\ and\ \citenamefont {Murray}}]{pmlr-v89-papamakarios19a}%
  \BibitemOpen
  \bibfield  {author} {\bibinfo {author} {\bibfnamefont {G.}~\bibnamefont
  {Papamakarios}}, \bibinfo {author} {\bibfnamefont {D.}~\bibnamefont
  {Sterratt}}, \ and\ \bibinfo {author} {\bibfnamefont {I.}~\bibnamefont
  {Murray}},\ }in\ \href
  {https://proceedings.mlr.press/v89/papamakarios19a.html} {\emph {\bibinfo
  {booktitle} {Proceedings of the Twenty-Second International Conference on
  Artificial Intelligence and Statistics}}},\ \bibinfo {series} {Proceedings of
  Machine Learning Research}, Vol.~\bibinfo {volume} {89},\ \bibinfo {editor}
  {edited by\ \bibinfo {editor} {\bibfnamefont {K.}~\bibnamefont {Chaudhuri}}\
  and\ \bibinfo {editor} {\bibfnamefont {M.}~\bibnamefont {Sugiyama}}}\
  (\bibinfo  {publisher} {PMLR},\ \bibinfo {year} {2019})\ pp.\ \bibinfo
  {pages} {837--848}\BibitemShut {NoStop}%
\bibitem [{\citenamefont {Brehmer}\ \emph
  {et~al.}(2020{\natexlab{b}})\citenamefont {Brehmer}, \citenamefont {Kling},
  \citenamefont {Espejo},\ and\ \citenamefont {Cranmer}}]{Brehmer:2019xox}%
  \BibitemOpen
  \bibfield  {author} {\bibinfo {author} {\bibfnamefont {J.}~\bibnamefont
  {Brehmer}}, \bibinfo {author} {\bibfnamefont {F.}~\bibnamefont {Kling}},
  \bibinfo {author} {\bibfnamefont {I.}~\bibnamefont {Espejo}}, \ and\ \bibinfo
  {author} {\bibfnamefont {K.}~\bibnamefont {Cranmer}},\ }\href {\doibase
  10.1007/s41781-020-0035-2} {\bibfield  {journal} {\bibinfo  {journal}
  {Comput. Softw. Big Sci.}\ }\textbf {\bibinfo {volume} {4}},\ \bibinfo
  {pages} {3} (\bibinfo {year} {2020}{\natexlab{b}})},\ \Eprint
  {http://arxiv.org/abs/1907.10621} {arXiv:1907.10621 [hep-ph]} \BibitemShut
  {NoStop}%
\bibitem [{\citenamefont {Stoye}\ \emph {et~al.}(2018)\citenamefont {Stoye},
  \citenamefont {Brehmer}, \citenamefont {Louppe}, \citenamefont {Pavez},\ and\
  \citenamefont {Cranmer}}]{Stoye:2018ovl}%
  \BibitemOpen
  \bibfield  {author} {\bibinfo {author} {\bibfnamefont {M.}~\bibnamefont
  {Stoye}}, \bibinfo {author} {\bibfnamefont {J.}~\bibnamefont {Brehmer}},
  \bibinfo {author} {\bibfnamefont {G.}~\bibnamefont {Louppe}}, \bibinfo
  {author} {\bibfnamefont {J.}~\bibnamefont {Pavez}}, \ and\ \bibinfo {author}
  {\bibfnamefont {K.}~\bibnamefont {Cranmer}},\ }in\ \href@noop {} {\emph
  {\bibinfo {booktitle} {{Machine Learning and the Physical Sciences Workshop
  at the 33rd Conference on Neural Information Processing Systems
  (NeurIPS)}}}}\ (\bibinfo {year} {2018})\ \Eprint
  {http://arxiv.org/abs/1808.00973} {arXiv:1808.00973 [stat.ML]} \BibitemShut
  {NoStop}%
\bibitem [{\citenamefont {Louppe}\ \emph {et~al.}(2016)\citenamefont {Louppe},
  \citenamefont {Kagan},\ and\ \citenamefont {Cranmer}}]{Louppe:2016ylz}%
  \BibitemOpen
  \bibfield  {author} {\bibinfo {author} {\bibfnamefont {G.}~\bibnamefont
  {Louppe}}, \bibinfo {author} {\bibfnamefont {M.}~\bibnamefont {Kagan}}, \
  and\ \bibinfo {author} {\bibfnamefont {K.}~\bibnamefont {Cranmer}},\
  }\href@noop {} {\  (\bibinfo {year} {2016})},\ \Eprint
  {http://arxiv.org/abs/1611.01046} {arXiv:1611.01046 [stat.ML]} \BibitemShut
  {NoStop}%
\bibitem [{\citenamefont {Kasieczka}\ and\ \citenamefont
  {Shih}(2020)}]{Kasieczka:2020yyl}%
  \BibitemOpen
  \bibfield  {author} {\bibinfo {author} {\bibfnamefont {G.}~\bibnamefont
  {Kasieczka}}\ and\ \bibinfo {author} {\bibfnamefont {D.}~\bibnamefont
  {Shih}},\ }\href {\doibase 10.1103/PhysRevLett.125.122001} {\bibfield
  {journal} {\bibinfo  {journal} {Phys. Rev. Lett.}\ }\textbf {\bibinfo
  {volume} {125}},\ \bibinfo {pages} {122001} (\bibinfo {year} {2020})},\
  \Eprint {http://arxiv.org/abs/2001.05310} {arXiv:2001.05310 [hep-ph]}
  \BibitemShut {NoStop}%
\bibitem [{\citenamefont {Price-Whelan}\ \emph {et~al.}(2018)\citenamefont
  {Price-Whelan} \emph {et~al.}}]{Price-Whelan:2018hus}%
  \BibitemOpen
  \bibfield  {author} {\bibinfo {author} {\bibfnamefont {A.~M.}\ \bibnamefont
  {Price-Whelan}} \emph {et~al.},\ }\href {\doibase 10.3847/1538-3881/aabc4f}
  {\bibfield  {journal} {\bibinfo  {journal} {Astron. J.}\ }\textbf {\bibinfo
  {volume} {156}},\ \bibinfo {pages} {123} (\bibinfo {year} {2018})},\ \Eprint
  {http://arxiv.org/abs/1801.02634} {arXiv:1801.02634} \BibitemShut {NoStop}%
\bibitem [{\citenamefont {Robitaille}\ \emph {et~al.}(2013)\citenamefont
  {Robitaille} \emph {et~al.}}]{Robitaille:2013mpa}%
  \BibitemOpen
  \bibfield  {author} {\bibinfo {author} {\bibfnamefont {T.~P.}\ \bibnamefont
  {Robitaille}} \emph {et~al.} (\bibinfo {collaboration} {Astropy}),\ }\href
  {\doibase 10.1051/0004-6361/201322068} {\bibfield  {journal} {\bibinfo
  {journal} {Astron. Astrophys.}\ }\textbf {\bibinfo {volume} {558}},\ \bibinfo
  {pages} {A33} (\bibinfo {year} {2013})},\ \Eprint
  {http://arxiv.org/abs/1307.6212} {arXiv:1307.6212 [astro-ph.IM]} \BibitemShut
  {NoStop}%
\bibitem [{\citenamefont {Lewis}(2019)}]{Lewis:2019xzd}%
  \BibitemOpen
  \bibfield  {author} {\bibinfo {author} {\bibfnamefont {A.}~\bibnamefont
  {Lewis}},\ }\href {https://getdist.readthedocs.io} {\  (\bibinfo {year}
  {2019})},\ \Eprint {http://arxiv.org/abs/1910.13970} {arXiv:1910.13970
  [astro-ph.IM]} \BibitemShut {NoStop}%
\bibitem [{\citenamefont {{Perez}}\ and\ \citenamefont
  {{Granger}}(2007)}]{PER-GRA:2007}%
  \BibitemOpen
  \bibfield  {author} {\bibinfo {author} {\bibfnamefont {F.}~\bibnamefont
  {{Perez}}}\ and\ \bibinfo {author} {\bibfnamefont {B.~E.}\ \bibnamefont
  {{Granger}}},\ }\href {\doibase 10.1109/MCSE.2007.53} {\bibfield  {journal}
  {\bibinfo  {journal} {Computing in Science and Engineering}\ }\textbf
  {\bibinfo {volume} {9}},\ \bibinfo {pages} {21} (\bibinfo {year}
  {2007})}\BibitemShut {NoStop}%
\bibitem [{\citenamefont {Kluyver}\ \emph {et~al.}(2016)\citenamefont {Kluyver}
  \emph {et~al.}}]{Kluyver2016JupyterN}%
  \BibitemOpen
  \bibfield  {author} {\bibinfo {author} {\bibfnamefont {T.}~\bibnamefont
  {Kluyver}} \emph {et~al.},\ }in\ \href@noop {} {\emph {\bibinfo {booktitle}
  {ELPUB}}}\ (\bibinfo {year} {2016})\BibitemShut {NoStop}%
\bibitem [{\citenamefont {Hunter}(2007)}]{Hunter:2007}%
  \BibitemOpen
  \bibfield  {author} {\bibinfo {author} {\bibfnamefont {J.~D.}\ \bibnamefont
  {Hunter}},\ }\href@noop {} {\bibfield  {journal} {\bibinfo  {journal}
  {Computing In Science \& Engineering}\ }\textbf {\bibinfo {volume} {9}},\
  \bibinfo {pages} {90} (\bibinfo {year} {2007})}\BibitemShut {NoStop}%
\bibitem [{\citenamefont {Chen}\ \emph {et~al.}(2020)\citenamefont {Chen} \emph
  {et~al.}}]{10.1145/3399579.3399867}%
  \BibitemOpen
  \bibfield  {author} {\bibinfo {author} {\bibfnamefont {A.}~\bibnamefont
  {Chen}} \emph {et~al.},\ }in\ \href {\doibase 10.1145/3399579.3399867} {\emph
  {\bibinfo {booktitle} {Proceedings of the Fourth International Workshop on
  Data Management for End-to-End Machine Learning}}},\ \bibinfo {series and
  number} {DEEM'20}\ (\bibinfo  {publisher} {Association for Computing
  Machinery},\ \bibinfo {address} {New York, NY, USA},\ \bibinfo {year}
  {2020})\BibitemShut {NoStop}%
\bibitem [{\citenamefont {Durkan}\ \emph {et~al.}(2020)\citenamefont {Durkan},
  \citenamefont {Bekasov}, \citenamefont {Murray},\ and\ \citenamefont
  {Papamakarios}}]{nflows}%
  \BibitemOpen
  \bibfield  {author} {\bibinfo {author} {\bibfnamefont {C.}~\bibnamefont
  {Durkan}}, \bibinfo {author} {\bibfnamefont {A.}~\bibnamefont {Bekasov}},
  \bibinfo {author} {\bibfnamefont {I.}~\bibnamefont {Murray}}, \ and\ \bibinfo
  {author} {\bibfnamefont {G.}~\bibnamefont {Papamakarios}},\ }\href {\doibase
  10.5281/zenodo.4296287} {\enquote {\bibinfo {title} {{nflows}: normalizing
  flows in {PyTorch}},}\ } (\bibinfo {year} {2020})\BibitemShut {NoStop}%
\bibitem [{\citenamefont {Harris}\ \emph {et~al.}(2020)\citenamefont {Harris}
  \emph {et~al.}}]{harris2020array}%
  \BibitemOpen
  \bibfield  {author} {\bibinfo {author} {\bibfnamefont {C.~R.}\ \bibnamefont
  {Harris}} \emph {et~al.},\ }\href {\doibase 10.1038/s41586-020-2649-2}
  {\bibfield  {journal} {\bibinfo  {journal} {Nature}\ }\textbf {\bibinfo
  {volume} {585}},\ \bibinfo {pages} {357} (\bibinfo {year}
  {2020})}\BibitemShut {NoStop}%
\bibitem [{\citenamefont {McKinney}(2010)}]{pandas:2010}%
  \BibitemOpen
  \bibfield  {author} {\bibinfo {author} {\bibfnamefont {W.}~\bibnamefont
  {McKinney}},\ }in\ \href@noop {} {\emph {\bibinfo {booktitle} {Proceedings of
  the 9th Python in Science Conference}}},\ \bibinfo {editor} {edited by\
  \bibinfo {editor} {\bibfnamefont {S.}~\bibnamefont {van~der Walt}}\ and\
  \bibinfo {editor} {\bibfnamefont {J.}~\bibnamefont {Millman}}}\ (\bibinfo
  {year} {2010})\ pp.\ \bibinfo {pages} {51 -- 56}\BibitemShut {NoStop}%
\bibitem [{\citenamefont {Defferrard}\ \emph {et~al.}(2017)\citenamefont
  {Defferrard}, \citenamefont {Martin}, \citenamefont {Pena},\ and\
  \citenamefont {Perraudin}}]{michael_defferrard_2017_1003158}%
  \BibitemOpen
  \bibfield  {author} {\bibinfo {author} {\bibfnamefont {M.}~\bibnamefont
  {Defferrard}}, \bibinfo {author} {\bibfnamefont {L.}~\bibnamefont {Martin}},
  \bibinfo {author} {\bibfnamefont {R.}~\bibnamefont {Pena}}, \ and\ \bibinfo
  {author} {\bibfnamefont {N.}~\bibnamefont {Perraudin}},\ }\href {\doibase
  10.5281/zenodo.1003158} {\enquote {\bibinfo {title} {{PyGSP: Graph Signal
  Processing in Python}},}\ } (\bibinfo {year} {2017})\BibitemShut {NoStop}%
\bibitem [{\citenamefont {Bingham}\ \emph {et~al.}(2019)\citenamefont {Bingham}
  \emph {et~al.}}]{bingham2019pyro}%
  \BibitemOpen
  \bibfield  {author} {\bibinfo {author} {\bibfnamefont {E.}~\bibnamefont
  {Bingham}} \emph {et~al.},\ }\href {http://jmlr.org/papers/v20/18-403.html}
  {\bibfield  {journal} {\bibinfo  {journal} {J. Mach. Learn. Res.}\ }\textbf
  {\bibinfo {volume} {20}},\ \bibinfo {pages} {28:1} (\bibinfo {year}
  {2019})}\BibitemShut {NoStop}%
\bibitem [{\citenamefont {Paszke}\ \emph {et~al.}(2019)\citenamefont {Paszke}
  \emph {et~al.}}]{NEURIPS2019_9015}%
  \BibitemOpen
  \bibfield  {author} {\bibinfo {author} {\bibfnamefont {A.}~\bibnamefont
  {Paszke}} \emph {et~al.},\ }in\ \href
  {http://papers.neurips.cc/paper/9015-pytorch-an-imperative-style-high-performance-deep-learning-library.pdf}
  {\emph {\bibinfo {booktitle} {Advances in Neural Information Processing
  Systems 32}}},\ \bibinfo {editor} {edited by\ \bibinfo {editor}
  {\bibfnamefont {H.}~\bibnamefont {Wallach}}, \bibinfo {editor} {\bibfnamefont
  {H.}~\bibnamefont {Larochelle}}, \bibinfo {editor} {\bibfnamefont
  {A.}~\bibnamefont {Beygelzimer}}, \bibinfo {editor} {\bibfnamefont
  {F.}~\bibnamefont {d\textquotesingle Alch\'{e}-Buc}}, \bibinfo {editor}
  {\bibfnamefont {E.}~\bibnamefont {Fox}}, \ and\ \bibinfo {editor}
  {\bibfnamefont {R.}~\bibnamefont {Garnett}}}\ (\bibinfo  {publisher} {Curran
  Associates, Inc.},\ \bibinfo {year} {2019})\ pp.\ \bibinfo {pages}
  {8024--8035}\BibitemShut {NoStop}%
\bibitem [{\citenamefont {Fey}\ and\ \citenamefont
  {Lenssen}(2019)}]{Fey/Lenssen/2019}%
  \BibitemOpen
  \bibfield  {author} {\bibinfo {author} {\bibfnamefont {M.}~\bibnamefont
  {Fey}}\ and\ \bibinfo {author} {\bibfnamefont {J.~E.}\ \bibnamefont
  {Lenssen}},\ }in\ \href@noop {} {\emph {\bibinfo {booktitle} {ICLR Workshop
  on Representation Learning on Graphs and Manifolds}}}\ (\bibinfo {year}
  {2019})\ \Eprint {http://arxiv.org/abs/1903.02428} {arXiv:1903.02428 [cs.LG]}
  \BibitemShut {NoStop}%
\bibitem [{\citenamefont {Falcon}\ \emph {et~al.}(2020)\citenamefont {Falcon}
  \emph {et~al.}}]{william_falcon_2020_3828935}%
  \BibitemOpen
  \bibfield  {author} {\bibinfo {author} {\bibfnamefont {W.}~\bibnamefont
  {Falcon}} \emph {et~al.},\ }\href {\doibase 10.5281/zenodo.3828935} {\enquote
  {\bibinfo {title} {Pytorchlightning/pytorch-lightning: 0.7.6 release},}\ }
  (\bibinfo {year} {2020})\BibitemShut {NoStop}%
\bibitem [{\citenamefont {Waskom}\ \emph {et~al.}(2017)\citenamefont {Waskom}
  \emph {et~al.}}]{seaborn}%
  \BibitemOpen
  \bibfield  {author} {\bibinfo {author} {\bibfnamefont {M.}~\bibnamefont
  {Waskom}} \emph {et~al.},\ }\href {\doibase 10.5281/zenodo.883859} {\enquote
  {\bibinfo {title} {mwaskom/seaborn: v0.8.1 (september 2017)},}\ } (\bibinfo
  {year} {2017})\BibitemShut {NoStop}%
\bibitem [{\citenamefont {Tejero-Cantero}\ \emph {et~al.}(2020)\citenamefont
  {Tejero-Cantero} \emph {et~al.}}]{tejero-cantero2020sbi}%
  \BibitemOpen
  \bibfield  {author} {\bibinfo {author} {\bibfnamefont {A.}~\bibnamefont
  {Tejero-Cantero}} \emph {et~al.},\ }\href {\doibase 10.21105/joss.02505}
  {\bibfield  {journal} {\bibinfo  {journal} {Journal of Open Source Software}\
  }\textbf {\bibinfo {volume} {5}},\ \bibinfo {pages} {2505} (\bibinfo {year}
  {2020})}\BibitemShut {NoStop}%
\bibitem [{\citenamefont {Pedregosa}\ \emph {et~al.}(2011)\citenamefont
  {Pedregosa} \emph {et~al.}}]{JMLR:v12:pedregosa11a}%
  \BibitemOpen
  \bibfield  {author} {\bibinfo {author} {\bibfnamefont {F.}~\bibnamefont
  {Pedregosa}} \emph {et~al.},\ }\href
  {http://jmlr.org/papers/v12/pedregosa11a.html} {\bibfield  {journal}
  {\bibinfo  {journal} {Journal of Machine Learning Research}\ }\textbf
  {\bibinfo {volume} {12}},\ \bibinfo {pages} {2825} (\bibinfo {year}
  {2011})}\BibitemShut {NoStop}%
\bibitem [{\citenamefont {{Virtanen}}\ \emph {et~al.}(2020)\citenamefont
  {{Virtanen}} \emph {et~al.}}]{2020SciPy-NMeth}%
  \BibitemOpen
  \bibfield  {author} {\bibinfo {author} {\bibfnamefont {P.}~\bibnamefont
  {{Virtanen}}} \emph {et~al.},\ }\href {\doibase
  https://doi.org/10.1038/s41592-019-0686-2} {\bibfield  {journal} {\bibinfo
  {journal} {Nature Methods}\ } (\bibinfo {year} {2020}),\
  https://doi.org/10.1038/s41592-019-0686-2}\BibitemShut {NoStop}%
\bibitem [{\citenamefont {da~Costa-Luis}\ \emph {et~al.}(2021)\citenamefont
  {da~Costa-Luis} \emph {et~al.}}]{casper_da_costa_luis_2021_5517697}%
  \BibitemOpen
  \bibfield  {author} {\bibinfo {author} {\bibfnamefont {C.}~\bibnamefont
  {da~Costa-Luis}} \emph {et~al.},\ }\href {\doibase 10.5281/zenodo.5517697}
  {\enquote {\bibinfo {title} {{tqdm: A fast, Extensible Progress Bar for
  Python and CLI}},}\ } (\bibinfo {year} {2021})\BibitemShut {NoStop}%
\end{thebibliography}%

\end{document}